\documentclass[twocolumn,preprintnumbers,amsmath,amssymb]{revtex4}

\usepackage{amsmath}    
\usepackage{graphicx}   

\begin{document}

\title{Time-delay of classical and quantum scattering processes: a conceptual overview and a general definition}
\author{Massimiliano Sassoli de Bianchi}
\affiliation{Laboratorio di Autoricerca di Base, 6914 Carona, Switzerland}\date{\today}
\email{autoricerca@gmail.com}   

\begin{abstract}
\noindent
We present a step by step introduction to the notion of time-delay in classical and quantum mechanics, with the aim of clarifying its foundation at a conceptual level. In doing so, we motivate the introduction of the concepts of ``fuzzy'' and ``free-flight'' sojourn times, that we use to provide the most general possible definition for the quantum time-delay, valid for simple and multichannel scattering systems, with or without conditions on the observation of the scattering particle, and for incoming wave packets whose energy can be smeared out or sharply peaked (fixed energy). We conclude our conceptual analysis by presenting what we think is the right interpretation of the concepts of sojourn and delay times in quantum mechanics, explaining why, in ultimate analysis, they shouldn't be called ``times.''  
\end{abstract}

\maketitle


\section{Introduction}

Time-delay is a classical concept everyone is familiar with. For instance, everybody agree in saying that a train has some delay when it doesn't reach  destination in time, where ``in time'' means ``the arrival time written in the train schedule.'' This means that the concept of time-delay hints to a measure of an arrival time at some given place in space, and its comparison with a reference arrival time which, by definition, is assumed to correspond to a situation of zero delay or, which is equivalent, of zero advance. In other terms, time-delay is a relative quantity. It is also a conventional quantity by reason of the infinitely many a priori possible different choices for a comparison reference time.

In physics one usually associates a time-delay to a scattering particle moving in presence of a force field. Time-delay then measures the excess or defect of time the particle spends in the interaction region, when its movement is compared to that of a free particle, subject to similar initial or final conditions. 

The knowledge of time-delay clearly informs us about the nature of the interaction. Generally speaking, a positive time-delay corresponds to an effect of deceleration; a positive large time-delay corresponds to the formation of a metastable quasi-bound state (resonance); an infinite positive time-delay corresponds to the capture of the particle by the interaction; finally, a negative time-delay indicates that the particle has been accelerated by the effects of the interaction. 

The main historical motivation in studying time-delay was to develop a formalism allowing for a general description and analysis of the different resonant scattering phenomena, as well as to provide an information that would be complementary with respect to that contained in the usual differential scattering cross-section (or transmission and reflection probabilities, in simple one-dimensional systems). 

Another important motivation was to clarify, by means of a proper time-delay operator, the status of so-called time-energy uncertainty relations, whose derivation was notoriously unsatisfactory because of the lack of a proper self-adjoint time-operator in (conventional) quantum mechanics (let us however note that this historical difficulty has been overcome by approaches justifying the association of time with a quantum observable by exploiting the more general properties of maximal Hermitian operators, in the case of continuous energy spectra, and of quasi-self-adjoint operators, in the case of discrete energy spectra~\cite{Allc, Wern, Kong, Muga-Lea, Olk, Olk2, Olk3, Rec, Rec2}; see also our comments in Sec.~\ref{The quantization problem}).

The first attempt to formalize the concept of time-delay in quantum scattering can be traced back to a seminal paper of Wigner~\cite{Wig} (who also refers to an earlier unpublished thesis by Eisenbud~\cite{Eis}), who in the ambit of a single channel scattering process, and considering wave packets very narrow in energy (quasimonochromatic), succeeded to heuristically derive a formula connecting the time-delay concept with the energy derivative of the scattering phase shift. 

This formula, today known as the \textsl{Eisenbud-Wigner} formula, was subsequently generalized by Smith~\cite{Smi} to a multichannel scattering context. However, the definition adopted by Smith only used the unnormalizable steady-state solutions of the time-independent Schroedinger equation, and therefore was not fully transparent from a physical point of view, seeing that the stationary formalism is obtained from the time-dependent one by precisely discarding the time variable.  

To have available a physically meaningful and mathematically precise definition of time-delay, in terms of proper normalizable wave-packets of arbitrary shape, one has to wait the time-dependent methods introduced by Goldberger and Watson~\cite{Gold} and then further developed by Jauch and collaborators~\cite{Jau-Mar, Jau-Sin-Mis}. In these years, the formal equivalence between the time-dependent and time-independent formulations was also established by Boll\'e and Osborn~\cite{Bol-Osb}, and following these important preliminary works an entire line of research took hold, with the aim of better understanding this important notion. 

There were essentially two distinct problems. The first one was to establish the most general conditions for the existence of the time-delay, defined as the difference between \emph{sojourn times}, i.e., between the time spent by the interacting and free reference particles inside a given spatial region, in the limit where the latter covers the entire physical space. Such a limit, and its equivalence to Eisenbud-Wigner expression, presented a very challenging mathematical problem, that gave birth to an entire line of research in mathematical physics. 

The other problem was the one of finding  generalizations of the time-delay definition, by considering more general scattering systems, like multichannel processes, as well as more general conditions of observation of the scattering particles, as they occur when the scattering is observed by counters in a differential cross-section measurement (angular time-delay); see Refs.~\cite{Mar2,Sas1} and the references cited therein; see also Refs.~\cite{Ger, Rich} for two examples of more recent works on the subject. 

In parallel to this rather abstract and mathematically oriented line of research, other physicists analyzed time-delay and related notions by employing a more pragmatic approach: the utilization of clocks. Again, it is Wigner, together with Salecker, that initiated this line of investigation, with his analysis of the accuracy limits in the reading of a microscopic clock, as a consequence of the quantum nature of the clock pointer's variable~\cite{Sale}. 

Probably, the better known example of a microscopic quantum clock is the so-called Larmor clock, that was firstly introduced by Baz'~\cite{Baz}, who had the idea to exploit the mechanism of spin precession in a homogeneous magnetic field, as a mean to equip quantum particles with a portable stopwatch, that would be activated when they enter a (weak homogeneous) magnetic field, and subsequently deactivated when they leave it. By reading the spin-clock, one could measure then, at least in principle, the amount of time a particle has sojourned inside the field region, and this could obviously be used to calculate sojourn times and, consequently, time delays. 

The preliminary analysis of Baz' was later followed by a number of works, through which a number of different clock models have been proposed and analyzed. This research developed rather impressively in more recent times, stimulated by the necessity of better understanding the electronic transport properties of semiconductor heterostructures, also in view of the possibility of designing high-speed devices based on the tunneling time phenomenon. 

In fact, since an old paper by Hartman~\cite{Har}, it was clear that the tunneling phenomenon can take place in an extremely short time. But how much time does exactly a tunneling particle spend in the barrier region? This puzzling and controversial question, known as the tunneling time problem, was at the origin of a great theoretical effort which resuscitated, in the last decades, the very old and apparently academic study of  one-dimensional quantum scattering systems (seeing that semiconductor mesoscopic structures can be modelized by effective one-dimensional systems). 

Being extremely difficult to identify only a few representative articles in this huge literature, we refer the interested reader to the review papers~\cite{Hau, Olk, Olk2, Olk3, Lan}, as well as to two recent multi-author volumes on the subject~\cite{Muga, Muga-vol2}. However, as we will explain in this paper, the notion of tunneling time (and more generally of transmission time) is rather different from the one of time-delay (and more specifically of angular time-delay), which was the main concern of the more mathematically oriented line of research we have previously mentioned and that, apparently, has been pretty much overlooked in the tunneling time controversy, although with some notable exceptions~\cite{JawWard, JawWard2, JawWard3, JawWard4, Sas2,Mar3}. 

The purpose of the present work is certainly not to illustrate, in a comprehensive way, this broad and multifaceted subject, that we have just outlined in this introduction. For this, many volumes would be required. What we shall do, instead, is to present a pedestrian, step by step introduction to the notion of time-delay, with the aim of providing the reader the necessary background to address this field of research with a clear mind regarding its conceptual foundations. In doing so we shall also present some new concepts and results, like the definition of the concepts of ``fuzzy'' and ``free-flight'' sojourn times, that we use to provide the most general possible definition for the quantum time-delay. The article will be organized as follows. 

In Sec.~\ref{Time-delay as a difference of arrival times}, we consider a classical point-like particle and define the notion of \textit{global time-delay} as the limit of a difference of arrival times, and derive for it an explicit formula. In Sec.~\ref{Time-delay as a difference of sojourn times}, we consider an alternative definition, using sojourn times instead of arrival times, then show their equivalence. 

In Sec.~\ref{Invariance under space translations}, we study the invariance of global time-delay under space translations. This will allow us to derive more general formulae, making explicit the inherent conventional character of the time-delay notion. 

In Sec.~\ref{A further definition of global time-delay}, we propose an equivalent definition for the global time-delay, in terms of a newly introduced ``free-flight'' reference sojourn time, and in Sec.~\ref{A simple interpretation} we conclude our classical analysis by giving a simple interpretation of the global time-delay formula, as a difference of two specific finite arrival times. 

In Sec.~\ref{The quantization problem}, we address the problem of quantization of the classical time-delay formula. We show that a quantum analogue can be obtained straightaway, by applying the standard quantization rule to the classical expression. However, such a procedure is not without conceptual difficulties, due to the absence of a self-adjoint arrival time operator in quantum mechanics (QM). 

For this reason, in Sec.~\ref{Sojourn time operators} we come back to sojourn times, that contrary to arrival times can be represented in QM by bona fide self-adjoint operators. A quantum sojourn time has to be understood as an average quantity, defined as a sum over probabilities of presence. Such a definition is perfectly consistent and reduces to the usual classical definition when the particle's dynamics is known, as we explicitly show in a simple one-dimensional example. 

In Sec.~\ref{Physical clocks}, we further investigate the concept of sojourn time by studying its relation to physical clocks. We consider three different paradigmatic examples of clocks: the Larmor clock, the dissipative clock and the energy clock (the latter, as far as we know, hasn't been considered so far in the literature) and show that the reading of all of them is in perfect agreement with the abstract sojourn time definition. 

Having analyzed the conceptual foundation of the quantum sojourn time operator, we then proceed  by studying it in more explicit terms. In Sec.~\ref{Free sojourn time}, after showing that the free sojourn time operator commutes with the free Hamiltonian, and therefore does not participate to Heisenberg uncertainty relations, we derive an explicit expression for it in the one-dimensional case. 

In Sec.~\ref{Interferences}, we pursue our analysis of the quantum free sojourn time showing that, contrary to the classical case, it allows for the manifestation of interference effects. 

In Sec.~\ref{Interaction sojourn time}, we analyze the sojourn time in presence of an interaction, and always in the one-dimensional context we derive an explicit formula, that we then use, in Sec.~\ref{The time-delay limit}, to study the global time-delay limit and its relation to the Eisenbud-Wigner formula, which we then use in Sec.~\ref{Breit-Wigner} to illustrate the relation between time-delay and the energy-width of a resonance.

In Sec.~\ref{Fuzzy sojourn times}, we introduce the new concept of ``fuzzy sojourn time,'' which we show is free from the ``troublesome'' oscillating interference terms that are present in the standard definition of sojourn time, thus allowing to consistently derive the time-delay limit also at fixed energy. The main difference between standard and fuzzy sojourn time definitions is that the former asks the question regarding the localization of the quantum entity in very sharp terms (only admitting a ``yes'' or a ``no'' as possible answers), whereas the latter allows for an entire range of intermediate responses, according to the degree of certainty one can ascertain the belonging or non-belonging of the entity to a given region of space.

In Sec.~\ref{Time-delay in multichannel scattering}, we analyze the notion of time-delay in the more general context of multichannel systems. For this, following a brief introduction to the time-dependent formalism, we consider, in Sec.~\ref{Sojourn time on the quasi-energy shell}, the paradigmatic example of scattering by a symmetric time-periodic potential and derive an explicit formula for the sojourn time operator on the quasi-energy shell. We then use it, in Sec.~\ref{Time-delay on the quasi-energy shell}, to show that the existence of the time-delay limit necessitates (contrary to the static case) a symmetrized free reference sojourn time. 

In Sec.~\ref{Contitional time-delay}, we discuss the notion of conditional time-delay and the conceptual problems it presents. We show that although we cannot make sense of a  notion of conditional sojourn time in conventional quantum mechanics, we can nevertheless give a proper meaning to the one of conditional time-delay. We also show that, contrary to the unconditional case, such a definition necessarily requires an outgoing free reference sojourn time to remain consistent, then we derive for it an explicit formula that generalizes the one of Eisenbud-Wigner. 

In Sec.~\ref{The most general definition}, we provide what we believe is the most general possible definition of time-delay in quantum mechanics, valid for simple and multichannel scattering systems, with or without conditions on the observation of the scattering particle in the distant future, and for incoming wave packets whose energy can be smeared out or, instead, sharply peaked (fixed energy). For such a general definition to remain fully consistent, the use of a newly introduced notion of ``fuzzy free-flight'' sojourn time will be shown to be crucial. 

Finally, in Sec.~\ref{Discussion}, we present some concluding reflections regarding the conceptual status of the notion of sojourn time in quantum mechanics and of its physical interpretation, showing that, in ultimate analysis, it shouldn't be understood as a ``time'' of permanence, but as a quantifier of the \textit{total (spatial) availability} of a quantum entity in a given region of space. Accordingly, time-delay has to be re-interpreted as the \textit{total availability shift} of a quantum entity, as a consequence of the ``switching on'' of the interaction.

\section{Classical global time-delay}\label{Classical global time-delay}

\subsection{Time-delay as a difference of arrival times}\label{Time-delay as a difference of arrival times}

In classical mechanics, a notion of trajectory is available. Therefore, it is natural to define the time-delay associated to a scattering particle in terms of a difference of arrival times, exactly as we would do for the macroscopic entities populating our everyday life. For this, one can consider the times at which the scattering particle arrives at a distance $r$ from the origin. Taking $r$ sufficiently large, this happens exactly twice: once in the past, before it enters the interaction region, and once in the future, when it emerges from it. 

Let us denote these two arrival times by $t^{-}(r)$ and $t^{+}(r)$, respectively. Then, one can choose to compare them to the times associated with a free reference particle having  same (initial) condition as the scattering particle in the remote past, that we shall denote by $t_{\text{in}}^{0,-}(r)$ and $t_{\text{in}}^{0,+}(r)$, respectively. Accordingly, we can define the following time-differences, or \textit{local time-delays}:
\begin{subequations}
\begin{align}
\label{local-classical-arrival-in-}
\tau_{\text{in}}^-(r)&=t^-(r) - t_{\text{in}}^{0,-}(r) \\
\label{local-classical-arrival-in+}
\tau_{\text{in}}^+(r)&=t^+(r) - t_{\text{in}}^{0,+}(r).
\end{align}
\end{subequations}
In a similar way, one can also take as a reference the movement of a particle having same (final) condition as the scattering particle in the distant future, and denote by $t_{\text{out}}^{0,-}(r)$ and $t_{\text{out}}^{0,+}(r)$ the corresponding arrival times at a distance $r$ from the origin. Again, we can define the  following local time-delays:
\begin{subequations}
\begin{align}
\label{local-classical-arrival-out-}
\tau_{\text{out}}^-(r)&=  t_{\text{out}}^{0,-}(r)- t^-(r)\\
\label{local-classical-arrival-out+}
\tau_{\text{out}}^+(r)&= t_{\text{out}}^{0,+}(r)-t^+(r).
\end{align}
\end{subequations}

The next step is to study the limit of the above four quantities, as the distance $r\rightarrow\infty$, in order to obtain $r$-independent \textit{global} (instead of \textit{local}) time-delays. This limit is of course meaningful, as we are dealing here with scattering trajectories that behave as free trajectories far away from the interaction region. 

To this end, let $\left\{\textbf{q}(t),\textbf{p}(t)\right\}$ be the position and momentum of the scattering particle of mass $m$ (three-vectors are in bold type), where $\textbf{p}(t)=m\dot{\textbf{q}}(t)$, and $\textbf{q}(t)$ is the unique solution of the Newton's equation of motion $m\ddot{\textbf{q}}(t)=\textbf{F}(\textbf{q}(t))$, with  asymptotic form:
\begin{equation}
\label{asymptotic-classical}
\textbf{q}(t)=
\begin{cases}
\textbf{q}^0_{\text{in}}(t)=\textbf{q}_-+\textbf{p}_-t/m, & \text{$t\rightarrow -\infty$} \\
\textbf{q}^0_{\text{out}}(t)=\textbf{q}_++\textbf{p}_+t/m, & \text{$t\rightarrow +\infty$.} 
\end{cases}
\end{equation}
In (\ref{asymptotic-classical}) we have defined the asymptotic momenta $\textbf{p}_\pm=\lim_{t\rightarrow\pm\infty}\textbf{p}(t)$, and  $|\textbf{p}_-|=|\textbf{p}_+|=\sqrt{2mE}$, by energy conservation. 

To determine the arrival times $t^{\pm}(r)$, we set $|\textbf{q}(t)|=r$ in (\ref{asymptotic-classical}), and solve for $t$ as a function of $r$. In the limit $r\rightarrow\infty$ ($|t|\rightarrow\infty$), we find: 
\begin{subequations}
\begin{align}
\label{asymptotic-arrival-classical1}
t^{\pm}(r)&= \sqrt{\frac{m}{2E}}\,\hat{\textbf{p}}_{\pm}\left(r\hat{\textbf{q}}(t)-\textbf{q}_{\pm}\right)+ o(1) \\
\label{asymptotic-arrival-classical2}
&= \frac{1}{v}\left(\pm r-\hat{\textbf{p}}_{\pm}\textbf{q}_{\pm}\right)+ o(1),
\end{align}
\end{subequations}
where we have defined the unit length vectors $\hat{\textbf{p}}_\pm =\textbf{p}_\pm /|\textbf{p}_\pm|$, $\hat{\textbf{q}}(t)=\textbf{q}(t)/r$, $v=\sqrt{2E/m}$ is the scalar velocity, and we have used the fact that $\hat{\textbf{p}}_{\pm}\hat{\textbf{q}}(t)\rightarrow\pm1$ as $t\rightarrow\pm\infty$. 

Proceeding in the same way for the free incoming trajectory $\textbf{q}^0_{\text{in}}(t)$, we find:
\begin{equation}
\label{asymptotic-arrival-classical-free1}
t_{\text{in}}^{0,\pm}(r)= \frac{1}{v}\left(\pm r-\hat{\textbf{p}}_{-}\textbf{q}_{-}\right)+ o(1),
\end{equation}
whereas for the outgoing free trajectory $\textbf{q}^0_{\text{out}}(t)$, we get:
\begin{equation}
\label{asymptotic-arrival-classical-free2}
t_{\text{out}}^{0,\pm}(r)= \frac{1}{v}\left(\pm r-\hat{\textbf{p}}_{+}\textbf{q}_{+}\right)+ o(1).
\end{equation}
Inserting (\ref{asymptotic-arrival-classical2}), (\ref{asymptotic-arrival-classical-free1}) and (\ref{asymptotic-arrival-classical-free2}) into (\ref{local-classical-arrival-in-}), (\ref{local-classical-arrival-in+}), (\ref{local-classical-arrival-out-}) and (\ref{local-classical-arrival-out+}), we obtain:
\begin{equation}
\label{zero-time-delay}
\lim_{r\rightarrow\infty}\tau_{\text{in}}^-(r)=\lim_{r\rightarrow\infty}\tau_{\text{out}}^+(r)=0
\end{equation}
and
\begin{equation}
\label{global-classical-time-delay-limit}
\lim_{r\rightarrow\infty}\tau_{\text{in}}^+(r)=\lim_{r\rightarrow\infty}\tau_{\text{out}}^-(r)=\tau,
\end{equation}
with
\begin{equation}
\label{global-classical-time-delay-formula}
\tau=-\frac{1}{v}\left(\hat{\textbf{p}}_+\textbf{q}_+-\hat{\textbf{p}}_-\textbf{q}_-\right).
\end{equation}

The reason why the local time-delay $\tau_{\text{in}}^-(r)$ [respectively $\tau_{\text{out}}^+(r)$] tends to zero, as $r\rightarrow\infty$, is that, by definition, the incoming (respectively, the outgoing) free trajectory coincides with the scattering trajectory in the remote past (respectively, distant future) so that, as $r\rightarrow\infty$, it reaches the distance $r$ from the origin in the past (respectively, in the future) at the same time as the scattering particle.

On the other hand, to understand why the local time-delays $\tau_{\text{in}}^+(r)$ and $\tau_{\text{out}}^-(r)$ converge both to the same global time-delay limit $\tau$, as $r\rightarrow\infty$, one only needs to observe that if a time reversal transformation $t\rightarrow -t$ is performed, then the free outgoing trajectory becomes the free incoming trajectory, and vice versa, so that apart from a sign change [which is duly taken into account in the definition of $\tau_{\text{out}}^\pm(r)$] the outgoing and incoming free evolving particles do play an equivalent role in the definition of time-delay (this however, as we shall see in Sec.~\ref{Contitional time-delay}, is not any more true when dealing with the concept of conditional time-delay).

\subsection{Time-delay as a difference of sojourn times}\label{Time-delay as a difference of sojourn times}

Following the above analysis, we now introduce alternative, but equivalent, definitions of global time-delay, that will prove their usefulness in the sequel, when considering the quantum case. 

To start with, we observe that, as it is the case with position, we never measure time instants in absolute terms, but always in relative terms (i.e., we measure durations). When for instance we tell somebody that a train will arrive at the railway station at, say, 16:00, what we mean is that a time interval of 16 hours will have elapsed between the following two events: ``our watch indicates 00:00'' and ``the train arrives at the railway station.'' 

The reason why we usually forget to mention the first event is that we assume that our interlocutor's clock is duly synchronized with our, so that we share a same time origin. If, on the contrary, we suspect this not being the case, then we certainly need to make a more precise statement, making for instance explicit our time zone, or giving whatever other relevant information allowing the other observer to unambiguously determine the time origin with respect to which we have measured the train's arrival time. 

In the same way, when we say that a particle emerging from the scattering region arrives at the distance $r$ from the origin at time $t^+(r)$, what we truly mean, more exactly, is that a time interval $\Delta t^+(r) =t^+(r)-0$ has elapsed between the following two events: ``the laboratory clock indicates 00:00'' and ``the particle arrives at a distance $r$ from the origin, after having interacted with the force field.'' 

Of course, the choice of the time origin of the laboratory clock is completely arbitrary, and we are free to change it according to our preferences. For instance, we can choose to set the zero of the laboratory's clock in coincidence with the instant the scattering particle arrives at a distance $r$ from the origin ($r$ large) before it enters the interaction region. 

This choice amounts to consider a new inertial frame, specified by the following time-shift transformation: $t\rightarrow t-t_-(r)$. In this new inertial frame, the arrival time $t^+(r)$ becomes $t^+(r)\rightarrow t^+(r)-t^-(r)$. In other terms, in a frame of reference having the time origin at $t_-(r)$, the arrival time $t^+(r)$ becomes equal to the sojourn time: 
\begin{equation}
\label{sojourn time classical}
T(B_r)= t^+(r)-t^-(r),
\end{equation}
namely to the time spent by the scattering particle inside a ball $B_r$ of radius $r$, centered at the origin of the spatial system of coordinates. 

Obviously, time-delay being itself a difference of arrival times, it cannot be affected by a shift of the time origin. Therefore, we can also write for (\ref{local-classical-arrival-in+}):
\begin{eqnarray}
\tau_{\text{in}}^+(r)&=& \left[t^+(r)-t^-(r)\right]-\left[t_{\text{in}}^{0,+}(r)-t^-(r)\right]\nonumber \\
&=&T(B_r)-\left[t_{\text{in}}^{0,+}(r)-t_{\text{in}}^{0,-}(r)\right]\nonumber \\
& &+\left[t^-(r)-t_{\text{in}}^{0,-}(r)\right]\nonumber \\
\label{local-classical-arrival-in+bis}
&=&T(B_r)-T^0_{\text{in}}(B_r) +\tau_{\text{in}}^-(r),
\end{eqnarray}
where 
\begin{equation}
\label{free sojourn time classical in}
T^0_{\text{in}}(B_r)=t_{\text{in}}^{0,+}(r)-t_{\text{in}}^{0,-}(r)
\end{equation}
is the time spent by the free evolving incoming particle inside the ball $B_r$. According to (\ref{local-classical-arrival-in+bis}), we can now introduce another local time-delay
\begin{eqnarray}
\label{local-classical-sojourn-in}
\tau_{\text{in}}(r)&=&T(B_r)-T^0_{\text{in}}(B_r)\\
&=&\tau_{\text{in}}^+(r)-\tau_{\text{in}}^-(r),
\end{eqnarray}
defined as a difference of sojourn times, instead of a difference of arrival times. 

In a similar way, we can also define the local time-delay 
\begin{eqnarray}
\label{local-classical-sojourn-out}
\tau_{\text{out}}(r)&=&T(B_r)-T^0_{\text{out}}(B_r)\\
&=&\tau_{\text{out}}^-(r)-\tau_{\text{out}}^+(r),
\end{eqnarray}
defined with reference to the outgoing free sojourn time
\begin{equation}
\label{free sojourn time classical out}
T^0_{\text{out}}(B_r)=t_{\text{out}}^{0,+}(r)-t_{\text{out}}^{0,-}(r).
\end{equation}
According to (\ref{zero-time-delay}) and (\ref{global-classical-time-delay-limit}), it immediately follows that
\begin{equation}
\label{global-classical-time-delay-limit-bis}
\lim_{r\rightarrow\infty}\tau_{\text{in}}(r)=\lim_{r\rightarrow\infty}\tau_{\text{out}}(r)=\tau.
\end{equation}

In other terms, we have shown that the local time-delays (\ref{local-classical-sojourn-in}) and (\ref{local-classical-sojourn-out}), defined in terms of sojourn times, are classically equivalent to the local time-delays (\ref{local-classical-arrival-in+}) and (\ref{local-classical-arrival-out-}), defined in terms of arrival times, in the global time-delay limit $r\to\infty$. 

Of course, we can easily construct as many equivalent local time-delays expressions as we like, all converging to the same global time-delay limit (\ref{global-classical-time-delay-formula}), by simply combining together, in different ways, the four expressions (\ref{local-classical-sojourn-in}), (\ref{local-classical-sojourn-out}),(\ref{local-classical-arrival-in+}),(\ref{local-classical-arrival-out-}). For later purpose, we also define the following symmetrized local time-delay:
\begin{eqnarray}
\tau_\text{s}(r)&=& \frac{1}{2}\left[\tau_{\text{in}}(r)+\tau_{\text{out}}(r)\right]\nonumber \\
\label{local-classical-sojourn-in/out}
&=&T(B_r)-T^0_\text{s}(B_r),
\end{eqnarray}
where we have defined the symmetrized free reference sojourn time:
\begin{equation}
\label{free sojourn time classical in/out}
T^0_\text{s}(B_r)=\frac{1}{2}\left[T^0_{\text{in}}(B_r)+T^0_{\text{out}}(B_r)\right].
\end{equation}
Clearly, also in this case we have:
\begin{equation}
\label{global-classical-time-delay-limit-tris}
\lim_{r\rightarrow\infty}\tau_\text{s}(r)=\tau.
\end{equation}

\subsection{Invariance under space translations}\label{Invariance under space translations}

In Sec.~\ref{Time-delay as a difference of sojourn times}, we have observed that a translation of the time origin cannot affect the value taken by the time-delay. Let us now consider a translation of the origin of the spatial coordinate system to a point $\textbf{a}$, i.e., $\textbf{x}\rightarrow \textbf{x}-\textbf{a}$. Is time-delay affected by this transformation? 

To answer this question, we observe that when performing a spatial translation, momentums remain unchanged but positions are affected, so that $\textbf{q}_\pm\rightarrow\textbf{q}_\pm -\textbf{a}$, and (\ref{global-classical-time-delay-formula}) transforms to: 
\begin{equation}
\label{global-classical-time-delay-formula-shifted-origin}
\tau\rightarrow -\frac{1}{v}\left[\hat{\textbf{p}}_+\left(\textbf{q}_+-\textbf{a}\right)-\hat{\textbf{p}}_-\left(\textbf{q}_--\textbf{a}\right)\right].
\end{equation}
In other terms, the global time-delay formula (\ref{global-classical-time-delay-formula}) is not invariant under a spatial translation. 

This however should  not surprise us, as when we have defined the time-delay we have only considered arrival times at a distance $r$ from the \textit{origin} or, which is equivalent, times of sojourn in balls $B_r$ of radius $r$, centered at the \textit{origin}. But there are infinitely many other possible choices, i.e. different conventions, we can alternatively adopt. 

For instance, instead of considering arrival times at a distance $r$ from the origin, we can consider arrival times at a distance $r$ from an arbitrary given point $\textbf{c}$. In terms of sojourn times, this amounts using balls $B_r^\textbf{c}$ centered at a point $\textbf{c}\neq 0$ in space, instead of the origin. 

Then, considering any one of the five equivalent local time-delay expressions we have previously derived, and taking the global time-delay limit $r\rightarrow\infty$, it is a simple matter to check that the global time-delay becomes in this case:
\begin{equation}
\label{global-classical-time-delay-formula-shifted-point}
\tau(\textbf{c})=-\frac{1}{v}\left[\hat{\textbf{p}}_+\left(\textbf{q}_+-\textbf{c}\right)-\hat{\textbf{p}}_-\left(\textbf{q}_--\textbf{c}\right)\right].
\end{equation}

That the usual definition of time-delay in classical mechanics wasn't invariant under spatial translations was first noted in~\cite{Narnhofer}, then further clarified in ~\cite{Sas1}. Contrary to (\ref{global-classical-time-delay-formula}), formula (\ref{global-classical-time-delay-formula-shifted-point}) is a more general and consistent expression, duly taking into account all possible different choices for the spatial point $\textbf{c}$ from which the arrival or sojourn times are defined. Indeed, if $\textbf{x}\rightarrow\textbf{x}-\textbf{a}$, then not only $\textbf{q}_\pm\rightarrow\textbf{q}_\pm -\textbf{a}$, but $\textbf{c}\rightarrow\textbf{c}-\textbf{a}$, so that (\ref{global-classical-time-delay-formula-shifted-point}) remains clearly invariant under a spatial translation, as it should by its very definition. 

However, let us observe  that not even (\ref{global-classical-time-delay-formula-shifted-point}) corresponds to the more general possible situation. Indeed, one could as well decide to estimate the interacting and free reference (arrival or sojourn) times from two different points in space. For instance, one could compare the arrival time of the interacting particle at a distance $r$ from a point $\textbf{c}$, to the arrival time of the free particle at a distance $r$ from another point $\textbf{c}_0\neq\textbf{c}$. Adopting such a mixed convention, one finds that (\ref{local-classical-arrival-in-}) and (\ref{local-classical-arrival-out+}) do not converge any more to zero, but to:
\begin{equation}
\label{local-classical-arrival-in-double-point}
\tau_{\text{in}}^-(\textbf{c},\textbf{c}_0)=-\frac{1}{v}\hat{\textbf{p}}_-\left(\textbf{c}_0-\textbf{c}\right)
\end{equation}
\begin{equation}
\label{local-classical-arrival-out+double-point}
\tau_{\text{out}}^+(\textbf{c},\textbf{c}_0)=-\frac{1}{v}\hat{\textbf{p}}_+\left(\textbf{c}-\textbf{c}_0\right).
\end{equation}
Also, the local time-delays (\ref{local-classical-arrival-in+}) and (\ref{local-classical-arrival-out-}) do not converge any more to the same limit (\ref{global-classical-time-delay-formula-shifted-point}), but each one to a different value:
\begin{equation}
\label{local-classical-arrival-in+double-point}
\tau_{\text{in}}^+(\textbf{c},\textbf{c}_0)= -\frac{1}{v}\left[\hat{\textbf{p}}_+\left(\textbf{q}_+-\textbf{c}\right)-\hat{\textbf{p}}_-\left(\textbf{q}_--\textbf{c}_0\right)\right]
\end{equation}
\begin{equation}
\label{local-classical-arrival-out-double-point}
\tau_{\text{out}}^-(\textbf{c},\textbf{c}_0)= -\frac{1}{v}\left[\hat{\textbf{p}}_+\left(\textbf{q}_+-\textbf{c}_0\right)-\hat{\textbf{p}}_-\left(\textbf{q}_--\textbf{c}\right)\right].
\end{equation}

On the other hand, considering local time-delays defined in terms of sojourn times, one can also choose a ball $B_r^{\textbf{c}}$ centred at $\textbf{c}$ for the interacting particle, and a different ball $B_r^{\textbf{c}_0}$ centred at $\textbf{c}_0$ for the free reference particle. However, seeing that the free sojourn times $T^0_{\text{in}}(B_r^{\textbf{c}_0})$ and $T^0_{\text{out}}(B_r^{\textbf{c}_0})$ behave asymptotically as $2r/v$, one finds in this case that (\ref{local-classical-sojourn-in}), (\ref{local-classical-sojourn-out}) and (\ref{local-classical-sojourn-in/out}) converge all to the same limit (\ref{global-classical-time-delay-formula-shifted-point}), independently of the choice of $\textbf{c}_0$.

In the sequel, unless otherwise stated, we shall limit ourselves to the choice $\textbf{c}=\textbf{c}_0=0$, which is the most simple and natural one, especially in the case of a symmetric potential centered at the origin. However, it is important to keep in mind that this is only one among an infinite number of possible different conventions, and that the most general expression for the classical global time-delay is not (\ref{global-classical-time-delay-formula}), but (\ref{global-classical-time-delay-formula-shifted-point}), or even (\ref{local-classical-arrival-in+double-point}) and (\ref{local-classical-arrival-out-double-point}).

\subsection{A further definition of global time-delay}\label{A further definition of global time-delay}

As we have seen in the previous sections, classical global time-delay can be equivalently defined in terms of sojourn or arrival times. In all these definitions, the idea is to \textit{subtract} from the interacting (sojourn or arrival) time, a suitable reference time, associated to a free particle whose motion is synchronized with the interacting particle, either in the remote past or in the distant future (or a combination of both). 

We want now to provide a slightly different definition, the idea of which is to simply \textit{extract} from the interacting time its \textit{free-flight component}. More precisely, we observe from (\ref{asymptotic-arrival-classical2}) and (\ref{sojourn time classical}) that:
\begin{equation}
\label{classical-sojourn-time-asymptotic}
T(B_r)=\frac{2r}{v}+\tau + o(1).
\end{equation}
Thus, the divergent component of the interacting sojourn time is only given by its free-flight contribution, which grows linearly with $r$. Therefore, we can define the local time-delay (the subscript `ff' stands for `free-flight'): 
\begin{equation}
\label{classical-local-time-delay-new}
\tau_{\text{ff}}(r)=T(B_r)-T^0_{\text{ff}}(B_r),
\end{equation}
where 
\begin{equation}
\label{free sojourn time-new}
T^0_{\text{ff}}(B_r)\equiv r\left[\lim_{r'\to\infty}\frac{T(B_{r'})}{r'}\right]=\frac{2r}{v},
\end{equation}
so that
\begin{equation}
\label{classical-local-time-delay-new-limit}
\lim_{r\to\infty}\tau_{\text{ff}}(r)=\tau.
\end{equation}

As we shall see in the foregoing, the ``free-flight'' sojourn time (\ref{free sojourn time-new}) is the only reference time that remains fully consistent in the most general situation: multichannel scattering and arbitrary conditions of observation of the scattering particle.

\subsection{A simple interpretation}\label{A simple interpretation}

The classical global time-delay formula (\ref{global-classical-time-delay-formula}) has a simple and direct interpretation. To see this, we define the following two arrival times~\cite{Narnhofer}: 
\begin{equation}
\label{arrival-time-origin}
t^0_{\text{out}}=-\frac{1}{v}\hat{\textbf{p}}_+\textbf{q}_+, \  t^0_{\text{in}}=-\frac{1}{v}\hat{\textbf{p}}_-\textbf{q}_-,
\end{equation}
and observe that we can simply express the global time-delay as the difference:
\begin{equation}
\label{global-classical-time-delay-formula2}
\tau=t^0_{\text{out}}-t^0_{\text{in}},
\end{equation}
where the \textit{finite} arrival time $t^0_{\text{out}}$ (respectively $t^0_{\text{in}}$) is the time at which the outgoing free particle $\textbf{q}^0_{\text{out}}(t)$ (respectively, the incoming free particle $\textbf{q}^0_{\text{in}}(t)$) intersects the plan passing from the origin, orthogonal to the direction of  movement $\hat{\textbf{p}}_+$ (respectively, $\hat{\textbf{p}}_-$).

\section{Quantum global time-delay}\label{Quantum global time-delay}

\subsection{The quantization problem}\label{The quantization problem}

We want now to generalize the concept of global time-delay to the case of a quantum scattering particle. The main difficulty in comparison to the classical case is that a notion of trajectory is not any more available. But, as we shall see, this doesn't constitute a major problem. 

In (non-relativistic) quantum mechanics, a $n$-dimensional scattering particle is described by a vector $|\psi_t\rangle$ belonging to a Hilbert space ${\cal H}=L^2(\mathbb{R}^n)$ of square integrable wave functions, obeying the Schroedinger equation $i\hbar d/dt|\psi_t\rangle =H|\psi_t\rangle$, where $H=H_0+V$ is the total Hamiltonian, $H_0=\textbf{P}^2/2m$ the free Hamiltonian and $V$ the potential (we shall use capital letters to distinguish quantum operators from classical variables). 

If $|\psi_t\rangle$ describes a scattering solution, its asymptotic behavior is of the form: 
\begin{equation}
\label{asymptotic-quantum}
|\psi_t\rangle=
\begin{cases}
e^{-\frac{i}{\hbar}H_0t}|\varphi_-\rangle, & \text{$t\rightarrow -\infty$} \\
e^{-\frac{i}{\hbar}H_0t}|\varphi_+\rangle, & \text{$t\rightarrow +\infty$.} 
\end{cases}
\end{equation}
The state $|\varphi_-\rangle$ is the so-called incoming state at time $t=0$, whereas $|\varphi_+\rangle$ is the outgoing state at time $t=0$. 

The incoming state is mapped into the outgoing one by the scattering operator: $|\varphi_+\rangle=S|\varphi_-\rangle$. For later convenience, we also introduce the (isometric) wave operators:
\begin{equation}
\label{wave-operators}
\Omega_\pm = {\text{s}}\!\!-\!\!\!\!\!\lim_{t\to\pm\infty}e^{\frac{i}{\hbar}Ht}e^{-\frac{i}{\hbar}H_0t}, 
\end{equation}
that we assume to exist (as strong limits) and to be complete, so that the scattering operator $S=\Omega_+^\dagger\Omega_-$ is unitary. The scattering state at time $t$ can then be written as $|\psi_t\rangle = e^{-\frac{i}{\hbar}Ht}\Omega_-|\varphi_-\rangle = \Omega_-e^{-\frac{i}{\hbar}H_0t}|\varphi_-\rangle$, where for the last equality we have used the intertwining property of the wave operators, $H\Omega_\pm = \Omega_\pm H_0$, from which it also follows that the scattering operator is compatible with the free evolution, i.e., $H_0S=SH_0$.
  
Now, if we apply the standard quantization rule that consists in replacing in a classical expression the position and momentum variables by the corresponding position and momentum operators, $\textbf{q}\rightarrow\textbf{Q}$, $\textbf{p}\rightarrow\textbf{P}$, then symmetrizing all products of non-commuting observables~\cite{Temple}, we immediately get from (\ref{arrival-time-origin}) the following candidate for a quantum mechanical arrival time operator, relative to a free evolving particle: 
\begin{equation}
\label{time operator}
T_0=-\frac{1}{4}\left(H_0^{-1}\textbf{P}\textbf{Q}+\textbf{Q}\textbf{P} H_0^{-1}\right).
\end{equation}

Taking then the expectation value of (\ref{time operator}) over the incoming and outgoing states $|\varphi_-\rangle$ and $|\varphi_+\rangle$, respectively, we obtain the following formal quantum analogues of the classical arrival times (\ref{arrival-time-origin}):
\begin{equation}
\label{quantum-arrival-time-origin}
t^0_{\text{out}}(\varphi_+)=\left\langle\varphi_+ |T_0 |\varphi_+\right\rangle,   \ \ t^0_{\text{in}}(\varphi_-)=\left\langle\varphi_- |T_0| \varphi_-\right\rangle.
\end{equation}
Using $|\varphi_+\rangle=S|\varphi_-\rangle$, we can thus write for the quantum global time-delay:
\begin{eqnarray}
\tau_{\varphi^-}&=& t^0_{\text{out}}(\varphi_+)-t^0_{\text{in}}(\varphi_-)\\
&=&\left\langle\varphi_- |S^\dagger T_0 S|\varphi_-\right\rangle - \left\langle\varphi_- |T_0 |\varphi_-\right\rangle\nonumber\\
\label{global-quantum-time-delay-formula}
&=&\left\langle\varphi_- |S^\dagger \left[T_0,S\right]|\varphi_-\right\rangle
\end{eqnarray}
In other terms, the quantum analogue of the classical global time-delay expression (\ref{global-classical-time-delay-formula}) can be obtained by simply taking the expectation value of the \textit{global time-delay operator}
\begin{equation}
\label{timedelay operator}
\tau = S^\dagger \left[T_0,S\right]
\end{equation}
over the incoming state $|\varphi_-\rangle$.

Despite its immediacy, the above formal procedure of obtaining the quantum global time-delay is not without difficulties. Indeed, the arrival times (\ref{quantum-arrival-time-origin}) have no simple interpretation in standard quantum mechanics. One of the reasons is that the free arrival time operator $T_0$ is not self-adjoint. 

This can be easily shown using the canonical commutation relations between position and momentum, to prove that:
\begin{equation}
\label{canonical commutation}
\left[H_0,T_0\right]=i\hbar \mathbb{I}.
\end{equation}

Since $T_0$ obeys the canonical commutation relation with the free Hamiltonian, if it would be self-adjoint, then $\exp\left(i\alpha T_0\right)$ would be a unitary representation of the group of energy translations, and since we can translate both to the right and to the left, the very existence of such a representation would be in contradiction with the boundedness from below of the spectrum of $H_0$. Thus, $T_0$ cannot be self-adjoint and $H_0$ doesn't possess a canonically conjugate operator (this is a famous argument due to Pauli; see for instance the discussion in~\cite{Hilgevoord}).

Let us open a brief parenthesis, to recall that in physics a system is described in terms of its properties, and that in quantum mechanics properties correspond to orthogonal projectors, whose expectation values over the state of the system give the \textit{a priori} probabilities for the properties being confirmed by an experiment, and this independently of the specificities of the measuring apparatus. 

The request to represent physical observables by (densely defined) self-adjoint operators then follows from the spectral theorem, which allows to \textit{uniquely} decompose a self-adjoint operator by means of a projection-valued measure, and therefore to unambiguously relate the measure of the observable to the properties of the system. 

Thus, since $T_0$ is not self-adjoint, but only symmetric, one cannot easily interpret the quantum global time-delay (\ref{global-quantum-time-delay-formula}) as a difference of arrival times, at least not within the standard interpretational frame requiring physical observables to be represented by self-adjoint operators. 

The difficulty of not having a self-adjoint operator for arrival time observables is usually believed to be related to the essentially different role that time would play in quantum physics in comparison to classical physics. However, as it has been lucidly pointed out by Hilgevoord~\cite{Hilgevoord}, the problem is only apparent and results from a confusion between the time coordinate (the partner of the space coordinate of the space-time reference frame), which needs not be quantized, and the time variables, which are ordinary dynamical variables, measured by specific instruments, called \textit{clocks}. 

Of course, quantum mechanics being not classical mechanics (for instance, there is no notion of trajectory in quantum mechanics), one must be prepared to encounter situations where quantum dynamical time variables cannot always be defined as in the corresponding classical situation. This is exactly what happens when dealing with arrival times, which cannot be defined in terms of self-adjoint operators, but only in terms of symmetric operators. 

This means that one has to renounce to decompose the time operator in terms of projection-valued measures, using instead more general positive operator-valued measures. The price to be paid is that then the projection postulate no longer holds and arrival time observables cannot anymore be uniquely defined and will in general depend on the detailed description of the experimental apparatus used to carry out the measure. 

Although the study of non self-adjoint arrival time observables is per se an interesting and important field of investigation (see for instance~\cite{Allc, Wern, Kong, Muga-Lea, Olk, Olk2, Olk3, Rec} and the references cited therein, and particularly the very recently published work~\cite{Rec2}, which presents an interesting methodical and conceptual review on time as a quantum observable), our concern in this article is to confine ourselves within the usual direct correspondence between physical observables and self-adjoint operators. More precisely, our goal is to motivate a general, physically transparent and self-consistent formula for the quantum mechanical global time-delay, making only use, from the beginning, of self-adjoint operators.

It is worth emphasizing that with this strategy we do not want to imply that the above mentioned approaches, in terms of non-self-adjoint arrival time operators, wouldn't be important, or secondary, or that somehow our approach would be in a sense superior. Also, a comparison between our results and those obtained using the concept of arrival time instead of sojourn time, as a primary classical concept to be quantized, is certainly of interest, but would go beyond the scope of the present essay. Therefore, we refer the interested reader to the above mentioned references (and those cited therein) and the review papers mentioned in the Introduction.  

Said this, we can start noticing that although the operator $T_0$ is not self-adjoint, the global time-delay operator (\ref{timedelay operator}), also referred to in the literature as the \textit{Eisenbud-Wigner time-delay operator}, is in fact a bona fide self-adjoint operator (for a proof see Refs.~\cite{AJS, Je}). Therefore, the following question arises: although expression (\ref{timedelay operator}) doesn't possess a direct unambiguous interpretation in terms of a difference of finite arrival times in the ambit of standard quantum mechanics, can we nevertheless provide a general and physically sound justification for its use? In the next section, we shall give a positive answer to this interrogative.

\subsection{Quantum sojourn times}\label{Sojourn time operators}

As we have seen in Sec.~\ref{Time-delay as a difference of arrival times} and \ref{Time-delay as a difference of sojourn times}, classically speaking one can define the time-delay in terms of a difference of arrival times or, equivalently, in terms of a difference of sojourn times. On the other hand, as we discussed in the previous section, arrival times do not possess a simple interpretation in standard quantum mechanics. What about sojourn times? 

It would be natural to guess that as there is no place in quantum theory for arrival times as self-adjoint observables, the same should be true for notions which are classically related to them, like for instance sojourn times which, in principle, should be defined as a difference (or sum of differences) of arrival times. Fortunately, as we shall see, this is not true, as one can make sense of a notion of sojourn time without making any explicit reference to a notion of arrival time. 

As we did in (\ref{sojourn time classical}), it is very natural to define the time of sojourn of a classical (point-like) particle in the spatial region $B_r\subset \mathbb{R}^n$, as the difference between the times the particle leaves and enters the region (if we assume $r$ large enough, it will enter and leave it only once). 

However, one can also adopt a probabilistic perspective and consider the probability ${\cal P}_t(B_r)$ for the classical particle being inside $B_r$ at time $t$. Integrating this probability of presence inside $B_r$ over all time instants, we can calculate the average time the particle spends in total inside $B_r$ by: 
\begin{equation}
\label{sojourn-time-definition}
T(B_r) = \int_{-\infty}^{\infty}dt\, {\cal P}_t(B_r).
\end{equation} 

Defining the sojourn time as a sum over probabilities of presence is a natural procedure if one only possesses a statistical knowledge of the particle's trajectory, as it is the case for instance when its initial (or final) conditions are described by a probability distribution $\rho(x,p)$ in phase space. Then, the time-dependent probability ${\cal P}_t(B_r)$ can be entirely expressed in terms of $\rho$, of the measure preserving dynamical transformations describing the particles' dynamics, and the characteristic function of the spatial region $B_r$ (see for instance \cite{Sas2} for the details).

However, definition (\ref{sojourn-time-definition}) makes  full sense also when the particle's dynamics is perfectly known, and thus constitutes an alternative, more general definition for the sojourn time, which is in fact equivalent to the one given in Sec.~\ref{Time-delay as a difference of sojourn times} in terms of a difference of arrival times, when a notion of trajectory is available. 

Let us show this more explicitly, and for sake of simplicity let us limit ourselves to the one-dimensional case. Then, the ball $B_r$ of radius $r$, centered at the origin, reduces to the interval $[-r,r]$, and the particle's probability of being present inside $B_r$ is equal to $1$ if $q(t)\in [-r,r]$, and zero otherwise. 

More precisely: ${\cal P}_t(B_r)=\chi_r[q(t)]$, where $\chi_r(x)$ is the characteristic function of the interval $[-r,r]$. Thus, the sojourn time $T(B_r)$ is given by:
\begin{eqnarray}
\label{sojourn-time-definition2}
T(B_r) &=& \int_{-\infty}^{\infty}dt\, \chi_r[q(t)] \\
\label{two-integrals}
&=&\int_{-\infty}^{-s}dt\, \chi_r[q(t)]+\int_{s}^{\infty}dt\, \chi_r[q(t)]\\
\label{last-integral}
&+&\int_{-s}^{s}dt\, \chi_r[q(t)],
\end{eqnarray} 
where $s$ is an arbitrary positive number. 

If one takes the radius $r$ to be large enough, then $q(s)$ and $q(-s)$ belong to the interval $[-r,r]$, and integral (\ref{last-integral}) becomes equal to $2s$. Furthermore, if $s$ is also chosen large enough, we can replace $q(t)$ in the two integrals (\ref{two-integrals}) by the free asymptotic forms $q^0_{\text{in}}(t)=q_-+v_-t$, and $q^0_{\text{out}}(t)=q_+ +v_+t$, respectively ($v_\pm =p_\pm /m$). Then, performing the change of variables $\alpha =q_- + v_-t$, in the first integral of (\ref{two-integrals}), and assuming $v_->0$ (the particle comes from the left), we find: 
\begin{eqnarray}
\label{first-integral}
\int_{-\infty}^{-s}dt\, \chi_r[q(t)]&\approx& \int_{-\infty}^{-s}dt\, \chi_r(q_-+v_-t)\\ 
\label{first-integral2}
&=&\frac{1}{v_-}\int_{-\infty}^{q_--v_-s}d\alpha\, \chi_r(\alpha)\\
\label{first-integral3}
&=&\frac{q_- +r}{v_-} -s.
\end{eqnarray} 

In the same way, performing the change of variables $\alpha =q_+ + v_+t$, in the second integral of (\ref{two-integrals}), and assuming for instance  that $v_+>0$ (the particle is transmitted), one obtains: 
\begin{eqnarray}
\label{second-integral}
\int_{s}^{\infty}dt\, \chi_r[q(t)]&\approx& \int_{s}^{\infty}dt\, \chi_r(q_++v_+t)\\ 
\label{second-integral2}
&=&\frac{1}{v_+}\int_{q_++v_+s}^{\infty}d\alpha\, \chi_r(\alpha)\\
\label{second-integral3}
&=&\frac{r-q_+}{v_+} -s.
\end{eqnarray} 
Thus, for a sufficiently large radius $r$, the sojourn time $T(B_r)$ converges to ($v=v_-=v_+$):
\begin{eqnarray}
\label{sojourntime1dimensionalclassical}
T(B_r)&\approx& \frac{2r}{v} -\frac{1}{v}\left(q_+-q_-\right)\\ 
\label{sojourntime1dimensionalclassical2}
&=&\underbrace{\frac{1}{v}\left(r-q_+\right)}_{\approx t^+(r)}- \underbrace{\frac{1}{v}\left(-r-q_-\right)}_{\approx t^-(r)}.
\end{eqnarray} 

In other terms, defining the sojourn time as a time-integral over probabilities of presence, or as a difference between an exit and entrance time, is in fact equivalent when a trajectory is available. Of course, we can repeat the same reasoning as above for the case of a reflected particle, and the result can easily be generalized to more than one spatial dimension~\cite{Narnhofer}. 

Let us now come back to our  concern, which is the proper definition of global time-delay in quantum mechanics. Thanks to definition (\ref{sojourn-time-definition}), we can bypass the mentioned difficulty of a lack of a self-adjoint arrival time operator and define the quantum mechanical sojourn time as an integral over presence probabilities. 

In fact, in quantum mechanics the probability of presence of a particle inside a given region of space is a perfectly well defined quantity. More precisely, to the property ``The particle is inside the spatial ball $B_r$,'' we can associate an orthogonal projection operator $P_r$, such that if $|\psi_t\rangle =e^{-\frac{i}{\hbar}Ht}\Omega_-|\varphi\rangle$ is the state describing the scattering particle at time $t$ (we have set $|\varphi\rangle\equiv |\varphi_-\rangle$), then
\begin{eqnarray}
{\cal{P}}_{\psi_t}(B_r)&=& \left\|P_r\psi_t\right\|^2=\left\langle \psi_t|P_r|\psi_t\right\rangle \nonumber \\
\label{probability of presence}
&=&\int_{B_r} d^nx\, \left|\psi_t(\textbf{x})\right|^2
\end{eqnarray}
is the probability for the quantum particle to be found inside the ball $B_r$, at time $t$, following a measurement. Thus, using (\ref{sojourn-time-definition}), we can define the quantum sojourn time by:
\begin{equation}
\label{sojourn-time-definition-quantum}
T_\varphi(B_r)=\int_{-\infty}^{\infty}dt\, {\cal{P}}_{\psi_t}(B_r)=\int_{-\infty}^{\infty}dt\, \int_{B_r} d^nx\, \left|\psi_t(\textbf{x})\right|^2.
\end{equation}

Let us observe that the conceptual validity of the definition (\ref{sojourn-time-definition-quantum}) depends only on the conceptual validity of the probability (\ref{probability of presence}). And since the latter possesses a proper meaning in quantum mechanics, the same must also be true for the average (\ref{sojourn-time-definition-quantum}). 

In other terms, as a purely probabilistic statement, definition (\ref{sojourn-time-definition-quantum}) is independent of the details of the theory which underlies (\ref{probability of presence}), and in particular of the existence or not of a classical notion of trajectory~\footnote{This last statement is of course only partially true, as the very definition of time has been historically motivated by the observation of entities moving in space along trajectories (think for example to our sun, as a first rudimental clock). And in that sense, time is primarily a classical concept, pertaining to the realm of the macro-objects.}.

A possible objection could be that (\ref{sojourn-time-definition-quantum}) is not just a probabilistic statement, as its physical interpretation also depends on the interpretation one attaches to the time parameter $t$, in quantum physics. There are indeed two different concepts of time incorporated in (\ref{sojourn-time-definition-quantum}), which should not be mixed. 

On one hand, we have the time variable $t$, which is here to be understood as a simple classical parameter, having the only function of ordering the different measurement projects an experimenter can possibly do in his laboratory, in order to define and attribute properties and states to the different physical entities. On the other hand, we have the time observable $T_\varphi(B_r)$, associated to a specific entity, which is a purely dynamical observable that can be linked to a certain class of measurements performed by very specific instruments, called clocks (see the next section). 

The ordering time-parameter $t$ and the dynamical  time-observable $T_\varphi(B_r)$, although linked together by formula (\ref{sojourn-time-definition-quantum}), are of course very different quantities from a conceptual point of view, and are not to be confused.

\subsection{Physical clocks}\label{Physical clocks}

In the previous section we have shown that the concept of ``probability of presence'' of a quantum particle can be used to obtain a general definition for a quantum sojourn time, that reduces to the usual classical definition (in terms of a difference of arrival times) when a notion of trajectory is available. In this section we want to further motivate  the sojourn time definition (\ref{sojourn-time-definition}), considering the possibility of measuring it by means of physical clocks.

Generally speaking, a clock is a physical system exhibiting a dynamical variable (which we can call the \textit{clock variable}), the evolution of which is known and sufficiently regular. The observation of the clock variable (like the hand position of a watch) corresponds then to a measure (or realization) of time. Here we shall limit ourselves to \textit{idealized clocks}, such that the clock variable $C$ behaves under time translations similarly to the time coordinate $t$ (see also the discussion in~\cite{Hilgevoord}):
\begin{equation}
\label{covariance of clock variable}
C(t)-C(t_0)= t-t_0.
\end{equation}

A comment is in order to elucidate the meaning of (\ref{covariance of clock variable}) that, similarly to (\ref{sojourn-time-definition-quantum}), is an expression containing two different kinds of time: the ordering time-parameter $t$ and the clock dynamical variable $C(t)$. 

Generally speaking, the time coordinate $t$ can be associated to the reading of an idealized classical laboratory clock (or ensemble of duly synchronized laboratory clocks), so that every expression indexed by the parameter $t$, like for instance the state vector $|\psi_t\rangle$, does implicitly refer to the reading of such an idealized classical clock:  vector $|\psi_t\rangle$ is the state of the system at the time instant $t$ indicated by the laboratory clock, which is the instrument used by the experimenter to properly order the different happenings of his laboratory. 

This means that (\ref{covariance of clock variable}) is just to be considered as a consistency relation, expressing the request that idealized clocks are instruments that have to measure  the same time intervals, and therefore deliver fully compatible readings.

Said this, we want to use a suitable clock as a chronometer, to measure the amount of time spent by a particle inside the spatial region $B_r$. Therefore, we will have to find a way to start the clock (i.e., allow the clock variable to evolve) when the particle enters $B_r$, and then stop it when it leaves it. 

For this, we obviously need to couple the two systems, inevitably causing a certain non zero amount of perturbation to the particle's motion. In the following, we shall consider three different paradigmatic examples of idealized clocks -- the spin-clock, the dissipative-clock and the energy-clock -- and will show that they all provide the same answer (\ref{sojourn-time-definition}).

\subsubsection{The spin-clock}\label{The spin-clock}

The spin-clock (also known as the Larmor clock), originally introduced in Refs.~\cite{Baz, Ryba}, exploits the well-known mechanism of the uniform precession of a spin in a homogeneous magnetic field. 

The idea is to locally apply a constant magnetic field in the region of interest, to activate and deactivate the particle's spin precession at the entry and exit of the field region, respectively. In the limit of an infinitesimal field strength (i.e., in the limit of a minimal perturbation of the spin-clock on the particle's movement), the total accumulated angle of the outgoing spin (with respect to the incoming one) is then expected to be proportional to the time spent by the particle inside the field region. 

More precisely, if the magnetic field points in the $z$-direction, one needs to consider the perturbed Hamiltonian $H(\omega) = H + \omega WS_z$, acting on the Hilbert space $L^2(\mathbb{R}^n)\otimes\mathbb{C}^{2s+1}$, where $s$ is the spin, $\omega = -\mu B$ ($\mu$ is the magnetic moment and $B$ the intensity of the field), $S_z$ the $z$-component of the spin operator vector $\textbf{S}=(S_x,S_y,S_z)$, and $W=\int d^nx\, \textsc{w}(\textbf{x})|\textbf{x}\rangle\langle\textbf{x}|$ the multiplication operator by the local bounded function $\textsc{w}(\textbf{x})$, whose support determines the spatial region where the field is applied.

To simplify the discussion, we can assume that the particle is neutral (for instance, for a neutron, $S_i=\frac{\hbar}{2}\sigma_i$, $i=x,y,z$, where the $\sigma_i$ are the $2\times 2$ Pauli matrices and $\mu$ is negative). Let $S_\pm=S_x\pm iS_y$, and $|\Psi_t\rangle$ be the scattering state at time $t$. Using the commutation relation $[S_z,S_\pm]=\pm\hbar S_\pm$, and the fact that $|\Psi_t\rangle$ obeys the Schroedinger equation $i\hbar d/dt|\Psi_t\rangle = H(\omega)|\Psi_t\rangle$, it is straightforward to show that the average $\langle S_\pm\rangle_t = \langle \Psi_t|S_\pm |\Psi_t\rangle$ obeys the differential equation:
\begin{equation}
\label{spin-equation-differential}
\frac{d}{dt}\langle S_\pm\rangle_t =\pm i\omega\langle S_\pm W\rangle_t,
\end{equation}
or, in integral form:
\begin{equation}
\label{spin-equation-integral}
\langle S_\pm\rangle_{t}=\langle S_\pm\rangle_{t_0}\pm i\omega\int_{t_0}^{t}dt'\langle S_\pm W\rangle_{t'}.
\end{equation}

Consider first  the case where magnetic field is constant and fills the entire three-dimensional space, i.e., $W=\mathbb{I}$. Then, the spin clock is uncoupled to the spatial degrees of freedom of the particle and (\ref{spin-equation-integral}) becomes: 
\begin{equation}
\label{spin-equation-uncoupled}
\langle S_\pm\rangle_{t}=\langle S_\pm\rangle_{t_0}e^{\pm i\omega (t-t_0)},
\end{equation}
which corresponds to a uniform rotation of the spin vector in the plane perpendicular to the field direction, with angular speed $\omega$ (the so-called Larmor precession frequency). Therefore, because of the uniform spin precession, 
\begin{equation}
\label{spin-clock-variable}
C(t)\equiv\frac{1}{\pm i\omega}\ln \langle S_\pm\rangle_{t}
\end{equation}
is a bona fide clock variable, obeying the consistency relation (\ref{covariance of clock variable}), and can be used to properly measure time. 

Consider now the case where $\textsc{w}(\textbf{x})$ is the characteristic function $\chi_r(\textbf{x})$ of the ball $B_r$, i.e., $\chi_r(\textbf{x})=1$, if $|\textbf{x}|\leq 1$, and $\chi_r(\textbf{x})=0$, otherwise, so that $W =\int_{B_r}d^nx |\textbf{x}\rangle\langle\textbf{x}| =P_r$ is the projection operator into the subspace of states spatially localized inside $B_r$. 

The spin-clock is then coupled to the particle's translational movement and the spin will be set into precession only when the particle is inside the constant field region. Hence, the difference $C(t) - C(t_0)$ will not anymore be equal to $t-t_0$, but to the amount of time $T(B_r; t_0,t;\omega)$ the particle has remained inside $B_r$, during the time interval $[t_0,t]$. 

In fact, it would be so only provided the magnetic field would cause no perturbation to the particle's evolution, which in general cannot be true because of the well-known phenomena of reflections at the field boundaries and Stern-Gerlach splitting of the spin components~\cite{Martin-spin}. 

However, in the limit of a zero field ($\omega\to 0$), one can expect the perturbation to be the weakest possible and the spin-clock to deliver a proper measure of the time spent by the particle inside $B_r$. 

To see this, let $|\Psi_t\rangle=e^{-\frac{i}{\hbar}H(\omega)t}\Omega_-(\omega)|\Phi\rangle$ be the scattering state of the particle in presence of the magnetic field, with past asymptotic form $|\Psi_t\rangle \to |\Phi_t\rangle$, as $t\to -\infty$, where $|\Phi_t\rangle = e^{-\frac{i}{\hbar}H_0t}|\Phi\rangle$, and $|\Phi\rangle = |\varphi\rangle\otimes |\xi\rangle$, where $|\xi\rangle$ is the incoming spin state, and $|\varphi\rangle \equiv |\varphi_-\rangle$. 

Clearly, $|\Psi_t\rangle = |\psi_t\rangle\otimes |\xi\rangle + O(\omega)$, with $|\psi_t\rangle=e^{-\frac{i}{\hbar}Ht}\Omega_-|\varphi\rangle$ being the scattering state for the problem without spin (or without magnetic field), and we have:
\begin{align}
\label{spin-clock-reading}
T_\varphi &(B_r; t_0,t)=\lim_{\omega\to 0}T_\Phi(B_r; t_0,t;\omega)\nonumber\\
&=\lim_{\omega\to 0}\frac{1}{\pm i\omega}\ln\left( 1\pm\frac{i\omega}{\langle S_\pm\rangle_{t_0}}\int_{t_0}^t dt' \langle S_\pm P_r\rangle_{t'}\right)\nonumber\\
&=\int_{t_0}^t dt' \langle\psi_{t'} |P_r |\psi_{t'}\rangle \nonumber\\
&=\int_{t_0}^t dt' \int_{B_r}d^nx |\psi_{t'}(\textbf{x})|^2.
\end{align}

Finally, taking the limits $t_0\to -\infty$ and $t\to \infty$, we observe that the reading of the spin clock coincides, in the zero field limit, with the sojourn time (\ref{sojourn-time-definition-quantum}), defined in terms of presence probabilities~\cite{JawWard3,Mar3,JawWard5}.

\subsubsection{The dissipative-clock}\label{The dissipative-clock}

We now consider a different example of clock that exploits the constant rate of absorption of a dissipative medium (the idea of which was first proposed by Golub et al.~\cite{Gol}). 

This can be modelized by adding a purely dissipative interaction term in the Hamiltonian: $H(\lambda) = H +i\lambda W$, where $\lambda$ is a real coupling constant and $W$ is defined as per above. Then, the evolution operator is not anymore unitary, but given by a (strongly continuous) semi-group of contractions~\cite{Mar1}: 
\begin{equation}
\label{contraction}
U(t,t_0)=
\begin{cases}
e^{-\frac{i}{\hbar}H(\lambda)(t-t_0)}, & t>t_0 \\
e^{-\frac{i}{\hbar}H^\ast(\lambda)(t-t_0)}, & t<t_0, 
\end{cases}
\end{equation}
and the scalar product ${\cal P}_t=\langle\Psi_t |\Psi_t\rangle$ can be interpreted as the probability that the particle is still present (i.e., that it has not been absorbed) at time $t$. 

For $t>t_0$, it clearly obeys the integral equation: 
\begin{equation}
\label{dissipative-equation-integral}
{\cal P}_t={\cal P}_{t_0}-2 \frac{\lambda}{\hbar}\int_{t_0}^{t}dt'\langle \Psi_{t'}|W|\Psi_{t'}\rangle,
\end{equation}
which in the homogeneous case $W=\mathbb{I}$ can be readily integrated to give the exponential law: 
\begin{equation}
\label{exponential-law}
{\cal P}_t={\cal P}_{t_0}e^{-2 \frac{\lambda}{\hbar}(t-t_0)}.
\end{equation}
Therefore, the clock variable obeying (\ref{covariance of clock variable}) is now given by
\begin{equation}
\label{dissipative-clock-variable}
C(t)\equiv-\frac{\hbar}{2\lambda}\ln {\cal P}_t.
\end{equation}

As we did for the spin-clock, we consider the case $W=P_r$ and take the limit of a dissipative interaction of zero stength: $\lambda\to 0$. Then, if $|\Psi_t\rangle = U(t,0)\Omega_-(\lambda)|\Phi\rangle$ is the scattering state in presence of dissipation, we have $|\Psi_t\rangle= |\psi_t\rangle + O(\lambda)$, where $|\psi_t\rangle =e^{-\frac{i}{\hbar}Ht}\Omega_- |\varphi\rangle$ is the scattering state for the problem without dissipation, and we have: 
\begin{align}
\label{dissipative-clock-reading}
T_\varphi &(B_r; t_0,t)=\lim_{\lambda\to 0}T_\Phi(B_r; t_0,t;\lambda)\nonumber\\
&=\lim_{\lambda\to 0}\frac{-\hbar}{2\lambda}\ln\left(1 - \frac{2\lambda}{\hbar{\cal P}_{t_0}}\int_{t_0}^t dt' \langle\Psi_{t'}|P_r|\Psi_{t'}\rangle\right)\nonumber\\
&=\int_{t_0}^t dt' \int_{B_r}d^nx |\psi_{t'}(\textbf{x})|^2.
\end{align}

Taking the limits $t_0\to -\infty$ and $t\to \infty$, we thus find that the dissipative-clock fully agrees with the spin-clock, as it also yields the permanence time (\ref{sojourn-time-definition-quantum}).

\subsubsection{The energy-clock}\label{The energy-clock}

As a last paradigmatic example of an idealized clock, we can add to the particle's Hamiltonian a time-dependent perturbation growing linearly with time: $H(\lambda t) = H +\lambda t W$. Then, the average energy of the particle $\langle E\rangle_t = \langle \Psi_t |H(\lambda t)|\Psi_t\rangle$ is not anymore conserved, but obeys the integral equation: 
\begin{equation}
\label{energy-equation-integral}
\langle E\rangle_t=\langle E\rangle_{t_0}+\lambda \int_{t_0}^{t}dt'\langle \Psi_{t'}|W|\Psi_{t'}\rangle,
\end{equation}
which in the homogeneous case $W=\mathbb{I}$ can be easily integrated to give: 
\begin{equation}
\label{linear-law}
\langle E\rangle_t-\langle E\rangle_{t_0}=\lambda (t-t_0).
\end{equation}
Thus, in this case the clock variable verifying (\ref{covariance of clock variable}) is simply
\begin{equation}
\label{energy-clock-variable}
C(t)=\frac{1}{\lambda}\langle E\rangle_t. 
\end{equation}

Again, let us consider the case $W=P_r$ and take the limit of an infinitesimal time-dependent perturbation: $\lambda\to 0$. Observing once more that $|\Psi_t\rangle= |\psi_t\rangle + O(\lambda)$, we obtain: 
\begin{align}
\label{energy-clock-reading}
T_\varphi &(B_r; t_0,t)=\lim_{\lambda\to 0}T_\Phi(B_r; t_0,t;\lambda)\nonumber\\
&=\lim_{\lambda\to 0}\frac{1}{\lambda}\left[\langle E\rangle_t - \langle E\rangle_{t_0}\right]\nonumber\\
&=\int_{t_0}^t dt' \int_{B_r}d^nx\, |\psi_{t'}(\textbf{x})|^2.
\end{align}

Similarly to the case of the spin and dissipative clocks, in the limits $t_0\to -\infty$ and $t\to \infty$, we recover once more the sojourn time expression (\ref{sojourn-time-definition-quantum}).

\subsubsection{Linear response to a perturbation}\label{Linear response to a perturbation}

From a mathematical point of view, the reason why the spin, absorption and energy clocks, all measure the same quantum sojourn time (\ref{sojourn-time-definition-quantum}), is related to the general role played by the sojourn time operator in the linear response of a scattering system to a perturbation. 

If to an Hamiltonian $H= H_0 + V$ we add a constant perturbation $\lambda\mathbb{I}$, so that the perturbed Hamiltonian becomes $H(\lambda) = H + \lambda\mathbb{I}$, the evolution operator in the interaction picture can be written as
\begin{align}
\label{evolution operator}
U_I(t,t_0;\lambda)&=e^{\frac{i}{\hbar}H_0t}e^{-\frac{i}{\hbar}H(\lambda)(t-t_0)}e^{-\frac{i}{\hbar}H_0t_0}\nonumber\\
&=e^{\frac{i}{\hbar}H_0t}e^{-\frac{i}{\hbar}Ht}\tilde{U}_I(t,t_0;\lambda)e^{\frac{i}{\hbar}H t_0}e^{-\frac{i}{\hbar}H_0t_0},
\end{align}
where
\begin{equation}
\label{evolution operator2}
\tilde{U}_I(t,t_0;\lambda)\equiv e^{\frac{i}{\hbar}Ht}e^{-\frac{i}{\hbar}H(\lambda)(t-t_0)}e^{-\frac{i}{\hbar}Ht_0}
\end{equation}
obeys the differential equation 
\begin{equation}
\label{differential equation}
i\hbar\frac{\partial}{\partial t} \tilde{U}_I(t,t_0;\lambda) = -\lambda \tilde{U}_I(t,t_0;\lambda)
\end{equation}
and is therefore a pure phase factor:
\begin{equation}
\label{solution differential equation}
\tilde{U}_I(t,t_0;\lambda)=e^{-\frac{i}{\hbar}\lambda (t-t_0)}\mathbb{I}. 
\end{equation}

Inserting (\ref{solution differential equation}) into (\ref{evolution operator}), we thus obtain
\begin{equation}
\label{solution differential equation bis}
U_I(t,t_0;\lambda)=U_I(t,t_0;0)e^{-\frac{i}{\hbar}\lambda (t-t_0)},
\end{equation}
which can also be written in the form
\begin{equation}
\label{differential equation for lambda}
i\hbar U_I^\dagger(t,t_0;\lambda) \frac{\partial U_I(t,t_0;\lambda)}{\partial \lambda}=t-t_0.
\end{equation}

In other terms, the linear response of the evolution operator in the interaction picture, to a constant perturbation, is a clock variable. Therefore, if  we restrict the action of the perturbation to the finite spatial region $B_r$, i.e., $H(\lambda) = H + \lambda P_r$, the scattering particle will only be affected by it for a finite amount of time, and we can expect the average of the left hand side of (\ref{differential equation for lambda}) to converge to a finite value in the limit $t\to\infty$ and $t_0\to -\infty$. 

Furthermore, taking the zero-field limit $\lambda \to 0$, we can expect this value to correspond to the average time spent by the particle inside $B_r$. To see that this is indeed the case, we can use the Dyson's series for $\tilde{U}_I(t,t_0;\lambda)$:
\begin{equation}
\label{Dyson}
\tilde{U}_I(t,t_0;\lambda) =\mathbb{I} -\frac{i\lambda}{\hbar}\int_{t_0}^t ds \, e^{\frac{i}{\hbar}Hs} P_r e^{-\frac{i}{\hbar}Hs}+O(\lambda^2).
\end{equation}
Observing that, by definition, the perturbed scattering operator $S(\lambda)$ is nothing but the (strong) limit of $U_I(t,t_0;\lambda)$, as $t\to\infty$ and $t_0\to -\infty$, if we derive (\ref{Dyson}) with respect to $\lambda$, then  take the infinite time limits and use (\ref{wave-operators}), we obtain:
\begin{equation}
\label{linear response for S}
i\hbar \lim_{\lambda\to 0}\langle\varphi| S^\dagger(\lambda) \frac{dS(\lambda)}{d\lambda}|\varphi\rangle = T_\varphi(B_r),
\end{equation}
where $T_\varphi(B_r)$ is the quantum sojourn time (\ref{sojourn-time-definition-quantum}). (For a rigorous proof of this result, we refer the interested reader to Refs.~\cite{Mar3, JawWard3, JawWard5}). 

Seeing the above intimate relation between the quantum sojourn time and the linear response of the scattering operator to an additional external perturbation, in the limit where its strength tends to zero, and considering that all clock models require the perturbation of the scattering particle by an infinitesimal local field, to conveniently couple its motion to the clock variable, it becomes clear why, independently of the specific nature of the perturbation (i.e., the specific model chosen for the clock), it will necessarily give the sojourn time (\ref{sojourn-time-definition-quantum}).

\subsection{Free sojourn time}\label{Free sojourn time}

Having shown that the quantum sojourn time (\ref{sojourn-time-definition-quantum}) is consistent with the reading of idealized physical clocks, let us now study it in some detail, starting with the simple situation of a free evolving particle. Setting $|\psi_t\rangle = |\varphi_t\rangle = e^{-\frac{i}{\hbar}H_0t}|\varphi\rangle$ into (\ref{sojourn-time-definition-quantum}), with $|\varphi\rangle\equiv|\varphi_-\rangle$, we can write:
\begin{equation}
\label{free-sojourn-time-operator}
T^0_{{\text{in}},\varphi}(B_r)=\int_{-\infty}^{\infty}dt\, {\cal{P}}_{\varphi_t}(B_r)=\left\langle \varphi |T^0(B_r)|\varphi\right\rangle,
\end{equation}
where $T^0(B_r)$ is the so-called  \textit{free sojourn time operator}:
\begin{equation}
\label{free-sojourn-time-operator2}
T^0(B_r)=\int_{-\infty}^{\infty}dt\, e^{\frac{i}{\hbar}H_0t}P_r e^{-\frac{i}{\hbar}H_0t}.
\end{equation}

We can observe that:
\begin{eqnarray}
\label{free-sojourn-time-operator-commutation with free evolution}
e^{\frac{i}{\hbar}H_0\alpha}T^0(B_r)&=&\int_{-\infty}^{\infty}dt\, e^{\frac{i}{\hbar}H_0(t+\alpha)}P_r e^{-\frac{i}{\hbar}H_0t} \nonumber\\
&=&\int_{-\infty}^{\infty}dt\, e^{\frac{i}{\hbar}H_0t}P_r e^{-\frac{i}{\hbar}H_0(t-\alpha)} \nonumber\\
&=& T^0(B_r)e^{\frac{i}{\hbar}H_0\alpha}.
\end{eqnarray}
Deriving (\ref{free-sojourn-time-operator-commutation with free evolution}) with respect to $\alpha$, then setting $\alpha =0$, we thus obtain:
\begin{equation}
\label{free-commutation-relation}
\left[H_0,T^0(B_r)\right]=0.
\end{equation}

In other terms, contrary to the arrival time operator $T_0$, which formally obeys the canonical commutation relation with the free Hamiltonian, the free sojourn time operator $T^0(B_r)$ is fully compatible with the energy $H_0$ of the system and doesn't entertain with it an Heisenberg uncertainty relation. In particular, the Pauli's argument (mentioned in Sec.~\ref{The quantization problem}) doesn't apply, and in fact one can show that $T^0(B_r)$ is a bona fide self-adjoint operator~\cite{Jaw}. 

Let us also observe that the sojourn time (\ref{free-sojourn-time-operator}) is finite only if the probability density $\left|\varphi_t(\textbf{x})\right|^2$ decreases sufficiently rapidly as $t\to\pm\infty$. If, for example, we chose for the state $\varphi$ at time $t=0$ a Gaussian wave packet, one can easily show that $\left|\varphi_t(\textbf{x})\right|^2=O\left(t^{-n}\right)$, so that $T^0_\varphi(B_r)$ is finite for $n\geq 2$, but infinite for $n=1$, and this for any choice of the initial velocity of the wave packet. 

This difference between the one-dimensional and higher dimensional cases can be explained in terms of the spreading of the wave packet, a purely quantum phenomenon with no analogues in classical mechanics: contrary to the case $n\geq 2$, in the $n=1$ situation the spreading of the wave packet increases at the same linear rate $t$ as the distance covered by the particle. 

In general, one can show that $T^0_\varphi(B_r)$ is a bounded operator for $n\geq 2$, and an unbounded, but densely defined, operator for $n=1$, typically on the set of states having no components near the zero of energy~\cite{Lavine, Jaw, Dambo}. 

To explicitly calculate the free sojourn time (\ref{free-sojourn-time-operator}), it is useful to introduce the simultaneous improper eigenvectors of $H_0$ and $\hat{\textbf{P}}= \textbf{P}/|\textbf{P}|$:
\begin{equation}
\label{impropereigenvectors}
H_0|E,\hat{\textbf{k}}\rangle=E |E,\hat{\textbf{k}}\rangle, \ \hat{\textbf{P}}|E,\hat{\textbf{k}}\rangle =\hat{\textbf{k}}|E,\hat{\textbf{k}}\rangle,
\end{equation}
where $E\in [0,\infty)$ and $\hat{\textbf{k}}\in {\cal S}^{n-1}$ (the unit sphere). They obey the relations of completeness
\begin{equation}
\label{completeness}
\int_0^\infty dE\, \int_{{\cal S}^{n-1}} d{\hat k}|E,\hat{\textbf{k}}\rangle\langle E,\hat{\textbf{k}}| =\mathbb{I}
\end{equation}
and orthogonality
\begin{equation}
\label{orthogonaity}
\langle E,\hat{\textbf{k}}| E^{\prime},\hat{\textbf{k}}^{\prime}\rangle = \delta(E-E^{\prime})\delta(\hat{\textbf{k}}-\hat{\textbf{k}}^{\prime}),
\end{equation}
and their wave function is given by:
\begin{equation}
\label{planewaves}
\langle\textbf{x}|E,\hat{\textbf{k}}\rangle =\left(2\pi\hbar\right)^{-\frac{n}{2}}\sqrt{m}\left(2mE\right)^{\frac{n-2}{4}}e^{i\sqrt{2mE}\hat{\textbf{k}}\textbf{x}}.
\end{equation}

In the following, we shall denote by $\varphi(E)=\langle E|\varphi\rangle$ the vectors in $L^2({\cal S}^{n-1})$, at fixed energy, and by 
\begin{equation}
\label{fixed energy scalar product}
\langle\varphi(E)|\varphi^{\prime}(E)\rangle=\int_{{\cal S}^{n-1}}d\hat{k}\, \varphi^{\ast}(E,\hat{\textbf{k}})\varphi^{\prime}(E,\hat{\textbf{k}}),
\end{equation}
the corresponding scalar product, where $\varphi(E,\hat{\textbf{k}})=\langle E,\hat{\textbf{k}}|\varphi\rangle$. 

For simplicity, we shall limit ourselves to the one-dimensional case $n=1$. For a single spatial dimension, the unit sphere ${\cal S}^{0}$ is made only of two points, $\hat{\textbf{k}}=\pm 1$, and we can write $\hat{\textbf{P}}=P_+^0-P_-^0$, where
\begin{equation}
\label{one-dim proj-operators}
P_\pm^0=\int_0^\infty dE\, |E,\pm\rangle\langle E,\pm| \equiv |\pm\rangle\langle \pm|
\end{equation}
are the projection operators into the subspaces of states of positive $(+)$ and negative $(-)$ momentum, respectively. 

For a free particle coming, say, from the left, i.e., $P_+^0 |\varphi\rangle =|\varphi\rangle$, we have:
\begin{align}
\label{free sojourn time calculation}
&T^0_{{\text{in}},\varphi}(B_r)=\langle \varphi |T^0(B_r)|\varphi\rangle \nonumber\\
&=\!\!\int_0^\infty \!\!\!\! dE \int_0^\infty \!\!\!\! dE^{\prime}\, \varphi^{\ast}(E,+)\langle E,+|T^0(B_r)|E^{\prime},+\rangle\varphi(E^{\prime},+)\nonumber\\
&=\!\!\int_0^\infty \!\!\!\! dE \int_0^\infty \!\!\!\! dE^{\prime}\, \varphi^{\ast}(E,+)\langle E,+|P_r|E^{\prime},+\rangle\varphi(E^{\prime},+)\times\nonumber\\
&\times \int dt\, e^{\frac{i}{\hbar}(E-E^{\prime})t}\nonumber\\
&=\int_0^\infty dE\, \langle +|T_E^0(B_r)|+\rangle |\varphi(E,+)|^2,
\end{align}
where for the last equality we have used the identity $\int dt\exp[\frac{i}{\hbar}(E-E^{\prime})t]=2\pi\hbar\delta(E-E^{\prime})$, and we have defined:
\begin{align}
\label{on-shell free sojourn time}
&\langle +|T_E^0(B_r)|+\rangle = 2\pi\hbar \langle E,+|P_r|E,+\rangle \nonumber\\
&= 2\pi\hbar\int_{-r}^r dx\, |\langle x|E,+\rangle|^2\nonumber\\
&=2\pi\hbar\int_{-r}^r dx\,\left|\frac{1}{\sqrt{2\pi\hbar}}\sqrt{\frac{m}{\hbar k}}e^{ikx}\right|^2 = \frac{2r}{v},
\end{align}
with $v=\hbar k/m=\sqrt{2E/m}$. 

Finally, inserting (\ref{on-shell free sojourn time}) into (\ref{free sojourn time calculation}), and setting $\varphi(E)\equiv\varphi(E,+)$, we obtain that the one-dimensional quantum free sojourn time for a particle coming from the left is simply given by:
\begin{equation}
\label{one-dimensional free sojourn time}
T^0_{{\text{in}},\varphi}(B_r)=\int_0^\infty dE\, \frac{2r}{v} |\varphi(E)|^2.
\end{equation}
In the limit of an incoming wave packet sharply peaked about the energy $E$, i.e., in the limit 
\begin{equation}
\label{monoenergetic limit}
|\varphi(E^\prime)|^2\rightarrow \delta (E^\prime -E)
\end{equation} 
of a monoenergetic (but still square integrable!) incoming wave, we obtain that the quantum one-dimensional incoming free sojourn time is equal to the classical expression $2r/v$. The same result holds of course for a particle coming from the right, i.e., $\langle +|T_E^0(B_r)|+\rangle =\langle -|T_E^0(B_r)|-\rangle$.

\subsection{Interferences}\label{Interferences}

According to the above calculation, for a monoenergetic particle coming from the left (or from the right), the one-dimensional quantum free sojourn time coincides with the classical one. This however will not be true in general, because of the well known phenomenon of interference, which is typical of quantum mechanics but totally absent in classical mechanics. Let us show how interferences manifest in the ambit of the one-dimensional free sojourn time. 

For this, we recall that in quantum mechanics interference terms manifest as a consequence of the non compatibility of certain properties or, which is equivalent, of the non commutativity of certain observables. More precisely, consider two properties $a$ and $b$ and let $P_a$ and $P_b$ be the associated orthogonal projection operators. Let also $\bar{a}$ be the inverse property of $a$, associated to the projector $P_{\bar a}=\mathbb{I}-P_a$. Then, we can write:
\begin{eqnarray}
\label{projectors1}
P_b &=& \left(P_a+P_{\bar a}\right)P_b\left(P_a+P_{\bar a}\right)\nonumber\\
&=&P_aP_bP_a + P_{\bar a}P_bP_{\bar a}+P_aP_bP_{\bar a}+P_{\bar a}P_bP_a.
\end{eqnarray}

Taking the expectation value of (\ref{projectors1}) over a state $|\varphi\rangle$, we thus find that the probability ${\cal P}_{\varphi}(b)=\langle\varphi|P_b|\varphi\rangle$ can be written as:
\begin{eqnarray}
\label{projectors2}
{\cal P}_{\varphi}(b) &=& {\cal P}_{\varphi}(a \;\textnormal{and then} \; b) + {\cal P}_{\varphi}(\bar{a}\; \textnormal{and then}\; b)\nonumber\\
&+& 2 \Re \langle \varphi |P_aP_bP_{\bar a}|\varphi\rangle,
\end{eqnarray}
where
\begin{equation}
\label{jointprobability1}	
{\cal P}_{\varphi}(a \;\textnormal{and then} \; b)=\langle \varphi |P_aP_bP_a|\varphi\rangle
\end{equation}
is the expectation value (which lies between $0$ and $1$) of the self-adjoint operator $P_aP_bP_a$, which can be roughly interpreted as corresponding to a measure of property $a$ immediately followed by a measure of property $b$. Similarly, 
\begin{equation}
\label{jointprobability2}	
{\cal P}_{\varphi}(\bar{a} \;\textnormal{and then} \; b)=\langle \varphi |P_{\bar{a}}P_bP_{\bar{a}}|\varphi\rangle	
\end{equation}
is the expectation value of the self-adjoint operator $P_{\bar{a}}P_bP_{\bar{a}}$, that can be associated to a measure of property $\bar{a}$ immediately followed by a measure of property $b$. 

When properties $a$ and $b$ are compatible (i.e., the associated orthogonal projection operators commute), the last term in (\ref{projectors2}) is zero and one finds that:
\begin{equation}
\label{total probability}
{\cal P}_{\varphi}(b)={\cal P}_{\varphi}(a \;\textnormal{and then} \; b) + {\cal P}_{\varphi}(\bar{a} \;\textnormal{and then} \; b),
\end{equation}
which is the theorem of total probability of classical probability theory (see for istance the discussion in~\cite{Farina}). In this case ${\cal P}_{\varphi}(a \;\textnormal{and then} \; b)$ and  ${\cal P}_{\varphi}(\bar{a} \;\textnormal{and then} \; b)$ can be interpreted as the joint probabilities associated to the meet properties $a\wedge b$ and $\bar{a}\wedge b$, respectively. 

However, if $a$ and $b$ are not compatible, then the last term in (\ref{projectors2}), which is an interference term, will in general be different from zero, and one cannot anymore interpret ${\cal P}_{\varphi}(a \;\textnormal{and then} \; b)$ and ${\cal P}_{\varphi}(\bar{a} \;\textnormal{and then} \; b)$ as joint probabilities, at least not in the usual sense of classical probability theory.

\subsubsection{Incoming free sojourn time}\label{Incoming free sojourn time}

Let us show now how interferences can manifest at the level of the quantum free sojourn time. For this, let $b$ be the property ``The particle is inside the ball $B_r$,'' associated to the projector $P_r$, $a$ the property ``The particle has positive momentum,'' associated to the projector $P^0_+$, and $\bar{a}$ the property  ``The particle has negative momentum,'' associated to the projector $P^0_-$. Let also the state describing the particle at time $t=0$ be given by $|\varphi\rangle=(|\varphi_1\rangle+|\varphi_2\rangle)/\sqrt{2}$, where $|\varphi_1\rangle$ is a normalized state with only positive momentum, i.e., $P^0_+|\varphi_1\rangle=|\varphi_1\rangle$, and $|\varphi_2\rangle$ is a normalized state with only negative momentum, i.e., $P^0_-|\varphi_2\rangle=|\varphi_2\rangle$. (Being $|\varphi_1\rangle$ and $|\varphi_2\rangle$ orthognal, $|\varphi\rangle$ is duly normalized to $1$). 

Then, (\ref{projectors2}) becomes:
\begin{eqnarray}
\label{interference1}
{\cal P}_{\varphi}(B_r) &=& \frac{1}{2}\left[{\cal P}_{\varphi_1}(B_r) + {\cal P}_{\varphi_2}(B_r)\right]\nonumber\\
&+&  \Re \int_{-r}^r dx\, \varphi_1^\ast(x)\varphi_2(x),
\end{eqnarray}
which makes even more evident the interpretation of the last term in (\ref{interference1}) as an interference term. In the same way, considering a free evolving state $|\varphi_t\rangle =e^{-\frac{i}{\hbar}H_0t}|\varphi\rangle$, we find for the incoming free sojourn time:
\begin{eqnarray}
\label{interference2}
T^0_{{\text{in}},\varphi}(B_r) &=& \frac{1}{2}\left[T^0_{\varphi_1}(B_r) + T^0_{\varphi_2}(B_r)\right]\nonumber\\
&+&  \Re \langle\varphi_1 |T^0(B_r)|\varphi_2\rangle.
\end{eqnarray}

More explicitly, if for instance we choose for the initial state $|\varphi\rangle$ an odd function of the momentum~\footnote{If we consider the one-dimensional Schr\"odinger equation as a radial equation ($x\geq 0$), then $\varphi$ describes a spherical $3$-dimensional wave of zero angular momentum ($s$-wave).}, i.e., $\varphi_1(E,+)=-\varphi_2(E,-)\equiv g(E)$, then the last interference term of (\ref{interference2}) is given by the following oscillating contribution: 
\begin{align}
\label{interference3}
&\Re \langle\varphi_1 |T^0(B_r)|\varphi_2\rangle \nonumber\\
&= \Re \int_0^\infty \!\!\!\! dE\, \varphi_1^\ast(E,+)\langle +|T^0_E(B_r)|-\rangle \varphi_2(E,-)\nonumber\\
&= -\Re \int_0^\infty dE\, |g(E)|^2\int_{-r}^r dx\, \frac{m}{\hbar k}e^{-2ikx}\nonumber\\
&=-\int_0^\infty dE\, \frac{\hbar}{2E}\sin(2kr)|g(E)|^2.
\end{align}
Thus, the incoming free sojourn time becomes:
\begin{equation}
\label{free-sojourn-time-s-wave}
T^0_{{\text{in}},\varphi}(B_r)=\int_0^\infty dE\,\left(\frac{2r}{v}-\frac{\hbar}{2E}\sin(2kr)\right)|g(E)|^2.
\end{equation}

Finally, taking for $|g(E^\prime)|^2$ the monoenergetic limit (\ref{monoenergetic limit}), one finds that the free sojourn time at fixed energy $E$, for an incoming wave which is an odd function of the momentum, is given by the classical term $2r/v$, plus an interference oscillating contribution, with no classical analogue. 

However, if before taking the monoenergetic limit one considers very large regions, i.e., $r\rightarrow\infty$, because of the Riemann-Lebesgue Lemma (or more simply by an integration by parts if $|g(E)|^2$ is sufficiently smooth), the interference contribution will tend to zero, as the sinus becomes infinitely oscillating in the integral (\ref{interference3}). 

Of course, the vanishing of the interference term as  $r\rightarrow\infty$ is to be expected, as in this limit the projector $P_r\rightarrow\mathbb{I}$ (in the strong sense) and therefore becomes compatible with $P^0_+$ and $P^0_-$ (in fact, it becomes compatible with everything). Thus, because of the orthogonality of $|\varphi_1\rangle$ and $|\varphi_2\rangle$, the interference term in (\ref{interference1}) must vanish. 

In other terms, although  terms of interference can contribute to the quantum sojourn time, seeing that they oscillate with the radius $r$, they cannot contribute to the global time-delay limit (as we will show more clearly in the sequel).

\subsubsection{Outgoing and symmetric free sojourn times}\label{Outgoing and symetric free sojourn time}

Let us investigate a little further the interference phenomenon by also calculating the quantum outgoing and symmetric free sojourn times. 

The outgoing free sojourn time makes use, instead of the free evolving incoming state $e^{-\frac{i}{\hbar}H_0t}|\varphi\rangle$, of the free evolving outgoing one: $e^{-\frac{i}{\hbar}H_0t}S|\varphi\rangle$. This gives, for a particle coming from the left ($P_+^0|\varphi\rangle=|\varphi\rangle$):
\begin{align}
\label{free outgoing sojourn time calculation}
&T^0_{{\text{out}},\varphi}(B_r)=\left\langle \varphi |S^\dagger T^0(B_r)S|\varphi\right\rangle \nonumber\\
&=\int_0^\infty dE\, \langle +|S_E^\dagger T_E^0(B_r)S_E|+\rangle |\varphi(E,+)|^2,
\end{align}
where $S_E$ is the on-shell scattering matrix, which in the one-dimensional case corresponds to the $2\times 2$ unitary matrix:
\begin{equation}
\label{on-shell scattering matrix}
S_E =
\begin{pmatrix}
\langle +|S_E|+\rangle \hfill & \langle +|S_E|-\rangle \\
\langle -|S_E|+\rangle & \langle -|S_E|-\rangle \hfill 
\end{pmatrix} \nonumber \\
\equiv
\begin{pmatrix}
T_E \hfill & R_E \hfill \\
L_E \hfill & T_E \hfill
\end{pmatrix}.
\end{equation}

The matrix element $T_E=|T_E|e^{i\alpha_E^T}$ is the transmission amplitude (which is the same for a particle coming from the left or from the right), whereas $L_E=|L_E|e^{i\alpha_E^L}$ and $R_E=|R_E|e^{i\alpha_E^R}$ are the reflection amplitudes from the left and from the right, respectively. 

Because of the unitarity of $S_E$: $|L_E|=|R_E|$, $|L_E|^2 + |T_E|^2 =1$, and $T_E^\ast L_E + R_E^\ast T_E =0$. Writing
\begin{align}
\label{free outgoing sojourn time calculation2}
&\langle +|S_E^\dagger T_E^0(B_r)S_E|+\rangle \nonumber\\
&=\sum_{\sigma,\rho =\pm}\langle +|S_E^\dagger |\sigma\rangle \langle \sigma|T_E^0(B_r)|\rho\rangle 
\langle \rho|S_E|+\rangle
\end{align}
and observing that $\langle \sigma|T_E^0(B_r)|\rho\rangle = 2\pi\hbar \langle E,\sigma|P_r|E,\rho\rangle$, we can use (\ref{planewaves}) and the unitarity property of the scattering matrix, to obtain, after some calculations:
\begin{align}
\label{free outgoing sojourn time calculation3}
&\langle +|S_E^\dagger T_E^0(B_r)S_E|+\rangle \nonumber\\
&=\frac{2r}{v}+\frac{\hbar}{E}|T_EL_E|\cos (\alpha_E^L - \alpha_E^T)\sin (2kr).
\end{align}
Similarly, for the symmetric free sojourn time, one obtains:
\begin{align}
\label{free symetric sojourn time calculation2}
&\frac{1}{2}\left[\langle +|S_E^\dagger T_E^0(B_r)S_E|+\rangle +\langle +|T_E^0(B_r)|+\rangle\right]\nonumber\\
&=\frac{2r}{v}+\frac{\hbar}{2E}|T_EL_E|\cos (\alpha_E^L - \alpha_E^T)\sin (2kr).
\end{align}

So, we observe that when we use as a reference a free evolving state having both positive and negative momentum components, the free sojourn time will show interference terms, similar to those previously obtained for an odd function of the momentum. 

It is worth also noting that because of the term $|T_EL_E|/2E$, the oscillating interference terms in (\ref{free outgoing sojourn time calculation3}) and (\ref{free symetric sojourn time calculation2}) vanish when the transmission or reflection probabilities are equal to $1$, at resonance, or in the high energy limit. They also vanish if the potential is parity invariant, i.e., if $\textsc{v}(x)=\textsc{v}(-x)$, because in this case we have $L_E = R_E$, so that $\Re (T_E^\ast L_E)=0$, implying that the relative phase between the transmission and reflection amplitudes is $\pi /2$, and so the cosine terms in (\ref{free outgoing sojourn time calculation3}) and (\ref{free symetric sojourn time calculation2}) are zero.

\subsection{Interaction sojourn time}\label{Interaction sojourn time}

Let us now consider the sojourn time (\ref{sojourn-time-definition-quantum}) for a particle evolving in the presence of the interaction. Replacing $|\psi_t\rangle=e^{-\frac{i}{\hbar}H_0t}\Omega_-|\varphi\rangle$ into (\ref{sojourn-time-definition-quantum}), we can write:
\begin{equation}
\label{sojourn-time-operator}
T_\varphi(B_r)=\int_{-\infty}^{\infty}dt\, {\cal{P}}_{\psi_t}(B_r)=\left\langle \varphi |T(B_r)|\varphi\right\rangle,
\end{equation}
where $T(B_r)$ is the so-called (interaction) \textit{sojourn time operator}:
\begin{equation}
\label{sojourn-time-operator2}
T(B_r)=\int_{-\infty}^{\infty}dt\, e^{\frac{i}{\hbar}H_0t}\Omega_-^\dagger P_r\Omega_- e^{-\frac{i}{\hbar}H_0t}.
\end{equation}

Repeating the same argument as in (\ref{free-sojourn-time-operator-commutation with free evolution}), we find that it also commutes with the free evolution, i.e.,
\begin{equation}
\label{commutation-relation}
\left[H_0,T(B_r)\right]=0,
\end{equation}
and thus possesses on-shell matrix elements. 

As we did for the free evolving case, we limit our analysis to the simple one-dimensional scattering problem and of a particle coming from the left. As for (\ref{free sojourn time calculation}), we then obtain: 
\begin{equation}
\label{sojourn time calculation-a}
T_\varphi(B_r)=\int_0^\infty dE\, \langle +|T_E(B_r)|+\rangle |\varphi(E)|^2,
\end{equation}
where
\begin{align}
\label{sojourn time calculation}
\langle +|T_E(B_r)|+\rangle &= 2\pi\hbar \langle E,+|\Omega_-^\dagger P_r\Omega_- |E,+\rangle\nonumber\\
&= 2\pi\hbar \int_{-r}^r dx \left|\langle x|\Omega_- |E, +\rangle\right|^2\nonumber\\
&=\frac{m}{\hbar k}\int_{-r}^r dx \left|\psi_+ (E,x)\right|^2,
\end{align}
and $\psi_+ (E,x)=\hbar\sqrt{2\pi k/m}\langle x|\Omega_- |E, +\rangle$ is the solution of the stationary Schroedinger equation
\begin{equation}
\label{stationary Schroedinger equation}
\left\{\frac{\partial^2}{\partial x^2}+\frac{2m}{\hbar^2}\left[E- \textsc{v}(x)\right]\right\}\psi_+(E,x)=0
\end{equation}
with asymptotic condition 
\begin{equation}
\label{asymptotic condition from the left}
\psi_+ (E,x)=
\begin{cases}
e^{ikx}+L_E e^{-ikx}, & x\to -\infty\\
T_E e^{ikx}, & x\to +\infty.
\end{cases}
\end{equation}

To integrate (\ref{sojourn time calculation}), we derive (\ref{stationary Schroedinger equation}) with respect to energy
\begin{equation}
\label{stationary Schroedinger equation2}
\left\{\frac{\partial^2}{\partial x^2}+\frac{2m}{\hbar^2}\left[E- \textsc{v}(x)\right]\right\}\frac{\partial\psi_+}{\partial E}(E,x) + \frac{2m}{\hbar^2}\psi_+(E,x)=0,
\end{equation}
and observe that multiplying (\ref{stationary Schroedinger equation2}) by $\psi_+^\ast(E,x)$, then using once more (\ref{stationary Schroedinger equation}), we obtain the identity:
\begin{equation}
\label{squared stationary solution}
\left|\psi_+(E,x)\right|^2 = \frac{\hbar^2}{2m}\frac{\partial}{\partial x}\left(\frac{\partial\psi_+^\ast}{\partial x}\frac{\partial\psi_+}{\partial E}-\psi_+^\ast \frac{\partial^2\psi_+}{\partial x\partial E} \right)(E,x).
\end{equation}

Finally, inserting (\ref{squared stationary solution}) into (\ref{sojourn time calculation}), then using the asymptotic form (\ref{asymptotic condition from the left}), after some calculations we get, in the limit $r\to\infty$:
\begin{align}
\label{sojourn time calculation2}
\langle +| &T_E(B_r)|+\rangle = |T_E|^2\hbar\frac{d\alpha^T_E}{dE} + |L_E|^2\hbar\frac{d\alpha^L_E}{dE}\nonumber\\
&+ \frac{2r}{v}+\frac{\hbar}{2E}|L_E|\sin\left(\alpha^L_E +2kr\right) + o(1).
\end{align}

We observe that the last term in (\ref{sojourn time calculation2}) is again of an interference nature, and is due to the reflective power of the potential $\textsc{v}(x)$. Similarly to (\ref{free outgoing sojourn time calculation3}) and (\ref{free symetric sojourn time calculation2}), it vanishes at resonance and in the high energy limit, but doesn't vanish if the potential is parity invariant. 

We conclude this section  observing that if we multiply (\ref{sojourn time calculation2}) by $1/r$, then take the limit $r\to\infty$, the expression will tend to $2/v$. Therefore, the quantum ``free-flight'' reference time (\ref{free sojourn time-new}), 
\begin{equation}
\label{quantum free-flight reference time}
T^0_{\text{ff}}(B_r)=\int_0^\infty dE\, \frac{2r}{v} |\varphi(E)|^2,
\end{equation}
coincides with the incoming free sojourn time (\ref{one-dimensional free sojourn time}) and is free of interference terms.

\subsection{The time-delay limit}\label{The time-delay limit}

In this section we study the global time-delay limits (\ref{global-classical-time-delay-limit-bis}), (\ref{global-classical-time-delay-limit-tris}) and (\ref{classical-local-time-delay-new-limit}), in the quantum case. We start considering the local time-delay (\ref{local-classical-sojourn-in}), defined in terms of a free incoming state, and once more, for sake of simplicity, we limit our analysis to the one-dimensional case. 

Then, for a particle coming from the left, we can use the explicit formulae (\ref{sojourn time calculation2}) and (\ref{on-shell free sojourn time}), to obtain:
\begin{align}
\label{quantum local time-delay one dimension}
\tau_{\text{in},\varphi}(r) &=T_\varphi(B_r)-T_{\text{in},\varphi}^0(B_r)\nonumber\\
&=\int_0^\infty dE\, \langle +|\tau_{\text{in},E}(r)|+\rangle |\varphi(E)|^2,
\end{align}
where the on-shell diagonal matrix element of the local time-delay operator $\tau_{\text{in}}(r)$ is given by:
\begin{align}
\label{quantum local time-delay one dimension2}
\langle +|&\tau_{\text{in},E}(r)|+\rangle = |T_E|^2\hbar\frac{d\alpha^T_E}{dE} + |L_E|^2\hbar\frac{d\alpha^L_E}{dE}\\
\label{quantum local time-delay one dimension3}
&+ \frac{\hbar}{2E}|L_E|\sin\left(\alpha^L_E +2kr\right) + o(1).
\end{align}

As we observed for the oscillating term in (\ref{free-sojourn-time-s-wave}), the interference contribution (\ref{quantum local time-delay one dimension3}) vanishes in the limit $r\to\infty$, because of the Riemann-Lebesgue Lemma. Therefore, the one-dimensional quantum global time-delay for an initial state $|\varphi\rangle$ coming from the left, is: 
\begin{align}
\label{quantum global time-delay one dimension on-shell}
\tau_\varphi &=\lim_{r\to\infty} \tau_{\text{in},\varphi}(r)\nonumber\\
& = \int_0^\infty dE\, \langle +|\tau_{\text{in},E}|+\rangle |\varphi(E)|^2,
\end{align}
where
\begin{equation}
\label{quantum global time-delay one dimension on-shell2}
\langle +|\tau_{\text{in},E}|+\rangle =|T_E|^2\hbar\frac{d\alpha^T_E}{dE} + |L_E|^2\hbar\frac{d\alpha^L_E}{dE}.
\end{equation}

Of course, a similar calculation can be worked out for the case of a particle coming from the right, or for the more general case of an incoming state which is a superposition of states coming from the left and from the right. To do so, one also needs to consider the stationary solution from the right $\psi_- (E,x)=\hbar\sqrt{2\pi k/m}\langle x|\Omega_- |E, -\rangle$, with asymptotic form 
\begin{equation}
\label{asymptotic condition from the right}
\psi_- (E,x)=
\begin{cases}
T_E e^{-ikx}, & x\to -\infty\\
e^{-ikx}+R_E e^{ikx}, & x\to +\infty,
\end{cases}
\end{equation}
and the more general identity ($\sigma,\rho =\pm$): 
\begin{equation}
\label{squared stationary solution2}
\psi_\sigma^\ast\psi_\rho(E,x) = \frac{\hbar^2}{2m}\frac{\partial}{\partial x}\left(\frac{\partial\psi_\sigma^\ast}{\partial x}\frac{\partial\psi_\rho}{\partial E}-\psi_\sigma^\ast \frac{\partial^2\psi_\rho}{\partial x\partial E} \right)(E,x).
\end{equation}

One then obtains that: 
\begin{align}
\label{quantum global time-delay one dimension on-shell-general}
\tau_\varphi & =  \sum_{\sigma,\rho=\pm}\int_0^\infty dE\, \varphi^\ast(E,\sigma)\langle\sigma |\tau_E|\rho\rangle\varphi(E,\rho)\nonumber\\
&=\int_0^\infty dE\, \langle\varphi(E)|\tau_E|\varphi(E)\rangle,
\end{align}
where $\tau_E$ is the on-shell global (or Eisenbud-Wigner) time-delay operator, which can entirely be expressed in terms of the scattering matrix $S_E$ and its energy derivative, by the compact formula: 
\begin{equation}
\label{E-W}
\tau_E = -i\hbar S^\dagger_E\frac{dS_E}{dE}.
\end{equation}

Although we have here derived (\ref{E-W}) using the didactical guiding example of the one-dimensional problem, its validity is very general and goes beyond the mere one-dimensional context (see for instance Refs.~\cite{ACS, AC} and the references cited there). 

For example, when the potential is spherically symmetric, the scattering matrix is diagonal in the basis $\{|l,m\rangle, l=0,1,\ldots ; |m|\leq l\}$ of eigenvectors of the orbital momentum (the spherical harmonics), with matrix elements 
\begin{equation}
\label{scattering matrix spherically symetric}
\langle l,m|S_E|l',m'\rangle = e^{2i\delta_E^l}\delta_{l,l'}\delta_{m,m'}
\end{equation}
which are fully expressible in terms only of the phase shifts $\delta_E^l$. Then, the  global time-delay matrix $\tau_E$ is also diagonal in this basis, i.e.,
\begin{equation}
\label{time-delay matrix spherically symetric}
\langle l,m|\tau_E|l',m'\rangle = \tau_E^l\delta_{l,l'}\delta_{m,m'},
\end{equation}
and the diagonal elements  
\begin{equation}
\label{time-delay spherically symetric matrix element}
\tau_E^l = 2\hbar \frac{d\delta_E^l}{dE}
\end{equation}
correspond to the time-delays, at fixed energy $E$, for incoming waves of orbital momentum $l$.

The correspondence between the time-delay matrix (\ref{E-W}) and the operator (\ref{timedelay operator}), that we have formally derived in Sec.~\ref{The quantization problem}, can be easily established if one observes that the formal time operator (\ref{time operator}) acts as an energy derivative in the spectral representation of the free Hamiltonian, i.e.,
\begin{equation}
\label{time operator action}
\langle E|T_0|\varphi\rangle = -i\hbar \frac{d\varphi(E)}{dE},
\end{equation}
so that one can check that (\ref{E-W}) is the on-shell matrix of the time-delay operator $S^\dagger [T_0,S]=S^\dagger T_0 S - T_0$.

In other terms, the time-delay operator is nothing but the difference between the outgoing arrival time operator $S^\dagger T_0 S$ and the incoming one $T_0$ (see the discussion of Sec.~\ref{The quantization problem}). Even though the latter are not, taken individually, proper self-adjoint observables, their difference make full sense, as the operator $S^\dagger [T_0,S]$ not only can be shown to be self-adjoint~\cite{AJS, Je}, but it is also defined as the limit of a difference of self-adjoint operators. 

Therefore, in quantum mechanics we can understand the time-delay limit $\tau_\varphi =\lim_{r\to\infty} \tau_{\text{in},\varphi}(r)$ as a sort of rigorous \textit{inversion procedure} (both from the mathematical and conceptual point of view) that allows to switch from a time control variable to a space control variable, or, to say it differently, to switch from the probability of being in a given place at a given time, to the probability of arriving at a given time in a given place, thus justifying the utilization of the formal time-operator $T_0$ in (non-relativistic) quantum scattering theory, when considered as a difference.

The same holds true when we choose, instead of an incoming free reference time, an outgoing one, or a symmetric one, or the newly introduced ``free-flight'' reference time, as it can be easily shown that the corresponding time-delay limits all converge to the Eisenbud-Wigner time-delay formula. 

As we mentioned in the introduction, the rigorous study of the time-delay limit is a difficult mathematical problem, which has been and still is the object of investigation. Typically, what mathematical physicists try to do is to demonstrate the existence of the limit, and its identity with the Eisenbud-Wigner operator, for the greatest possible class of interactions and initial states. 

For the mathematically more oriented reader, let us briefly sketch what is the typical logic of this kind of proof. An important idea, which is due to Martin~\cite{Mar4}, consists in introducing an auxiliary quantity $\sigma_\varphi(r)=\langle \varphi|S^\dagger [\tilde{T}_0(B_r),S]|\varphi\rangle$, where 
\begin{equation}
\label{definition of tiledT}
\tilde{T}_0(B_r)=\int_0^\infty dt\, e^{\frac{i}{\hbar}H_0t}P_r e^{-\frac{i}{\hbar}H_0t},
\end{equation}
then show, using for instance Lebesgue's dominated convergence, that $\lim_{r\to\infty}[\tau_{\text{in},\varphi}(r)-\sigma_\varphi(r)]=0$ (this typically requires that the asymptotic conditions are integrable at infinity~\cite{Mar4}). 

Then, one only needs to study the asymptotic representation of $\sigma_\varphi(r)$, which in turn requires to investigate the asymptotic representation of the bounded operator $\tilde{T}_0(B_r)$. Under suitable regularity conditions to be imposed on the scattering operator and the incoming state (typically, they have to be sufficiently differentiable in all variables, energy and angles), one can show that, in the weak sense: 
\begin{equation}
\label{asymptotics of tiledT}
\tilde{T}_0(B_r)= T_0 +r\sqrt{\frac{m}{2H_0}} + o(1),
\end{equation}
from which the Eisenbud-Wigner result immediately follows. 
  
Before concluding this section, a last remark is in order. In the above quantum treatment we have only considered spherical regions of localization. But of course, instead of balls $B_r$ centered at the origin, we could as well have considered translated regions $B_r(\textbf{c})$, i.e., balls centered at an arbitrary spatial point $\textbf{c}$. 

In this case, as we have seen in Sec.~\ref{Invariance under space translations}, the classical time-delay formula acquires additional terms. The situation is similar in the quantum case. Indeed, using the transformation properties $S\to S_\textbf{c} = e^{\frac{i}{\hbar}\textbf{p}\textbf{c}}Se^{-\frac{i}{\hbar}\textbf{p}\textbf{c}}$, and $|\varphi\rangle\to |\varphi_\textbf{c}\rangle= e^{\frac{i}{\hbar}\textbf{p}\textbf{c}}|\varphi\rangle$, for the scattering operator and the incoming state under a translation of the spatial coordinate system to a point $\textbf{c}$, then inserting the translated quantities into the Eisenbud-Wigner formula, we obtain (using $d|\textbf{p}|/dH_0 = m/|\textbf{p}|$, with $m$ the mass of the particle): 
\begin{align}
\label{quantum translated time-delay}
\tau_\varphi(\textbf{c}) &=-i\hbar\langle \varphi_\textbf{c}|S^\dagger_\textbf{c}\frac{dS_\textbf{c}}{dH_0}|\varphi_\textbf{c}\rangle\nonumber\\
& = \tau_\varphi + \textbf{c}\: \langle \varphi|\frac{m}{|\textbf{p}|}\left(S^\dagger \hat{\textbf{p}}S-\hat{\textbf{p}}\right)|\varphi\rangle,
\end{align}
which is the quantum equivalent of (\ref{global-classical-time-delay-formula-shifted-point}). Formula (\ref{quantum translated time-delay}) has been firstly obtained in~\cite{JawWard}, in the one-dimensional scattering context, by dilation of intervals centered on an arbitrary point $c$, then subsequently generalized in~\cite{Sas1}, to more than one spatial dimension; see also the analysis in~\cite{Olk3}.

Let us also point out that in the analysis of~\cite{JawWard} one can also find (in the one-dimensional context) an explicit connection between two different approaches to time-delay in quantum mechanics: the first one exploiting the concept of sojourn time (making use of probability densities, as we have done in the approach of the present paper), and the other one exploiting the concept of arrival time (making use of flux probability densities, which are related to probability densities by a continuity relation). The preliminary analysis of~\cite{JawWard} has been further developed in~\cite{Olk3, Rec, Rec2} into a more general and systematic approach, based on the properties of maximal Hermitian operator, that we already mentioned in the Introduction, as well as in Sec.~\ref{The quantization problem}, and we refer the reader to these works for a comparison of these different approaches.

Said this, let us ask what if, instead of using spherical regions, we consider in the time-delay definition a sequence of arbitrary regions $\Sigma_r$, converging to the entire physical space, as $r\to\infty$? For a long time it has been believed that the time-delay limit should not depend on a particular choice of a sequence of localization regions, and that the usual choice of spherical regions was only a matter of convenience. This belief was mainly due to a result of Martin~\cite{Mar4}, that seemed to hold independently of the choice of the sequence $\Sigma_r$, but it was subsequently observed that its application to potential scattering in fact requires the use of spherical localization regions. 

There is in fact a more fundamental reason why one is compelled in using spherically shaped regions in the time-delay definition, as one can prove the following result~\cite{Sas1}: if $\Sigma_r$ is a sequence of regions obtained by dilation of a convex smooth initial region $\Sigma$, then the classical and quantum time-delay limit can only exist if $\Sigma$ is spherically symmetric~\footnote{Recently, this result has been further generalized to include sequences of regions that are not necessarily convex. It can then be shown that the time-delay limit can still exist if the dilated regions are \textit{star-shaped} i.e., such that $\Sigma_{r_1}\subset \Sigma_{r_2}$ if $r_1\leq r_2$. This however is only true for the symmetrized (or ``free-flight'') time-delay, and a new term is found to contribute to its limit, which vanishes in the case of spherical spatial regions~\cite{Ger}.}.

\subsection{Time-delay and resonances}\label{Breit-Wigner}

We want now to briefly illustrate the physical content of the (Eisenbud-Wigner) time-delay formula (\ref{quantum global time-delay one dimension on-shell2}) in relation to the phenomenon of resonance. Let us consider the case of a resonance $|T_{E_r}|^2=1$ in the transmission probability, where $E_r$ is the resonance energy. In its proximity, one can show that $|T_{E}|^2$ possesses the Lorentzian (Breit-Wigner) form: 
\begin{equation}
\label{Lorentzian}
|T_{E}|^2\approx \frac{(\Delta E)^2}{(E-E_r)^2 + (\Delta E)^2},
\end{equation}
where $\Delta E$ is the half-width of the resonance. Also, one can show that in the proximity of $E_r$ the phase of the transmission amplitude has the form: 
\begin{equation}
\label{phase at resonance}
\alpha^T_E \approx \beta_E + \arctan\left(\frac{E-E_r}{\Delta E}\right),
\end{equation}
where $\beta_E$ is the so-called background phase, which is typically a slowly varying function of energy if compared to the second term in (\ref{phase at resonance}), so that we can neglect it in the calculation of the time-delay (\ref{quantum global time-delay one dimension on-shell2}), which thus becomes ($|L_E|^2 \approx 0$ in the vicinity of $E_r$):
\begin{equation}
\label{quantum global time-delay at resonance}
\langle +|\tau_{\text{in},E}|+\rangle \approx \frac{\frac{\hbar}{\Delta E}}{\frac{E-E_r}{\Delta E}+1}.
\end{equation}
Hence, we find that $\langle +|\tau_{\text{in},E}|+\rangle$ is maximum at resonance and is approximately equal to:
\begin{equation}
\label{quantum global time-delay at resonance - bis}
\langle +|\tau_{\text{in},E}|+\rangle \approx \frac{\hbar}{\Delta E}.
\end{equation}

The physical meaning of (\ref{quantum global time-delay at resonance - bis}) is that the lifetime of the resonance is inversely proportional to its energy width. Therefore, (\ref{quantum global time-delay at resonance - bis}) expresses a relation of complementarity between time and energy, in the sense that the sharper the resonance's width, the longer its lifetime, and one can  understand the theory of time-delay as a way to give a proper meaning to this kind of relations.

\subsection{Classical-quantum correspondence}\label{Classical-quantum correspondence}

Before proceeding further in our analysis, by introducing in the next section the new concept of fuzzy sojourn time, we want to add a few more comments about the classical-quantum correspondence. For this, let us first recall the logical path we have followed, when going from classical to quantum systems. 

We have started by defining the time-delay of a classical particle, as one would intuitively do, as a difference of arrival times. Then, we have shown that a same definition holds in terms of a difference of sojourn times. And since classical sojourn times can also be expressed in terms of probabilities of presence, we have in this way identified a safe route to go from classical to quantum sojourn times, and therefore from classical to quantum time-delays. 

However, one may ask if it would be possible to establish a more explicit connection between classical and quantum time-delay formulae,  for instance by exhibiting a classical analogue of the quantum Eisenbud-Wigner time-delay expression  (\ref{quantum translated time-delay}), which for $\textbf{c}=0$ simply reads:
\begin{equation}
\label{E-W quantum c=0}
\tau_\varphi =-i\hbar\langle \varphi|S^\dagger\frac{dS}{dH_0}|\varphi\rangle.
\end{equation}

The answer is affirmative, and different approaches are possible to establish such a formal classical-quantum correspondence.  For instance, one can work in the framework of the Hilbert-space formulation of classical mechanics, as it has been done by Boll\'e and D'Hondt~\cite{Bol-Osb2}, and show that the classical global time-delay can be written in the form
\begin{equation}
\label{E-W classical c=0}
\tau_{\varphi_{\text{cl}}} = \langle \varphi_{\text{cl}}|S^\dagger_{\text{cl}}\frac{dS_{\text{cl}}}{dL_0}|\varphi_{\text{cl}}\rangle,
\end{equation}
which is clearly the analog of the quantum expression (\ref{E-W quantum c=0}). In (\ref{E-W classical c=0}), $|\varphi_{\text{cl}}\rangle$ denotes the initial state of the classical scattering system, which is an element of the Hilber-space $L^2(\Gamma)$, with $\Gamma$ the phase space of the system, $S_{\text{cl}}$ is the classical canonical scattering operator, or transformation, acting on $\Gamma$, and $L_0$ the (free) Liouville operator, which is related to the free Hamiltonian through the Poisson bracket relation. We shall not enter here into the details of this Hilbertian approach to classical mechanics, and simply refer the interested reader to the derivation in~\cite{Bol-Osb2} and the references cited therein.

Another possibility to point out the correspondence between quantum and classical time-delay formulae is to exploit the geometrical method of Narnhofer and Thirring~\cite{Narnhofer, Narn-Thir} (see also~\cite{Sas2}) to show that the classical time delay can be expressed, as in (\ref{time-delay spherically symetric matrix element}), in terms of the energy-derivative of the generator of the classical canonical scattering transformation, which in turn can be shown to correspond to the quasi-classical phase shift. 

One may also ask, more in the spirit of Ehrenfest's theorem, if it is possible to exhibit a correspondence between the quantum mechanical expectation values of sojourn time observables and the corresponding classical motions. To some extent, this is certainly possible. For instance, in the simple example of a one-dimensional quantum entity coming from the left, we have already shown that the expectation value $\langle T^0(B_r)\rangle$ of the free sojourn time operator (\ref{free-sojourn-time-operator2}) can be written in the classical-like form 
\begin{equation}
\label{Ehrenfest1}
\langle T^0(B_r)\rangle = 2r \langle \texttt{V}^{-1}\rangle,
\end{equation}
where $\texttt{V}\equiv\sqrt{2H_0/m}$ is the speed-operator. However, this expression doesn't possess a general validity, not even in the one-dimensional context, as if the incoming state is for instance an odd function of the momentum, there is the additional interference contribution (\ref{interference3}) to be added to the r.h.s. of (\ref{Ehrenfest1}), having no classical analogue. 

But even for a one-dimensional quantum scattering entity coming from the left, an interference contribution is also present as soon as one considers the interacting case, as evidenced by formula (\ref{sojourn time calculation2}). However, these contributions vanish in the limit $r\to\infty$, so that we can write, for the expectation value $\langle T(B_r)\rangle$ of the interaction sojourn time operator (\ref{sojourn-time-operator2}), the asymptotic formula:
\begin{equation}
\label{Ehrenfest2}
\langle T(B_r)\rangle = \langle \tau \rangle + 2r \langle \texttt{V}^{-1}\rangle + o(1),
\end{equation}
which again has a simple classical-like interpretation. 

In more than one spatial dimension, classical expressions are of course more complicated, as  sojourn times will depend not only on the incoming energy but also on the (incoming and outcoming) impact parameters. Nevertheless, in the limit of a large radius, the expressions simplify and one can show, using for instance the  methods outlined at the end of Sec.~\ref{The time-delay limit}, that the classical-like asymptotic formula (\ref{Ehrenfest2}) remains generally valid also for a $3$-dimensional entity. 

To conclude our digression on the classical-quantum correspondence, let us also consider the asymptotic expression (\ref{definition of tiledT}) of the free sojourn time operator restricted to positive times, which we can also write as:
\begin{equation}
\label{Ehrenfest3}
\langle \tilde{T}^0(B_r)\rangle = \langle T_0 \rangle + r \langle \texttt{V}^{-1}\rangle + o(1).
\end{equation}

This expression has again a simple classical-like interpretation. Indeed, for $r$ large enough, a classical point particle is with certainty contained in the ball $B_r$, at time $t=0$. Therefore, the time it will spend inside of it, during the semi-infinite time-interval $[0,\infty)$, is given by its arrival time at the center of the sphere (more precisely, one should say: the time it intersects the plan passing from the origin, orthogonal to the direction of its movement), which can be either positive or negative, according to the localization of the particle at time $t=0$ [corresponding to the first term in the r.h.s. of (\ref{Ehrenfest3})], plus the time needed to cover the distance from the center of the ball to its boundary [corresponding to the second free-flight term in the r.h.s. of (\ref{Ehrenfest3})], plus corrections vanishing as $r\to\infty$. 

The asymptotic expression (\ref{asymptotics of tiledT}), (\ref{Ehrenfest3}) establishes an interesting connection between the self-adjoint sojourn time operator $\tilde{T}^0(B_r)$ (here restricted to positive times) and the not self-adjoint arrival time operator $T_0$. As we already mentioned in the Introduction and Sec.~\ref{The quantization problem}, many authors consider that (maximal) Hermitian operators are sufficient to meaningfully define observables in quantum mechanics, and therefore base their analysis on generalized arrival time observables of the $T_0$ kind. In this way, they can easily derive generalized Ehrenfest-like relations between classical observables and the expectation values of quantum ones, similar to (\ref{Ehrenfest3}); see for instance~\cite{OlkA, OlkB}.

\section{Fuzzy sojourn times}\label{Fuzzy sojourn times}

In the previous sections we have considered the global limit, $r\to\infty$, of local time-delays $\tau_\varphi(r)$ [where $\tau_\varphi(r)$ stands here for either the time-delay defined in terms of incoming, outgoing, symmetric or ``free-flight'' reference sojourn times], and we have shown, in the simple one-dimensional case, its convergence to the Eisenbud-Wigner formula (\ref{quantum global time-delay one dimension on-shell-general})-(\ref{E-W}). 

Once the $r\to\infty$ limit is taken, one can of course also consider the limit of a monoenergetic wave-packet, $|\varphi(E')|^2\to\delta(E'-E)$, peaked at about a given energy $E$. In this way, one obtains on the energy shell formulae that do not depend any more on the details of the shape of the incoming wave. For instance, the monoenergetic limit of (\ref{quantum global time-delay one dimension on-shell}) yields for the time-delay the fixed energy expression (\ref{quantum global time-delay one dimension on-shell2}). 

The order of the two limits is however crucial: if the monoenergetic limit is performed in the first place, then the global time-delay limit doesn't converge anymore, because of the presence of the interference oscillating terms (\ref{quantum local time-delay one dimension3}). As we have seen in Sec.~\ref{Interferences}, these terms typically appear when the dynamics allows for the superposition of waves propagating in opposite directions.  

The presence of these interference contributions gave some headaches to the early scholars of time-delay, insomuch that they had to invent strange \textit{ad hoc} averaging in the definition of time-delay, to make these troublesome terms vanish (see the discussion in~\cite{Bol-Osb}). But their presence being inherent to the very definition of quantum sojourn time, there are of course no reasons to get rid of them by means of whatever artificial procedure. 

However, one may ask the following two natural questions: (1) What is exactly their origin? And: (2) Is it possible to find a consistent definition for the global time-delay that is free of them, and therefore allows to take the global time-delay limit also at fixed energy? As we shall see, both questions can be affirmatively answered by introducing a new kind of sojourn time operator, that we shall call \textit{fuzzy sojourn time operator}.

\subsection{Membership functions}\label{Membership functions}

For this end, we start by observing that the physical interpretation of the quantum sojourn time operator (\ref{sojourn-time-definition}) relies on the proper interpretation of ${\cal P}_t(B_r)=\langle\psi_t|P_r|\psi_t\rangle$ as the probability of finding the scattering particles, at time $t$, inside the ball $B_r$ of radius $r$. As we discussed in Sec.~\ref{The quantization problem}, this interpretation, in turn, relies on the fact that orthogonal projection operators are associated in quantum mechanics to properties, and that their expectation values yield the \textit{a priori} probabilities for these properties to be confirmed by an experiment. 

Properties are operationally defined by means of experimental tests (or, better, equivalence classes of experimental tests). An experimental test is a specific experimental procedure that allows  to answer a question. In the case of $P_r$ the question is the following: ``Is the particle inside the region $B_r$?'' And the answer is obviously ``yes,'' if the particle is found inside $B_r$, and ``no'' if not. 

This ``sharp'' alternative is made explicit in mathematical terms by the fact that $P_r$ is the multiplication operator by the two-valued characteristic funtion $\chi_r(\textbf{x})$, such that $\chi_r(\textbf{x})=1$, if $\textbf{x}\in B_r$, and $\chi_r(\textbf{x})=0$, otherwise.

However, although the above question is expressed in terms of a ``sharp'' predicate, allowing to unambiguously divide, at least in principle, the experimental outcomes into two disjoint classes (corresponding to the logical alternative of being or not being inside $B_r$), it is easy to imagine experimental situations where it is not always possible to exactly determine whether the particle is or isn't inside $B_r$, especially when it is close to the region's boundary. 

In these circumstances, the above question does not admit anymore a simple ``yes or no'' answer, which is exemplified by the two-valuedness of the characteristic function $\chi_r$, but an entire range of possible intermediate answers, that can be associated to all possible values in the interval $[0,1]$, according to the degree of certainty with which the belonging (or non-belonging) of the particle to $B_r$ can be determined. 

To describe this more realistic situation, one can for instance replace the characteristic function $\chi_r$ by a multivalued function, like for instance the following one:   
\begin{equation}
\label{fuzzy characteristic function}
\chi_{r,\rho}(\textbf{x})=
\begin{cases}
1, & |\textbf{x}|\leq r \\
g\left(\frac{|\textbf{x}|-r}{\rho}\right), & |\textbf{x}|> r, 
\end{cases}
\end{equation}
where $\rho >0$ is a parameter determining the sharpness (or fuzziness) with which the region's boundary can be experimentally delimited (the lowest is $\rho$ and the sharpest is the delimitation) and $g(x)$ is a sufficiently well behaved function with compact support, such that $g\in[0,1]$ and $g(0)=1$, that is used to quantify the ``membership degree'' of the particle to $B_r$. 

Clearly, $0\leq \chi_{r,\rho} \leq 1$, and we can say that a point $\textbf{x}$ belongs to $B_r$ if $\chi_{r,\rho}(\textbf{x})=1$, doesn't belong to $B_r$ if $\chi_{r,\rho}(\textbf{x})=0$, and uncertainly belongs to $B_r$ if  $0<\chi_{r,\rho}(\textbf{x})<1$. This uncertainty is to be considered small (and the point will be said to almost belong to $B_r$) if the value taken by $\chi_{r,\rho}$ is close to $1$, whereas the uncertainty is to be considered large (and the point will be said to almost not belong to $B_r$) if the value of $\chi_{r,\rho}$ is close to $0$. 

In other terms, we are now allowing for nuanced responses to the above question, that we characterize by means of a bounded positive self-adjoint operator $P_{r,\rho}$, which is the multiplication operator by the \textit{membership function} $\chi_{r,\rho}(\textbf{x})$. $P_{r,\rho}$ is clearly a generalization of the projection operator $P_r$, to which it tends when $\rho\to 0$.

Using $P_{r,\rho}$, we can therefore define the following weighted probability, which is the natural generalization of (\ref{probability of presence}):
\begin{equation}
\label{probability of presence fuzzy}
{\cal{P}}_{\psi_t}^\rho(B_r)= \left\langle \psi_t|P_{r,\rho}|\psi_t\right\rangle = \int d^nx\, \chi_{r,\rho}(\textbf{x}) \left|\psi_t(\textbf{x})\right|^2.
\end{equation}
Indeed, seeing that $\left|\psi_t(\textbf{x})\right|^2$ is the probability (density) of finding the particle at time $t$ at point $\textbf{x}$, it is clear that the above can be interpreted as a weighted sum over probabilities of presence, with $\chi_{r,\rho}$ playing the role of the weighting function, characterizing the degree of fuzziness with which the region $B_r$ is discriminated during the measurement. 

Then, proceeding similarly to what we have done with the sojourn time (\ref{sojourn-time-operator}), we can define the \textit{fuzzy sojourn time}: 
\begin{equation}
\label{fuzzy sojourn-time-operator}
T^\rho_\varphi(B_r)=\int_{-\infty}^{\infty}dt\, {\cal{P}}_{\psi_t}^\rho(B_r)=\left\langle \varphi |T^\rho(B_r)|\varphi\right\rangle,
\end{equation}
where $T^\rho(B_r)$ is the (interaction) \textit{fuzzy sojourn time operator}:
\begin{equation}
\label{fuzzy sojourn-time-operator2}
T^\rho(B_r)=\int_{-\infty}^{\infty}dt\, e^{\frac{i}{\hbar}H_0t}\Omega_-^\dagger P_{r,\rho}\Omega_- e^{-\frac{i}{\hbar}H_0t}.
\end{equation}
Obviously, repeating the same argument as in (\ref{free-sojourn-time-operator-commutation with free evolution}), we find that the fuzzy sojourn time operator  commutes with the free evolution, i.e.,
\begin{equation}
\label{commutation-relation2}
\left[H_0,T^\rho(B_r)\right]=0,
\end{equation}
and therefore possesses on-shell matrix elements.

\subsection{Asymptotics of the on-shell elements}\label{Asymptotics of the on-shell elements}

To study the on-shell matrix elements of the fuzzy sojourn time operator, let us consider the case of a spherically symmetric potential. Then, in the basis of eigenvectors of the angular momentum $L^2$ and $L_z$, we have  
\begin{equation}
\label{wave operator elements}
\langle \textbf{x}|\Omega_-|E,l,m\rangle = i^l\sqrt{\frac{2m}{\pi\hbar^2k}}\frac{1}{r}u^l_E(s)Y_l^m(\hat{\textbf{x}}),
\end{equation}
where the $Y_l^m(\hat{\textbf{x}})$ are the spherical harmonics and the $u^l_E(s)$, $s=|\textbf{x}|$, are the regular solutions of the radial Schroedinger equation:
\begin{equation}
\label{stationary radial Schroedinger equation}
\left\{\frac{\partial^2}{\partial s^2}-\frac{l(l+1)}{s^2}+\frac{2m}{\hbar^2}\left[E- \textsc{v}(s)\right]\right\}u^l_E(s)=0,
\end{equation}
with asymptotic behavior, as $s\to\infty$,
\begin{equation}
\label{asymptotic condition}
u^l_E(s)=e^{i\delta_E^l}\sin\left(ks - \frac{l\pi}{2} + \delta_E^l\right)+ o(1).
\end{equation}

Exploiting the orthogonality of the spherical harmonics, one then obtains:
\begin{equation}
\label{fuzzy sojourn-time-operator3}
T^\rho_\varphi(B_r)=\sum_{l,m}\int_{0}^{\infty}dE\, T^{\rho,l}_E(B_r)|\varphi_{l,m}(E)|^2, 
\end{equation}
where $\varphi_{l,m}(E)=\langle E,l,m|\varphi\rangle$ and
\begin{equation}
\label{on-shell fuzzy sojourn}
T^{\rho,l}_E(B_r)= \frac{4m}{\hbar k} \int_0^\infty ds\, \chi_{r,\rho}(s) |u^l_E(s)|^2
\end{equation}
is the on the energy shell component of the fuzzy sojourn time operator (\ref{fuzzy sojourn-time-operator2}), for the angular momentum $l$. 

Setting $\rho =0$ in (\ref{on-shell fuzzy sojourn}), we recover the on-shell components of the (conventional, sharp) sojourn time
\begin{equation}
\label{on-shell sojourn}
T^{l}_E(B_r)= \frac{4m}{\hbar k} \int_0^r ds\, |u^l_E(s)|^2,
\end{equation}  
and using a calculation similar to the one presented in Sec.~\ref{Interaction sojourn time}, we can show that, as $r\to\infty$, it has the asymptotic behavior
\begin{equation}
\label{on-shell sojourn asymptotic}
T^{l}_E(B_r)=  2\hbar\frac{\partial \delta_E^l}{\partial E} + \frac{2r}{v} - \frac{1}{2E}\sin(2kr -l\pi + 2\delta_E^l)+ o(1), 
\end{equation}  
exhibiting the typical interference oscillating terms. 

On the other hand, if the fuzziness parameter $\rho$ is kept different from zero, one can perform on (\ref{on-shell fuzzy sojourn}) an integration by parts. To this end, we observe that
$|u^l_E(s)|^2 =  dh^l_E(s)/ds$, with
\begin{equation}
\label{squared stationary solution radial}
h^l_E(s) = \frac{\hbar^2}{2m}\left(\frac{\partial{u_E^l}^\ast}{\partial s}\frac{\partial u_E^l}{\partial E}-{u^l_E}^\ast \frac{\partial^2u^l_E}{\partial s\partial E}\right)(s),
\end{equation}
so that, using the fact that $u^l_E(0)=0$, the integration by parts gives:
\begin{align}
\label{on-shell fuzzy sojourn2}
T^{\rho,l}_E(B_r)&= -\frac{4m}{\hbar k} \int_r^\infty ds\, g'\left(\frac{s-r}{\rho}\right)h^l_E(s)\nonumber\\
&=-\frac{4m}{\hbar k} \int_0^\infty ds\, g'(s)h^l_E(\rho s+r),
\end{align}
where $g'(s)=dg(s)/ds$. 

Using (\ref{squared stationary solution radial}), one can study the asymptotic behavior of $h^l_E(r)$, as $r\to\infty$. For this, one needs to assume that the potential is sufficiently regular and use the method of variation of constant to express the $u^l_E(s)$ as solutions of Volterra-type integral equations. 

We shall skip here the technical details and refer the interested reader to the Appendix B in~\cite{Mar3}. There, it is proved that if the potential fulfils the condition $\int_r^\infty ds\, s|\textsc{v}(s)|=N(r)<\infty$ (notice that $N(r)\to 0$ as $r\to\infty$), then 
\begin{align}
\label{squared stationary solution radial asymptotic}
h^l_E(r) = \frac{r}{2}&+\frac{\hbar^2 k}{2m}\frac{\partial \delta_E^l}{\partial E} -\frac{1}{4k}\sin(2kr -l\pi + 2\delta_E^l)\nonumber\\
&+ O(r^{-1}) + O[N(r)].
\end{align}

So, for $r$ large, we can insert the above asymptotic in (\ref{on-shell fuzzy sojourn2}). After an integration by parts of the linear term, and using $g(0)=1$, we obtain
\begin{align}
\label{squared stationary solution radial asymptotic2}
T^{\rho,l}_E&(B_r)= \frac{2mr}{\hbar k}+\frac{2m\rho}{\hbar k}\int_0^\infty ds\, g(s) + 2\hbar\frac{\partial\delta_E^l}{\partial E}\nonumber\\
&+ \frac{\hbar}{2E}\int_0^\infty ds\, g'(s)\sin(2k\rho s + 2kr -l\pi +2\delta_E^l)\nonumber\\
&+ \frac{\hbar}{2E}\int_0^\infty ds\, g'(s)\left\{O\left(\frac{1}{\rho s+r}\right)+O\left[N(\rho s+r)\right]\right\}.
\end{align}
Since $(\rho s+r)^{-1}<r^{-1}$, the last term of (\ref{squared stationary solution radial asymptotic2}) is $O(r^{-1})+ O[N(r)]$, for all $\rho$. Furthermore, an integration by parts of the fourth term immediately shows that it is less than
\begin{equation}
\label{estimate}
\frac{1}{k\rho}\frac{\hbar}{4E}\left(|g''(0)|+\int_0^\infty ds\, |g''(s)|\right) = O(\rho^{-1}).
\end{equation}
Therefore, as $r\to\infty$ and $\rho\to\infty$, we obtain that
\begin{align}
\label{on-shell fuzzy sojourn asymptotic}
T^{\rho,l}_E(B_r) &= 2\hbar\frac{\partial \delta_E^l}{\partial E}+\frac{2}{v}\int_0^\infty ds\, \chi_{r,\rho}(s) \nonumber\\
&+O(r^{-1})+ O(\rho^{-1})+ O(N(r)).
\end{align}

Comparing (\ref{on-shell fuzzy sojourn asymptotic}) with (\ref{on-shell sojourn asymptotic}), we observe that the on-shell fuzzy sojourn time has the remarkable property of being free from the interference oscillating terms. Therefore, considering the local time-delay
\begin{equation}
\label{fuzzy on-shell local time delay}
\tau^{\rho,l}_{\text{in},E}(B_r)=T^{\rho,l}_E(B_r)-T^{0,\rho,l}_{\text{in},E}(B_r),
\end{equation}
where $T^{0,\rho,l}_{\text{in},E}(B_r)$ is the fuzzy incoming free reference sojourn time, with asymptotic form
\begin{equation}
\label{on-shell fuzzy free sojourn asymptotic}
T^{0,\rho,l}_{\text{in},E}(B_r) = \frac{2}{v}\int_0^\infty ds\, \chi_{r,\rho}(s)+O(r^{-1})+ O(\rho^{-1}),
\end{equation}
we find that the global time delay limit
\begin{equation}
\label{fuzzy on-shell global time delay limit}
\lim_{r,\rho\to\infty}\tau^{\rho,l}_{\text{in},E}(B_r)= 2\hbar\frac{\partial \delta_E^l}{\partial E}
\end{equation}
now converges also at fixed energy! 

The above derivation can be easily extended to non-symmetric potentials, using for instance the methods developed in~\cite{Jau-Mar}. Also, the same result holds if one considers, instead of a fuzzy free incoming time, fuzzy outgoing or symmetric ones.

\subsection{Fuzzy free-flight sojourn time}\label{Fuzzy free-flight sojourn time}

On the other hand, if we want to use, as a reference, a ``free-flight'' time, then (\ref{free sojourn time-new}) needs to be modified in order to take into account that the characterization of the spatial region is now fuzzy, so that the ``free-flight'' component in the sojourn time cannot anymore be expected to grow linearly with $r$, as is clear from the nature of the second term in the r.h.s of (\ref{on-shell fuzzy sojourn asymptotic}), which after the change of variable $s=r + \rho x$, can be rewritten in the form: 
\begin{equation}
\label{modified linear term}
\int_0^\infty ds\, \chi_{r,\rho}(s)= r + \rho \int_0^\infty dx\, g(x)\equiv f(r,\rho). 
\end{equation}

Accordingly, a proper definition for a ``fuzzy free-flight'' sojourn time, convenient to be used as a reference time in the time-delay definition, is the following: 
\begin{equation}
\label{free sojourn time-new-fuzzy}
T^{0,\rho}_{\text{ff},\varphi}(B_r) \equiv f(r,\rho)\left[\lim_{r',\rho'\to\infty}\frac{T^{\rho'}_\varphi(B_{r'})}{f(r',\rho')}\right].
\end{equation}

It could be objected that defining the global time-delay in terms of fuzzy sojourn times can only yield an inaccurate measure of it. This however is not the case. Indeed, time-delay is a relative quantity and the fuzziness is simply ``cut off'' in the sojourn time difference. 

To put it in different terms, time-delay is not a question about how much time a particle spends in a region, but about the excess or defect time it spends in it, due to the interaction. Therefore, a sharp determination of when the particle enters and leaves the region is not a crucial aspect in its calculation, especially in the limit of infinitely extended regions. 

On that purpose, we can observe that for a classical particle, defining the time-delay in terms of sharp or fuzzy sojourn times is in fact perfectly equivalent (we leave the proof of this statement to the reader, as an exercise).

\subsection{Origin of the interference terms}\label{Origin of the interference terms}

Coming back to the two questions we have addressed at the beginning of this section, we certainly have  answered the second one, by providing a definition for the global time-delay that is free from the interference terms and therefore remains consistent also at fixed energy. 

Concerning the first question, about the origin of the interference terms, we can answer it by considering once more the Larmor clock. As we explained in Sec.~\ref{The spin-clock}, the spin clock can be used to measure the sojourn time of a particle in a ball $B_r$, by applying in it a weak homogeneous magnetic field and then observe how much the particle's spin has precessed while traversing the field region. This idealized situation requires a sharp determination of the spatial localization of the field, whose strength has therefore to be modulated by the characteristic function $\chi_r$ of $B_r$. 

In other terms, the spatial switching on and off of the field has to be abrupt (step function), thus idealizing a situation where the field strength varies on a much smaller scale than the de Broglie wavelength of the particle. Of course, such a circumstance is not very realistic, being that if the magnetic field is produced by a macroscopic device, its variations will occur on distances much larger than the latter. 

A more realistic situation is in fact modeled by the function (\ref{fuzzy characteristic function}). Then, the inverse of the parameter $\rho$ provides a measure of the field gradient in the transition region, as is clear from the fact that $d\chi_{s,\rho}/ds = O(\rho^{-1})$, for $s>r$. It is then an easy matter to check that if the field is shaped as $\chi_{r,\rho}$, then its reading agrees with the fuzzy sojourn time (\ref{fuzzy sojourn-time-operator})~\cite{Mar3}.

The spin-clock allows us to understand the origin of the oscillating interferences terms and the reason why they dissolve in the fuzzy sojourn time measurement. The essential difference between a field that is switched on and off abruptly and one which is switched on and off smoothly, resides in the fact that, even in the zero field limit, in the former case the particle's wave function can still be reflected at the sharp frontiers of the field, whereas in the latter case the weak field's reflective power vanishes. 

This means that it is the reflection mechanism at the field boundaries, far away from the interaction region, combined with the reflective power of the potential, that is truly at the origin of the interference oscillating terms in the on-shell sojourn time operator (see also the discussion in Refs.~\cite{Ver, Fal}).

\section{Time-delay in multichannel scattering}\label{Time-delay in multichannel scattering}

So far we have only considered  simple scattering systems. We want now to analyze the more general situation of a multichannel scattering theory. Contrary to a single channel system, a multichannel one is characterized by as many different free evolutions as is the number of possible channels generated by the dynamics. 

This means that, typically, the outgoing products emerging from the scattering region will propagate at different speeds in comparison to the incoming products. Therefore, one cannot expect anymore a time delay defined only in terms of incoming or outgoing free reference times to converge in the limit $r\to\infty$, as the divergent linear terms in $r$ will not anymore compensate in this case. 

This is the reasons why, when there is more than a single channel, it has been recognized that the natural choice for a reference time is the symmetric combination (\ref{free sojourn time classical in/out}).

The simplest example of a multichannel scattering system is the so-called \textit{optical model}, in which the effects of the new open channels are phenomenologically taken into account by a dissipative part of the interaction (as we have done in the dissipative clock example). We refer the reader to~\cite{Mar1} for a general study of the time-delay limit for dissipative interactions. Here we will consider another paradigmatic example of a multichannel scattering systems: the scattering by a time-periodic potential.

\subsection{Scattering by time-periodic potentials}\label{Scattering by time-periodic potentials}

Typically, time-dependent potentials arise in physics as approximate description of small subsystems, whose action on the larger part of the system (the motion of which is assumed to be known) can be neglected. This allows to describe the evolution of the subsystem in terms of an effective non-conservative force field (examples include interaction of electromagnetic waves with matter, thermal fluctuations, chemical reaction at surfaces, coupling of electrons with optical phonons and electron transport in presence of oscillating voltages). 

In other terms, the Hamiltonian describing the system, $H(t) = H_0+V(t)$, now explicitly depends on time via the time-dependence of the potential energy $V(t)$. As a consequence, a certain number of complications arise. In general $[H(t),H(t')]\neq 0$, for $t\neq t'$, so that the unitary evolution operator $U(t,t_0)$, $U(t_0,t_0)=\mathbb{I}$, solution of the Schroedinger equation
\begin{equation}
\label{Schroedinger time-dependent potential}
i\hbar\frac{\partial}{\partial t} U(t,t_0)= H(t)U(t,t_0),
\end{equation}
is no longer given by an exponential, as it is the case for a static Hamiltonian, but by its Dyson expansion, at least when it converges. Also, the evolution is no more invariant under time-translations, the energy of the system is in general not conserved, and one has to abandon the notion of stationary states. 

This however doesn't mean that one has to renounce to scattering theory, as is clear from the fact that the essential point for the characterization of scattering states, leaving any bounded region in configuration space as $t\to\pm\infty$, is not that the potential may or not be time-dependent, but that it decreases sufficiently rapidly in space, i.e., that it is sufficiently short-ranged.

More precisely, let $|\psi_{t_0}\rangle$ be the state of the system at time $t_0$. At time $t$ it becomes $|\psi_t\rangle = U(t,t_0)|\psi_{t_0}\rangle$. If the initial condition $|\psi_{t_0}\rangle$ is of the scattering type, $|\psi_t\rangle$ will behave in the distant future, and has behaved in the remote past, according to the free evolution. 

This means there exist free evolving outgoing and incoming states $|\varphi_{\pm,t}\rangle = e^{-\frac{i}{\hbar}H_0(t-t_0)}|\varphi_{\pm,t_0}\rangle$, such that the difference 
\begin{equation}
\label{asymptotic condition time-dependent}
|\psi_t\rangle -|\varphi_{\pm,t}\rangle =U(t,t_0)|\psi_{t_0}\rangle -e^{-\frac{i}{\hbar}H_0(t-t_0)}|\varphi_{\pm,t_0}\rangle
\end{equation}
tends to zero (in the Hilbert space norm) as $t\to\pm\infty$. 

Multiplying the asymptotic condition (\ref{asymptotic condition time-dependent}) from the left by $U^\dagger(t,t_0)$, one finds that it is equivalent to the existence (as strong limits) of the wave operators
\begin{equation}
\label{wave operators time-dependent potential}
\Omega_\pm(t_0)={\text{s}}\!\!-\!\!\!\!\!\lim_{t\to\pm\infty}U^\dagger(t,t_0)e^{-\frac{i}{\hbar}H_0(t-t_0)}.
\end{equation}
According to (\ref{asymptotic condition time-dependent}) and (\ref{wave operators time-dependent potential}), the scattering state at time $t_0$ is related to the incoming and outgoing states at time $t_0$ by 
\begin{equation}
\label{wave operators time-dependent potential bis}
|\psi_{t_0}\rangle =  \Omega_\pm(t_0)|\varphi_{\pm,t_0}\rangle,
\end{equation}
which yields the correspondence
\begin{equation}
\label{wave operators time-dependent potential tris}
|\varphi_{+,t_0}\rangle =  \Omega_+^\dagger(t_0)\Omega_-(t_0)|\varphi_{-,t_0}\rangle,
\end{equation}
between the outgoing and the incoming state, so defining the scattering operator
\begin{equation}
\label{scattering operator time-dependent potential}
S(t_0) =  \Omega_+^\dagger(t_0)\Omega_-(t_0)
\end{equation}
for an initial condition at time $t_0$.

The reason why we have redefined the wave and scattering operators, that we already defined in Sec.~\ref{The quantization problem}, is to make fully explicit the main difference between the static and time-dependent situation: the scattering process now depends on the choice of the initial condition $t_0$, as the evolution is not anymore invariant under time-translations. Therefore, we now dispose of an entire collection of wave and scattering operators, parameterized by the initial time-condition. 

According to (\ref{wave operators time-dependent potential bis}), these are related by the relations
\begin{equation}
\label{wave operators time-dependent potential quadris}
\Omega_\pm(t_0)=U(t_0,t_1)\Omega_\pm(t_1)e^{-\frac{i}{\hbar}H_0(t_0-t_1)},
\end{equation}
which in turn give, for the scattering operators
\begin{equation}
\label{scattering operator time-dependent potential bis}
S(t_0) =e^{-\frac{i}{\hbar}H_0(t_0-t_1)}S(t_1)e^{\frac{i}{\hbar}H_0(t_0-t_1)}.
\end{equation}
Eq. (\ref{scattering operator time-dependent potential bis}) makes explicit the fact that the free evolution $H_0$ doesn't commute anymore with the scattering operator, and therefore the process is not energy conserving. 

Here however, we are interested in considering the special case of a periodic time-dependence of the potential, i.e., $V(t)=V(t+T)$, where $T=2\pi/\omega$ is the period. Then, $U(t+T,T)=U(t,0)$, so that $\Omega_\pm(T)=\Omega_\pm(0)\equiv\Omega_\pm$, and $S(T)=S(0)\equiv S$. 

According to (\ref{scattering operator time-dependent potential bis}), the scattering operator $S$ (for the initial condition at time $t_0=0$) commutes with the free evolution over one period: $[S,e^{-\frac{i}{\hbar}H_0T}]=0$. This means that even though the energy is not conserved during the scattering, it can only be changed by discrete quanta $n\hbar\omega, n=0,\pm 1,\pm 2, \ldots$

To see this more explicitly, let $V(t)$ be the multiplication operator by a time-periodic function $\textsc{v}(\textbf{x},t)=\textsc{v}(s,t)$, $s=|\textbf{x}|$, of spherical symmetry. Then, the scattering operator $S$ is diagonal in the basis of the spherical harmonics, $\langle l,m|S|l',m'\rangle = S^l\delta_{l,l'}\delta_{m,m'}$, and the above commutation relation holds in every subspace of fixed angular momentum. 

More precisely, we have
\begin{align}
\label{quasi-energy commutation}
\langle E',l,m|&[S,e^{-\frac{i}{\hbar}H_0T}]|E,l,m\rangle\nonumber\\
&= \langle E'|S^l|E\rangle\left(e^{-\frac{i}{\hbar}ET}-e^{-\frac{i}{\hbar}E'T}\right)=0.
\end{align}
Equation (\ref{quasi-energy commutation}) implies that the kernel $\langle E'|S^l|E\rangle$ is zero except when $e^{-\frac{i}{\hbar}ET}=e^{-\frac{i}{\hbar}E'T}$, or, equivalently, when $E'-E=n\hbar\omega$, with $n=0,\pm 1,\pm 2,\ldots$ Therefore, writing the energy $E\in [0,\infty)$ as the sum $E=\epsilon + n\hbar\omega$, with $n\geq 0$ the entire part of $E/\hbar\omega$, and $\epsilon\in [0,\hbar\omega)$ the \textit{quasi-energy} (i.e., the energy modulo $\hbar\omega$), equality (\ref{quasi-energy commutation}) becomes (we set $|\epsilon,n\rangle\equiv |\epsilon +\hbar\omega\rangle$):
\begin{equation}
\label{quasi-energy commutation bis}
\langle \epsilon',n'|S^l|\epsilon,n\rangle\left(e^{-\frac{i}{\hbar}\epsilon T}-e^{-\frac{i}{\hbar}\epsilon'T}\right)=0.
\end{equation}
Hence, since the difference in the brackets can be zero if and only if $\epsilon = \epsilon'$, the kernel in (\ref{quasi-energy commutation bis}) has the form
\begin{equation}
\label{quasi-energy commutation tris}
\langle \epsilon',n'|S^l|\epsilon,n\rangle=\langle n'|S^l_\epsilon|n\rangle\delta(\epsilon - \epsilon'),
\end{equation}
showing that the quasi-energy $\epsilon$ is conserved during the scattering process. 

The operator $S^l_\epsilon$ is called the scattering matrix on the quasi-energy shell, and the physical interpretation of the amplitudes $\langle n'|S^l_\epsilon|n\rangle$ is simple: $|\langle n'|S^l_\epsilon|n\rangle|^2$ is the probability for an incoming wave of energy $\epsilon + n\hbar\omega$ (and fixed angular momentum $l$ and $m$) to be scattered with energy $\epsilon + n'\hbar\omega$. 

To say it differently, it is the probability for an energy transfer of exactly $n'-n$ quanta of energy $\hbar\omega$, with the external field. Clearly, if the theory is complete, i.e., if the scattering operator is unitary, we have ${S^l_\epsilon}^\dagger S^l_\epsilon = S^l_\epsilon {S^l_\epsilon}^\dagger =\mathbb{I}$, implying that
\begin{equation}
\label{probability conservation}
\sum_{n'\geq 0}|\langle n'|S^l_\epsilon|n\rangle|^2 = \sum_{n\geq 0} |\langle n'|S^l_\epsilon|n\rangle|^2 = 1.
\end{equation}

\subsection{The quasi-stationary Schroedinger equation}\label{The quasi-stationary Schroedinger equation}

To study the time-delay limit for the (multichannel) scattering by a time-periodic and symmetric potential, we  first need to establish the connection between the elements of $S^l_\epsilon$ and the solutions of the quasi-stationary Schroedinger equation, which is the analogue of the stationary Schroedinger equation for a time-periodic interaction. 

Let $|\psi_t\rangle =U(t,0)\Omega_-|\varphi\rangle$ be the scattering state at time $t$, for an initial condition at time $t_0=0$. By definition, it obeys the time-dependent Schroedinger equation (\ref{Schroedinger time-dependent potential}). If $V(t)$ is periodic, then it admits the Fourier decomposition:
\begin{equation}
\label{Fourier potential}
V(t)=\sum_n V_n e^{-in\omega t}.
\end{equation}
Also, because of (\ref{wave operators time-dependent potential tris}), we have $|\psi_t\rangle =\Omega_-(t)e^{-\frac{i}{\hbar}H_0t}|\varphi\rangle$, with $\Omega_-(t)$ which is also time-periodic and therefore also admits a Fourier decomposition:
\begin{equation}
\label{Fourier wave operator}
\Omega_-(t)=\sum_n \Omega_n e^{-in\omega t}.
\end{equation}

Inserting (\ref{Fourier potential}) and (\ref{Fourier wave operator}) into (\ref{Schroedinger time-dependent potential}), then comparing the Fourier coefficients, one finds the operatorial identity
\begin{equation}
\label{quasi-stationary Schroedinger}
\left(H_0 + \sum_\nu V_{n-\nu}\Omega_\nu\right) = \Omega_n \left(H_0 + n\hbar\omega \mathbb{I}\right).
\end{equation}
Formally, we can multiply (\ref{quasi-stationary Schroedinger}) from the right by $|E,l,m\rangle \equiv |\epsilon, \rho,l,m\rangle$, with $E=\epsilon +\rho\hbar\omega$. Then, performing the change of variables $\mu=\rho + n$ and $\sigma = \rho +\nu$, we get
\begin{align}
\label{quasi-stationary Schroedinger bis}
H_0\Omega_{\mu - \rho}&|\epsilon, \rho,l,m\rangle  + \sum_\sigma V_{\mu -\sigma}\Omega_{\sigma - \rho}|\epsilon, \rho,l,m\rangle\nonumber\\
&= (\epsilon + \mu\hbar\omega)\Omega_{\mu - \rho}|\epsilon, \rho,l,m\rangle.
\end{align}

The potential being symmetric, we can formally multiply (\ref{quasi-stationary Schroedinger bis}) by $\langle \textbf{x}|$ from the left, then consider the separation of variables: 
\begin{equation}
\label{separation of variables}
\langle \textbf{x}|\Omega_{\mu - \rho}|\epsilon, \rho,l,m\rangle = i^l\sqrt{\frac{2m}{\pi\hbar^2\kappa_\rho}}\frac{1}{s}u_{\mu\rho}^l(\epsilon,s)Y_l^m(\hat{\textbf{x}}),
\end{equation}
where $s=|\textbf{x}|$ and $\hbar\kappa_\rho =\sqrt{2m(\epsilon +\rho\hbar\omega)}$. We then find that the functions $u_{\mu\rho}(\epsilon,s)$, $\rho \geq 0$, $\mu\in \mathbb{Z}$, are the regular solutions of the quasi-stationary radial equation
\begin{align}
\label{quasi-stationary Schroedinger tris}
\left[-\frac{\hbar^2\partial_s^2}{2m}+\frac{\hbar^2l(l+1)}{2ms^2}\right]&u_{\mu\rho}^l(\epsilon,s)+\sum_\sigma \textsc{v}_{\mu - \sigma}(s)u_{\sigma\rho}^l(\epsilon,s)\nonumber\\
&=(\epsilon +\mu\hbar\omega)u_{\mu\rho}^l(\epsilon,s).
\end{align}

Their asymptotic behavior for $s\to\infty$ can be obtained from the asymptotic conditions of the scattering state $|\psi_t\rangle$, plus a stationary phase argument. We skip the details (for the method see Refs.~\cite{Yaf,Sas}) and just give here the result:
\begin{equation}
\label{asymptotics of regular solutions for periodic potentials}
u_{\mu\rho}^l(\epsilon,s)=\frac{1}{2i}\left[{\cal A}_{\mu\rho}^l(\epsilon) e^{i(\kappa_\mu s-\frac{l\pi}{2})}-\delta_{\mu\rho}e^{-i(\kappa_\mu s-\frac{l\pi}{2})}\right]+o(1)
\end{equation}
as $s\to\infty$, where $\delta_{\mu\rho}$ is the Kroenecker symbol and $\hbar\kappa_\mu =i\sqrt{2m|\epsilon +\mu\hbar\omega|}$, for $\mu <0$. 

The terms with $\mu <0$ are exponentially decaying and do not contribute to the limit $s\to\infty$ (they are sometimes called quasi-bound states). On the other hand, for $\mu\geq 0$, we have the relation~\cite{Yaf,Sas}
\begin{equation}
\label{amplitudes and scattering matrix}
{\cal A}_{\mu\rho}^l(\epsilon)=\sqrt{\frac{\kappa_\rho}{\kappa_\mu}}\langle\mu|S^l_\epsilon|\rho\rangle,
\end{equation}
establishing the link between the quasi-stationary approach and the time-dependent one.

\subsection{Sojourn time on the quasi-energy shell}\label{Sojourn time on the quasi-energy shell}

We are now in a position to study the notion of sojourn time for a time-periodic short-range potential. The first thing we need to observe is that the definition (\ref{sojourn-time-operator})-(\ref{sojourn-time-operator2}) is not affected by a possible time-dependence of the interaction, as is clear from the fact that the probability ${\cal{P}}_{\psi_t}(B_r)$ maintains all its meaning also in this case. 

Considering however relation (\ref{wave operators time-dependent potential tris}), we now have for the sojourn time operator the following expression:
\begin{align}
\label{sojourn-time-operator time dependent potential}
T(B_r)&=\int_{-\infty}^{\infty}dt\, \Omega_-^\dagger U^\dagger(t,0)P_r U(t,0)\Omega_- \nonumber\\
&=\int_{-\infty}^{\infty}dt\, e^{\frac{i}{\hbar}H_0t}\Omega_-^\dagger(t) P_r\Omega_-(t) e^{-\frac{i}{\hbar}H_0t}.
\end{align}
Obviously, we cannot repeat anymore the same argument as in (\ref{free-sojourn-time-operator-commutation with free evolution}), to show that (\ref{sojourn-time-operator time dependent potential}) commutes with $H_0$. However, if the potential is time-periodic, we still have that
\begin{align}
\label{sojourn-time-operator-commutation}
&e^{-\frac{i}{\hbar}H_0T}T(B_r)=\int_{-\infty}^{\infty}dt\, e^{\frac{i}{\hbar}H_0(t-T)}\Omega_-^\dagger(t) P_r\Omega_-(t) e^{-\frac{i}{\hbar}H_0t} \nonumber\\
&=\int_{-\infty}^{\infty}dt\, e^{\frac{i}{\hbar}H_0t}\Omega_-^\dagger(t+T) P_r\Omega_-(t+T) e^{-\frac{i}{\hbar}H_0(t+T)} \nonumber\\
&= T(B_r)e^{-\frac{i}{\hbar}H_0T},
\end{align}
where for the last equality we have used $\Omega_-(t+T)=\Omega_-(t)$. 

In other terms, similarly to the scattering operator, the sojourn time operator commutes with the free evolution over one period, i.e., $[T(B_r),e^{-\frac{i}{\hbar}H_0T}]=0$, and therefore possesses on the quasi-energy shell elements. Consequently, its average over an incoming state $|\varphi\rangle\equiv |\varphi_-\rangle$, that we will assume, for sake of simplicity, being of fixed angular momentum $l$ and $m$, can be written
\begin{align}
\label{sojourn time matrix on quasi energy shell}
&T_\varphi(B_r) = \langle \varphi|T(B_r)|\varphi\rangle =\int_0^{\hbar\omega}d\epsilon\, \langle \varphi(\epsilon)|T_\epsilon(B_r)|\varphi(\epsilon)\rangle\nonumber\\
&=\sum_{\mu,\rho\geq 0}\int_0^{\hbar\omega}d\epsilon\, \varphi^*(\epsilon+\mu\hbar\omega)\langle \mu|T^l_\epsilon(B_r)|\rho\rangle \varphi(\epsilon+\rho\hbar\omega),
\end{align}
where $T_\epsilon^l(B_r)$ is the interaction sojourn time matrix on the quasi-energy shell, at fixed angular momentum. 

To study its behavior as $r\to\infty$, we use (\ref{sojourn-time-operator time dependent potential}), (\ref{Fourier wave operator}) and (\ref{separation of variables}) to write
\begin{align}
\label{sojourn time matrix on quasi energy shell bis}
&\langle \mu|T^l_\epsilon(B_r)|\rho\rangle = 2\pi\hbar \sum_\sigma\langle \epsilon, \mu,l,m|\Omega_{\sigma-\mu}^\dagger P_r\Omega_{\sigma-\rho}|\epsilon,\rho,l,m\rangle\nonumber\\
&=\frac{4m}{\hbar\sqrt{\kappa_\mu\kappa_\rho}}\sum_\sigma\int_0^r ds\, {u_{\sigma\mu}^l(\epsilon,s)}^*u_{\sigma\rho}^l(\epsilon,s),
\end{align}
which is clearly the quasi-stationary generalization of (\ref{on-shell sojourn}). Similarly to what we have done in the stationary case, we then derive the quasi-stationary Schroedinger equation with respect to $\epsilon$, to obtain the identity
\begin{equation}
\label{quasi stationary identity}
\sum_\sigma {u_{\sigma\mu}^l}^*u_{\sigma\rho}^l =\frac{\hbar^2}{2m}\sum_\sigma \partial_s\left(\partial_s {u_{\sigma\mu}^l}^*\partial_\epsilon  u_{\sigma\rho}^l -{u_{\sigma\mu}^l}^* \partial_s\partial_\epsilon u_{\sigma\rho}^l\right),
\end{equation}
that we can use to directly integrate (\ref{sojourn time matrix on quasi energy shell bis}). 

Exploiting the asymptotic form (\ref{asymptotics of regular solutions for periodic potentials}) and the identity (\ref{amplitudes and scattering matrix}), after a long calculation without difficulties, we find, in the limit $r\to\infty$,
\begin{align}
\label{sojourn time matrix on quasi energy shell asymptotics}
&\langle \mu|T^l_\epsilon(B_r)|\rho\rangle = \langle \mu|\tau^l_\epsilon|\rho\rangle \nonumber\\ 
&+ \sum_{\sigma\geq 0} \frac{r}{v_\sigma}\left(\delta_{\sigma\mu}\delta_{\sigma\rho} + \langle \sigma|S^l_\epsilon|\mu\rangle^*\langle \sigma|S^l_\epsilon|\rho\rangle\right)\nonumber\\
&-\sum_{\sigma\geq 0}\frac{\hbar}{2(\epsilon +\sigma\hbar\omega)}\frac{1}{2i}\left[\delta_{\sigma\mu}\langle \sigma|S^l_\epsilon|\rho\rangle e^{i(2\kappa_\sigma r-l\pi)}\right.\nonumber\\
&\left.-\delta_{\sigma\rho}\langle \sigma|S^l_\epsilon|\mu\rangle^*e^{-i(2\kappa_\sigma r-l\pi)}\right]+o(1),
\end{align}
where we have defined the scalar velocities $v_\sigma=\hbar\kappa_\sigma/m$, and $\langle \mu|\tau^l_\epsilon|\rho\rangle$ is the matrix element of the (Eisenbud-Wigner) time-delay operator (\ref{timedelay operator}) on the quasi-energy shell (here restricted to a subspace of fixed angular momentum):
\begin{equation}
\label{Eisenbud Wigner quasi stationary}
\tau^l_\epsilon = -i\hbar {S^l_\epsilon}^\dagger \frac{dS^l_\epsilon}{d\epsilon}, 
\end{equation}
so that
\begin{align}
\label{Eisenbud Wigner quasi stationary bis}
\langle \mu|\tau^l_\epsilon|\rho\rangle &= -i\hbar \sum_{\sigma\geq 0}\langle \mu|{S^l_\epsilon}^\dagger|\sigma\rangle\frac{d}{d\epsilon} \langle\sigma|S^l_\epsilon |\rho\rangle\nonumber\\
&= -i\hbar \sum_{\sigma\geq 0}\langle \sigma|{S^l_\epsilon}|\mu\rangle^*\frac{d}{d\epsilon} \langle\sigma |S^l_\epsilon |\rho\rangle.
\end{align}

\subsection{Time-delay on the quasi-energy shell}\label{Time-delay on the quasi-energy shell}

Expression (\ref{sojourn time matrix on quasi energy shell asymptotics}) is the quasi-stationary generalization of (\ref{on-shell sojourn asymptotic}). An important difference here, with respect to the single channel situation, is that the scattering amplitudes now also contribute to the ``free-flight'' linear term in $r$. 

As we already mentioned, this is due to the fact that the process being not conservative, but only quasi-conservative, the state emerging from the scattering region is a superposition of an infinite number of waves characterized by the different free velocities $v_\sigma=\hbar\kappa_\sigma/m$, $\sigma = 0,1,2,\ldots$ Therefore, seeing that the incoming free reference sojourn time only describes a conservative process, and therefore has on shell asymptotic elements
\begin{align}
\label{incoming free time quasi energy shell}
&\langle \mu|T^{0,l}_\epsilon(B_r)|\rho\rangle = \frac{2r}{v_\rho}\delta_{\mu\rho} \nonumber\\
&-\delta_{\mu\rho}\frac{\hbar}{2(\epsilon +\rho\hbar\omega)}\sin(2\kappa_\rho r-l\pi)+o(1), 
\end{align}
it clearly cannot compensate the $r$-divergent terms in (\ref{sojourn time matrix on quasi energy shell asymptotics}), as $r\to\infty$. 

This means that a definition of time-delay in multichannel scattering employing an incoming free reference  sojourn time would not be consistent. The situation doesn't change if, instead of an incoming, we consider an outgoing free reference sojourn time, with asymptotic elements
\begin{align}
\label{outgoing free time quasi energy shell}
&\langle \mu|{S_\epsilon^l}^\dagger T^{0,l}_\epsilon(B_r)S_\epsilon^l|\rho\rangle = \sum_{\sigma\geq 0}\frac{2r}{v_\sigma}\langle\sigma|S^l_\epsilon|\mu\rangle^*\langle\sigma|S^l_\epsilon|\rho\rangle\nonumber\\
&-\sum_{\sigma\geq 0}\langle\sigma|S^l_\epsilon|\mu\rangle^*\langle\sigma|S^l_\epsilon|\rho\rangle\frac{\hbar}{2(\epsilon +\sigma\hbar\omega)}\sin(2\kappa_\sigma  r-l\pi) +o(1),
\end{align}
as also in this case the totality of the $r$-divergent terms in (\ref{sojourn time matrix on quasi energy shell asymptotics}) cannot be compensated. 

On the other hand, if we consider the symmetric reference time (\ref{free sojourn time classical in/out}), i.e., a symmetric combination of incoming and outgoing states, we obtain a perfect compensation of the linear terms in $r$, so that the symmetrized local time delay has asymptotic form
\begin{align}
\label{symmetrized local time delay quasi stationary}
&\langle\mu|\tau_{\text{s},\epsilon}^l(r)|\rho\rangle \nonumber\\
&=\langle \mu| \left\{T^l_\epsilon(B_r)-\frac{1}{2}\left[T^{0,l}_\epsilon(B_r)+{S_\epsilon^l}^\dagger T^{0,l}_\epsilon(B_r)S_\epsilon^l\right]\right\}|\rho\rangle \nonumber\\
&=\langle \mu|\tau^l_\epsilon|\rho\rangle + \text{oscillating terms} + o(1).
\end{align}

As we discussed for the static (single-channel) case, when averaging over a sufficiently well behaved packet $|\varphi\rangle$, the oscillating terms in (\ref{symmetrized local time delay quasi stationary}) do not contribute to the $r\to\infty$ limit, because of the Riemann-Lebesgue Lemma. Therefore, we obtain that the symmetrized local time-delay duly converges to the multichannel generalization of the Eisenbud-Wigner formula (\ref{Eisenbud Wigner quasi stationary})-(\ref{Eisenbud Wigner quasi stationary bis}). 

To better appreciate the physical content of the latter, we can consider the monochromatic limit of an incoming wave (of fixed angular momentum $l$ and $m$) of energy $\epsilon +n\hbar\omega$, i.e., the limit $|\langle l,m,n',\epsilon'|\varphi\rangle|^2\to \delta_{n'n}\delta(\epsilon' - \epsilon)$. Then, $\tau_{\varphi}\to \langle n|\tau^l_\epsilon|n\rangle$, so that (\ref{Eisenbud Wigner quasi stationary bis}) reduces to   
\begin{equation}
\label{Eisenbud diagonal}
\langle n|\tau^l_\epsilon|n\rangle = \sum_{\sigma\geq 0}|\langle \sigma|S^l_\epsilon|n\rangle|^2\tau_{\epsilon;\sigma,n}^{l},
\end{equation}
where we have defined
\begin{equation}
\label{Eisenbud quasi energy conditional}
\tau_{\epsilon;\sigma,n}^{l}\equiv 2\hbar\frac{d\delta^{l}_{\epsilon;\sigma,n}}{d\epsilon}
\end{equation}
and $\delta^{l}_{\epsilon;\sigma,n}\equiv \frac{1}{2}\arg \langle \sigma|S^l_\epsilon|n\rangle$. 

Now, since $|\langle \sigma|S^l_\epsilon|n\rangle|^2$ is the probability for a scattering with energy transfer $\epsilon +n\hbar\omega\to \epsilon +\sigma\hbar\omega$, the global time-delay (\ref{Eisenbud diagonal}) is a weighted sum expressing a conditional average over the \textit{conditional time-delays} (\ref{Eisenbud quasi energy conditional}), i.e., the time delays for an incoming particle of energy $\epsilon +n\hbar\omega$, conditional to the fact that it will emerge from the interaction region with energies $\epsilon +\sigma\hbar\omega$, $\sigma\geq 0$.

We shall discuss in some detail the notion of conditional time-delay, and the conceptual problems it presents, in the next section. Here we conclude by observing that the above analysis can be easily extended to different typologies of multichannel scattering systems (see~\cite{Mar2} and the references cited therein). Also, we refer the reader to~\cite{Sas,Yaf,Mar5} for a  comprehensive introduction to the general
formalism of time-periodic scattering systems, and for the treatment of the specific one-dimensional case, where the additional transmission and reflection channels can also be distinguished. 

Let us also observe that, as we did in Sec.~\ref{Fuzzy sojourn times}, also for time-periodic potentials we could have defined fuzzy sojourn times, to consistently get rid of the oscillating terms even at fixed quasi-energy. And instead of a symmetric free reference sojourn time, we could as well have used the ``free-flight'' time (\ref{free sojourn time-new}), or its modified version (\ref{free sojourn time-new-fuzzy}), in the fuzzy case.

\section{Conditional time-delay}\label{Contitional time-delay}

In this Section we discuss the notion of \textit{conditional time-delay}. 

As opposed to global time-delay, which doesn't incorporate in its definition any specific condition of observation of the scattering particle, the notion of conditional time-delay answers the more specific question of what is the excess (or defect) of time spent by the scattering particle in the interaction region, \textit{conditional} to the fact that it will be ultimately observed in a given subspace $F\cal H$ of the Hilbert space $\cal H$, as $t\to\infty$, specified by a projector operator $F$, compatible with the free evolution, i.e., $[F,H_0]=0$. 

The conditional time-delay is also a global quantity, in the sense that it is also obtained as the $r\to\infty$ limit (or $r,\rho\to\infty$ limit, in the fuzzy case) of a local conditional time-delay. 

In classical mechanics, the notion of conditional time-delay presents no conceptual problems, as the classical global time-delay is in fact already a conditional time-delay, as is clear from the fact that a particle's final state is uniquely determined by its initial condition in the remote past. 

For instance, in the simple one-dimensional context, if the particle's incoming energy $E$ belongs to $\{E>0|E>\sup_x \textsc{v}(x)\}$, we know it will be finally transmitted, so that the classical global time-delay is \textit{de facto}, in this case, a \textit{transmission time-delay}, i.e., a time-delay conditional to the fact that the particle will be found far away on the  right hand side of the potential in the distant future (assuming it came from the left). 

On the other hand, for incoming energies belonging to $\{E>0|E<\sup_x \textsc{v}(x)\}$, it will be reflected back, and we have in this case a \textit{reflection time-delay}~\footnote{For the special value $E=\sup_x \textsc{v}(x)>0$, all the kinetic energy of the incoming particle is transformed into potential energy, so that it cannot emerge from the interaction region, giving rise to an infinite time-delay. In this case the particle is neither transmitted nor reflected, but instead captured.}.

In the quantum case the situation is different, as there are no means to determine in advance if the particle will be ultimately transmitted or reflected, without performing an appropriate measurement. Such a measurement, like typically all quantum measurements, is not an act of discovery of properties that are already present in the system, but a creation of new properties that are only potential before the measurement, and can possibly become  actual as a result of the interaction between the quantum entity and the measuring apparatus~\cite{Aer, Aer2, Sas3}. 

Also, in quantum mechanics the property of ``being transmitted'' and ``being reflected'' are not compatible with the property of ``being in $B_r$.'' Therefore, one cannot expect to give a proper meaning to the notions of transmission and reflection sojourn times, and more generally to a notion of conditional sojourn time. This however, as we shall see, will not prevent us to give a proper definition for the quantum notion of conditional time-delay. 

We start by observing that the probability ${\cal P}_F(\varphi)$ of finding the scattering state $|\psi_t\rangle =e^{-\frac{i}{\hbar}Ht}\Omega_-|\varphi\rangle$ in the subspace $F{\cal H}$, in the distant future, is given by
\begin{equation}
\label{probability F}
{\cal P}_F(\varphi)=\lim_{t\to\infty} \|\psi_t\|^2 =\lim_{t\to\infty}\|F e^{-\frac{i}{\hbar}H_0t}S\varphi\|^2 = \|FS\varphi\|^2, 
\end{equation}
where for the second equality we have used the asymptotic condition $|\psi_t\rangle\to e^{-\frac{i}{\hbar}H_0t}S|\varphi\rangle$, as $t\to\infty$, and for the last equality the fact that $H_0$ commutes with $F$. 

On the other hand, we can observe that this same probability can be written in the form
\begin{align}
\label{probability F bis}
{\cal P}_F(\varphi)&= \langle\varphi|S^\dagger FS|\varphi\rangle =\langle\varphi|\Omega_-^\dagger\Omega_+ F\Omega_+^\dagger\Omega_-|\varphi\rangle\nonumber\\
&=\langle\varphi|\Omega_-^\dagger e^{\frac{i}{\hbar}Ht}e^{-\frac{i}{\hbar}Ht}\Omega_+ F\Omega_+^\dagger e^{\frac{i}{\hbar}Ht}e^{-\frac{i}{\hbar}Ht}\Omega_-|\varphi\rangle\nonumber\\
&=\langle\psi_t|\Omega_+ F\Omega_+^\dagger |\psi_t\rangle = \| \mathbb{F}\psi_t\|^2.
\end{align}
For the second equality in (\ref{probability F bis}) we have used $S=\Omega_+^\dagger\Omega_-$, for the fourth the intertwining relation $H\Omega_+=\Omega_+H_0$ and the fact that $FH_0 = H_0F$, and in the last equality we have defined the ``asymptotic'' operator $\mathbb{F}\equiv \Omega_+ F\Omega_+^\dagger$, which is clearly an orthogonal projection since $\Omega_+$ is an isometry, i.e., $\Omega_+^\dagger\Omega_+ =\mathbb{I}$. 

What we have just shown is that, in the same way as we have a projector operator $P_r$ associated to the property ``the particle is localized in $B_r$,'' we can also exhibit a genuine projector operator $\mathbb{F}$ associated to the property ``the particle will have the property associated to $F$ in the distant future.''

\subsection{Pseudo joint probabilities}\label{Pseudo joint probabilities}

In classical mechanics these two properties are defined by compatible experimental tests, and therefore can be consistently jointly measured. On the other hand, they are associated to incompatibles tests in quantum mechanics, as is clear from the fact that in general $[P_r,\mathbb{F}]\neq 0$ (see~\cite{Sas2, Muga} and the references cited therein). 

Obviously, this represents an insurmountable obstacle for the definition of a meaningful notion of quantum conditional sojourn time, and this is one of the reasons for the longstanding controversy (that we have mentioned in the introduction) over the countless definitions that have been proposed for a tunneling time. 

The question of how much time a transmitted particle (incoming, say, from the left) spends in a given (finite) spatial region is simply an ill-defined one in standard quantum mechanics (or, better, it is a classical question with no quantum analogue), as it requires the joint measurement of two incompatible projection-observables: $P_r$ and $\mathbb{P}_+^0 \equiv \Omega_+ P_+^0\Omega_+^\dagger$ [with $P_+^0$ defined as in (\ref{one-dim proj-operators})].

The situation  is however different if one considers, instead of the notion of conditional sojourn time, the one of conditional time-delay. To see this, and for sake of clarity, we limit the discussion to the one-dimensional case and to the condition of being transmitted. 

Consider two properties $a$ and $b$, associated to the projections $P_a$ and $P_b$, respectively, and the corresponding inverse properties $\bar{a}$ and $\bar{b}$, associated to the projectors $P_{\bar a}=\mathbb{I}-P_a$ and $P_{\bar b}=\mathbb{I}-P_b$. As we observed in Sec.~\ref{Interferences}, if the properties are incompatibles, i.e., $[P_a,P_b]\neq 0$, then the presence of interference terms prevents us from writing the theorem of total probability of classical probability theory. 

However, let us apply the standard quantum rule that consists in symmetrizing all products of non-commuting observables, to define the operator
\begin {equation}
\label{auxiliary operator}
\tilde{P}_{x,y}=\frac{1}{2}\left(P_xP_y+P_yP_x\right),
\end{equation}
with $x,y\in \{a, b, \bar{a}, \bar{b}\}$. 

Clearly, although the $\tilde{P}_{x,y}$ are self-adjoint, they are not orthogonal projections (this is the reason for the ``tilde'' in the notation). Nevertheless, considering their average over a given state $|\varphi\rangle$, we can define the auxiliary function
\begin {equation}
\label{auxiliary function}
\tilde{\cal P}_\varphi(x,y)=\langle\varphi|\tilde{P}_{x,y}|\varphi\rangle.
\end{equation}
It is then an easy matter to check that it obeys the relations
\begin {eqnarray}
{\cal P}_{\varphi}(b)&=& \tilde{\cal P}_\varphi(a,b) + \tilde{\cal P}_\varphi(\bar{a},b)\nonumber\\
\label{relation}
{\cal P}_{\varphi}(a)&=& \tilde{\cal P}_\varphi(a,b) + \tilde{\cal P}_\varphi(a,\bar{b}),
\end{eqnarray}
which are exactly those we would expect from a genuine joint probability. However, since it can also take negative values, we cannot use (\ref{auxiliary function}) as a bona fide joint probability, to define conditional probabilities
\begin {equation}
\label{conditional probabilities}
\tilde{\cal P}_\varphi(x|y)=\frac{\tilde{\cal P}_\varphi(x,y)}{{\cal P}_{\varphi}(y)},
\end{equation}  
that in turn would allow us to define bona fide quantum conditional sojourn times.

\subsection{Transmission and reflection time-delays}\label{Transmission and reflection time-delays}

In the case of our concern, and for a particle coming from the left, we can let $a$ be the property of being transmitted (that we denote by ``tr''), associated to the projection operator $\mathbb{P}_+^0$, and $b$ the property of being in $B_r$ (that we simply denote ``$B_r$''), associated to the projection operator $P_r$. 

Then, if it wasn't for the fact that it can take negative values, 
\begin {equation}
\label{conditional probability transmission}
\tilde{\cal P}_{\psi_t}(B_r|\text{tr})=\frac{\tilde{\cal P}_{\psi_t}(B_r,\text{tr})}{{\cal P}_{\psi_t}(\text{tr})}=\Re \frac{\langle\psi_t|\mathbb{P}_+^0 P_r|\psi_t\rangle}{\|P_+^0 S\varphi\|^2}
\end{equation} 
would possess all the good structural properties that a probability of finding the particle in $B_r$ at time $t$, conditional to the fact that it will be ultimately transmitted, is expected to obey. Consequently, we could use it to define the following transmission sojourn time
\begin{align}
\label{conditional transmission sojourn time}
\tilde{T}_\varphi(B_r|\text{tr})&=\int_{-\infty}^\infty dt\,\tilde{\cal P}_{\psi_t}(B_r|\text{tr})\nonumber\\
&=\Re \frac{\langle\varphi|S^\dagger P_+^0 S T(B_r)|\varphi\rangle}{\|P_+^0 S\varphi\|^2},
\end{align}
with $T(B_r)$ the sojourn time operator (\ref{sojourn-time-operator2}). And because of (\ref{relation}), we would have the conditional average
\begin{equation}
\label{conditional average}
T_\varphi(B_r) = \|P_+^0 S\varphi\|^2 \tilde{T}_\varphi(B_r|\text{tr}) + \|P_-^0 S\varphi\|^2 \tilde{T}_\varphi(B_r|\text{re}),
\end{equation}
connecting the conditional transmission and reflection sojourn times to the unconditional sojourn time, $\tilde{T}_\varphi(B_r|\text{re})$ being the equivalent of (\ref{conditional transmission sojourn time}) for the reflection case, obtained by replacing $P_+^0$ by $P_-^0$ in (\ref{conditional transmission sojourn time}). 

But the above we cannot do since, as we said, (\ref{conditional probability transmission}) is not positive. However, when considering the notion of time-delay, our interest is in the global limit $r\to\infty$, and in this limit the projection operator $P_r\to \mathbb{I}$; therefore, it becomes compatible with all observables, and in particular with $\mathbb{P}_+^0$. 

This means that in this limit, the pseudo joint probability $\tilde{\cal P}_{\psi_t}(B_r,\text{tr})\to{\cal P}_{\psi_t}(\text{tr})\geq 0$, i.e., it becomes positive and thus recovers a consistent probabilistic interpretation. So, keeping in mind that at the end of the calculation we shall take the limit $r\to\infty$, we can use the above defined pseudo conditional sojourn time as a useful auxiliary function, allowing us to consistently extract information about the conditional time-delay.   

To do this, we have to study the asymptotic of (\ref{conditional transmission sojourn time}). Using (\ref{squared stationary solution2}), and after a long calculation without difficulties, we find that, as $r\to\infty$,
\begin{align}
\label{transmission time}
&\tilde{T}_\varphi(B_r|\text{tr})\nonumber\\
&=\frac{\int_0^\infty dE\, |T_E|^2|\varphi(E)|^2\left[\hbar\frac{d\alpha_E^T}{dE}+\frac{2r}{v}+A_E(r)\right]}{\int_0^\infty dE\, |T_E|^2|\varphi(E)|^2} +o(1),
\end{align} 
where the interference contribution $A_E(r)$ is given by
\begin {equation}
\label{interference contribution transmission time}
A_E(r)=\frac{\hbar}{2E}|L_E|\sin(\alpha_E^L-\alpha_E^T)\cos(\alpha^T_E + 2kr).
\end{equation} 

What is interesting to observe in (\ref{transmission time}) is that the linear term in $r$ is not of the form (\ref{one-dimensional free sojourn time}). Therefore, if we subtract from (\ref{transmission time}) the incoming free sojourn time (\ref{one-dimensional free sojourn time}), which is also a transmission time (as a free particle coming from the left is, by definition, transmitted to the right), then take the limit $r\to\infty$, the linear terms in $r$ do not compensate and the difference diverges. In other terms, a free reference time associated to the incoming particle is not suitable for defining a conditional time-delay. 

On the other hand, if we consider the outgoing state $S|\varphi\rangle$, we have
\begin{align}
\label{conditional free transmission sojourn time}
&\tilde{T}_{\text{out},\varphi}^0(B_r|\text{tr})=\Re \frac{\langle\varphi|S^\dagger P_+^0 T^0(B_r)S|\varphi\rangle}{\|P_+^0 S\varphi\|^2}\nonumber\\
&=\frac{\int_0^\infty dE\, |T_E|^2|\varphi(E)|^2\left[\frac{2r}{v}+A_E^0(r)\right]}{\int_0^\infty dE\, |T_E|^2|\varphi(E)|^2},
\end{align}
where the interference contribution $A_E^0(r)$ is now given by
\begin {equation}
\label{interference contribution transmission time2}
A_E^0(r)=-\frac{\hbar}{2E}|T_E|^2\cos(\alpha_E^L-\alpha_E^T)\sin(2kr).
\end{equation} 

This time (\ref{conditional free transmission sojourn time}) does correctly extract the linear divergence in (\ref{transmission time}), as $r\to\infty$. This is because the free outgoing state being coincident with the scattering state in the distant future, it allows for a same split in terms of transmitted and reflected products. 

So, the pseudo conditional local transmission time-delay
\begin {align}
\label{transmission local time-delay}
&\tilde{\tau}_\varphi(B_r|\text{tr})= \tilde{T}_{\varphi}(B_r|\text{tr})- \tilde{T}_{\text{out},\varphi}^0(B_r|\text{tr})\nonumber\\
&=\Re \frac{\langle\varphi|S^\dagger P_+^0 S \left[T(B_r)-S^\dagger T^0(B_r)S\right]|\varphi\rangle}{\|P_+^0 S\varphi\|^2}
\end{align} 
does conveniently converge to a finite (non-pseudo) global limit, as $r\to\infty$, and considering that the infinitely oscillating terms do not contribute, because of the Riemann-Lebesgue Lemma, we obtain
\begin{align}
\label{transmission time-delay}
\tau_\varphi^{\text{tr}} &= \lim_{r\to\infty}\tilde{\tau}_\varphi(B_r|\text{tr})\nonumber\\
&= \frac{\int_0^\infty dE\, \hbar\frac{d\alpha_E^T}{dE}|T_E|^2|\varphi(E)|^2}{\int_0^\infty dE\, |T_E|^2|\varphi(E)|^2}.
\end{align}

Bearing in mind that in the limit of infinitely extended regions the pseudo joint probability (\ref{conditional probability transmission}) recovers a proper meaning, (\ref{transmission local time-delay}) can be assumed to be a bona fide transmission time-delay, with a proper probabilistic interpretation. 

In the monoenergetic limit (\ref{monoenergetic limit}), the numerator and denominator simplify, and we obtain for the transmission time-delay at fixed energy $E$:
\begin{equation}
\label{transmission time-delay fixed energy}
\tau_E^{\text{tr}} = \hbar\frac{d\alpha_E^T}{dE}.
\end{equation}

Obviously, similar expressions hold for the reflection time-delays from the left and from the right, in terms of the energy-derivatives of the corresponding reflection phases.

Formula (\ref{transmission time-delay fixed energy}) is in perfect agreement with what can be obtained using more heuristic arguments, like the one that consists in following the position of the transmitted peak (by means of a stationary phase argument), or more generally the evolution of the mean position of the transmitted packet. It also agrees with the reading of physical clocks, like for instance the Larmor clock, when the spin precession is considered separately for the transmitted and reflected components~\cite{Hau}. 

However, it is important to emphasize that one cannot use clocks to define a proper transmission (or reflection) sojourn time and overcome the difficulty of the incompatibility of the projection observables $P_r$ and $\mathbb{P}_+^0$. 

In the ambit of the Larmor clock, the shortcoming manifests in the fact that when transmission and reflection amplitudes are considered separately, there is also a change of the spin component parallel to the field direction, so that an unambiguous precession angle cannot anymore be defined, as it was firstly emphasized by B$\ddot{\text{u}}$ttiker~\cite{But}.

\subsection{The most general definition}\label{The most general definition}

In the above, we have considered the specific case of transmission and reflection in the one-dimensional scattering, but the analysis readily generalizes to more general situations, like three-dimensional scattering systems and arbitrary conditions of observation of the scattering particle, specified by a generic projection operator $F$ commuting with the free evolution. 

Then, replacing $P_+^0$ in (\ref{transmission local time-delay}) by $F$, and observing that  the operatorial difference $T(B_r)-S^\dagger T^0(B_r)S$ tends (in the weak topology sense) to $S^\dagger[T_0,S]$, we have for the conditional time-delay the general formula~\cite{Sas2} 
\begin{equation}
\label{conditional time-delay}
\tau_\varphi^{F} = \lim_{r\to\infty}\tilde{\tau}_\varphi(B_r|F) = \Re \frac{\langle\varphi|S^\dagger F [T_0,S]|\varphi\rangle}{\|F S\varphi\|^2},
\end{equation} 
where for a scattering by a static interaction the operator $i[T_0,S]/\hbar$ corresponds to the energy derivative of the scattering matrix, or to its quasi-energy derivative in the time-periodic case.   

We conclude this section observing that we could as well have derived the conditional time-delay limit (\ref{conditional time-delay}) using, instead of the outgoing (pseudo conditional) free sojourn time (\ref{conditional free transmission sojourn time}), the ``free-flight'' one: 
\begin{equation}
\label{free sojourn time conditional free-flight}
\tilde{T}_{\text{ff},\varphi}^0(B_r|F)=r\left[\lim_{r'\to\infty}\frac{\tilde{T}_\varphi(B_{r'}|F)}{r'}\right].
\end{equation}

We want now to synthesize the analysis of the previous sections by providing what we believe is the most general possible definition for time-delay in the context of non-relativistic quantum scattering theory. 

We start by observing that in single channel scattering systems one can equivalently exploit, in the definition of time-delay, incoming and outgoing free reference sojourn times. On the other hand, when dealing with multichannel scattering systems, only a symmetrized version of the free reference time can correctly cancel the linear divergent terms. Finally, when dealing with conditional time-delays, only the outgoing free reference sojourn time allows to duly extract the linear divergence. 

Therefore, if one wants to define a consistent notion of conditional time-delay in a general multichannel context, one apparently falls short of a convenient reference time. This is the reason why we have also defined in this work a new kind of free reference sojourn time, that we have called ``free-flight,'' which has the nice property of remaining fully consistent in all the above mentioned contexts. 

On the other hand, we have observed that it was possible to make the time-delay limit uniform with respect to the choice of the incoming wave-packet, introducing for this the notion of ``fuzzy sojourn time,'' thanks to which the limit of infinitely extended spatial regions can be taken also when the energy of the incoming state is sharply peaked (i.e., at fixed energy). Accordingly, we have shown that it was possible to define a modified ``fuzzy free-flight'' reference time, to suitably extract the divergent terms also in this ambit. 

It is now time to put all this together, in a single and coherent general definition that encompasses all possible cases. Let $(H_0, H)$ be a scattering system, which can be simple, like in two-body scattering processes, or in the scattering of a single particle by a static external field, or multichannel-like, such as in scattering by a time-periodic potential~\cite{Mar2, Yaf, Sas}, $N$-body scattering~\cite{Smi, Mar2, Boll2}, scattering with dissipative interactions~\cite{Mar2, Mar1}, step potential scattering~\cite{Amr and Jac}, scattering in waveguides~\cite{Ald}, and so on. 

Then, if $|\psi_t\rangle = U(t,0)\Omega_-|\varphi\rangle$ is the scattering state at time $t$, and $F$ is a projection operator which is compatible with $H_0$, we define the local, pseudo conditional, fuzzy time-delay as the difference
\begin{equation}
\label{local pseudo conditional fuzzy time-delay}
\tilde{\tau}^\rho_\varphi(B_r|F)= \tilde{T}_\varphi^\rho(B_{r}|F) - f(r,\rho)\left[\lim_{r',\rho'\to\infty}\frac{\tilde{T}_\varphi^{\rho'}(B_{r'}|F)}{f(r',\rho')}\right],
\end{equation}
where the function $f$ is given by (\ref{modified linear term}) and 
\begin{equation}
\label{pseudo conditional fuzzy sojourn time}
\tilde{T}_\varphi^\rho(B_r|F)=\Re \frac{\langle\varphi|S^\dagger F S T^\rho(B_r)|\varphi\rangle}{\|P_+^0 S\varphi\|^2}
\end{equation}
is the auxiliary pseudo conditional fuzzy sojourn time, with
\begin{equation}
\label{fuzzy sojourn-time-operator general}
T^\rho(B_r)=\int_{-\infty}^{\infty}dt\, \Omega_-^\dagger U^\dagger(t,0) P_{r,\rho}U(t,0)\Omega_-
\end{equation}
the fuzzy sojourn time operator, and $P_{r,\rho}$ the multiplication operator by the membership function (\ref{fuzzy characteristic function}). 

Then, for sufficiently well behaved interactions and initial states, one can expect the limit 
\begin{equation}
\label{conditional time-delay general}
\tau_\varphi^F = \lim_{r,\rho\to\infty} \tilde{\tau}^\rho_\varphi(B_r|F)
\end{equation}
to exist and to be equal to the conditional time-delay formula
\begin{equation}
\label{conditional time-delay formula}
\tau_\varphi^F = \Re \frac{\langle\varphi|S^\dagger F [T_0,S]|\varphi\rangle}{\|F S\varphi\|^2},
\end{equation}
which, in the unconditional case $F=\mathbb{I}$, reduces to the usual Eisenbud-Wigner formula
\begin{equation}
\label{unconditional time-delay formula}
\tau_\varphi = \langle\varphi|S^\dagger [T_0,S]|\varphi\rangle.
\end{equation}

Of course, as we explained in the previous sections of this work, (\ref{conditional time-delay formula}) cannot be the most general formula, but only a special case corresponding to the choice of balls $B_r\equiv B_r(0)$ centered at the origin. 

Indeed, if in the time-delay limit we choose balls $B_r(\textbf{c})$, centered at an arbitrary spatial point $\textbf{c}\neq 0$, then, instead of (\ref{conditional time-delay formula}), we obtain the more general translation-invariant formula 
\begin{equation}
\label{quantum translated time-delay conditional}
\tau_\varphi^F(\textbf{c}) = \tau_\varphi^F + \textbf{c}\, \Re \frac{\langle \varphi|\frac{m}{|\textbf{p}|}S^\dagger FS\left(S^\dagger \hat{\textbf{p}}S-\hat{\textbf{p}}\right)|\varphi\rangle}{\|F S\varphi\|^2},
\end{equation}
which is clearly the conditional generalization of (\ref{quantum translated time-delay}).

An important property of the local quantity (\ref{local pseudo conditional fuzzy time-delay}) is that the limit (\ref{conditional time-delay general}) is expected to exist also for incoming wave-packets sharply peaked in energy. In other terms, it has the nice property of commuting with the monoenergetic limit (\ref{monoenergetic limit}):
\begin{equation}
\label{conditional time-delay general2}
\lim_{|\varphi|^2\to\delta}\lim_{r,\rho\to\infty} \tilde{\tau}^\rho_\varphi(B_r|F) = \lim_{r,\rho\to\infty} \lim_{|\varphi|^2\to\delta}\tilde{\tau}^\rho_\varphi(B_r|F).
\end{equation}

We conclude this section by showing how to detail (\ref{conditional time-delay formula}) in a specific example. We consider the one-dimensional scattering by a time-periodic potential, for a particle coming from the left, i.e., $P_+^0|\varphi\rangle = |\varphi\rangle$, whose state has energetic support in the interval $[m\hbar\omega,m\hbar\omega + \Delta E]$, with $\Delta E < \omega$. 

We ask what is the  particle's time-delay conditional to the fact that it will be ultimately transmitted with an energy transfer of exactly $n$ quanta of energy $\hbar\omega$. 

To answer this question we need to remember that the scattering process conserves the quasi-energy, so that if the outgoing state is observed with energy in the interval $I\equiv [(m+n)\hbar\omega,(m+n)\hbar\omega +\Delta E]$, we know that during the scattering the particle experienced a transfer of energy of exactly $n\hbar\omega$ quanta~\footnote{In fact, it is possible to decompose in very general terms the scattering operator as an infinite  sum of so-called ``sideband'' contributions, each one describing a specific scattering event with an emission (or absorption) of a given number of quanta of energy. We refer the interested reader to~\cite{Mar5}.}. 

Therefore, considering the projection operator
\begin{equation}
\label{Projection operator for transmission probability + energy transfer}
P_{+,I}^0\equiv\int_0^{\Delta E} d\epsilon\, |\epsilon,m+n,+\rangle\langle \epsilon,m+n,+|,
\end{equation}
we have that the probability of being transmitted with an energy transfer of $n\hbar\omega$ is given by
\begin{equation}
\label{transmission probability + energy transfer}
\|P_{+,I}^0S\varphi\|^2=\int_0^{\Delta E} d\epsilon\, |T_{\epsilon,n}|^2 |\varphi(\epsilon)|^2,
\end{equation}
where we have defined $T_{\epsilon,n}\equiv \langle m+n, +|S_\epsilon|m,+\rangle$, and $\varphi(\epsilon)\equiv \langle \epsilon,m,+|\varphi\rangle$. 

Thus, replacing $F$ with $P_{+,I}^0$ in (\ref{conditional time-delay formula}), we obtain
\begin{align}
\label{transmission time-delay + energy transfer}
\tau_{\varphi,n}^{\text{tr}}&=\Re \frac{\langle\varphi|S^\dagger P_{+,I}^0 [T_0,S]|\varphi\rangle}{\|P_{+,I}^0 S\varphi\|^2}\nonumber\\
&= \frac{\int_0^{\Delta E} d\epsilon\, \hbar\frac{d\alpha^{T}_{\epsilon,n}}{d\epsilon} |T_{\epsilon,n}|^2|\varphi(\epsilon)|^2}{\int_0^{\Delta E} d\epsilon\, |T_{\epsilon,n}|^2 |\varphi(\epsilon)|^2},
\end{align}
where $\alpha^{T}_{\epsilon,n} = \arg T_{\epsilon,n}$, and we have used the fact that $\langle \epsilon, n, +|T_0|\varphi\rangle = -i\hbar d/d\epsilon \langle \epsilon, n, +|\varphi\rangle$. 

Then, if we consider the limit of an incoming packet sharply peacked in energy, i.e., $|\varphi(\epsilon')|^2\to\delta(\epsilon'-\epsilon)$, we obtain that $\tau_{\varphi,n}^{\text{tr}}\to\tau_{\epsilon,n}^{\text{tr}}$, where
\begin{equation}
\label{transmission time-delay + energy transfer fixed quasi-energy}
\tau_{\epsilon,n}^{\text{tr}}=\hbar\frac{d\alpha^{T}_{\epsilon,n}}{d\epsilon}.
\end{equation}

Formula (\ref{transmission time-delay + energy transfer fixed quasi-energy}) is the multichannel generalization of (\ref{transmission time-delay fixed energy}), and is in full agreement with what can be derived by means of more heuristic approaches, like for instance the one consisting in applying a stationary phase argument to the outgoing scattering state~\cite{Sas}.

\section{A conceptual upgrade}\label{Discussion}

In the present work we have reviewed different aspects of the fundamental notion of time-delay in classical and quantum mechanics. In doing so, we have also introduced some new concepts, like the energy-clock, the fuzzy sojourn time operator and the free-flight reference time. 

In order to limit the length of the article, we clearly had to make choices, leaving out some important concepts related to time-delay, like for instance the fundamental notion of \textit{total time-delay}, obtained by taking the trace of the energy shell time-delay operator~\cite{Mar2}, which can be proven to be independent of the choice of the center of the expanding balls, and more generally to exist also when the time-delay limit is considered for arbitrary sequences of dilated regions. 

Also, we haven't discussed the important link between the notion of time-delay and Levinson's theorem, its connection to causal bonds, to the virial, and the form it takes at resonances (apart our brief mention in Sec.~\ref{Breit-Wigner}) or at low energy~\cite{Mar2, Sassoli}. 

In this last section, we want to present some final considerations regarding the conceptual status of the quantum notions of sojourn and delay times. As we have seen, time-delay can be consistently defined in standard quantum mechanics as a difference of sojourn times, in the limit of infinitely extended spatial balls. And sojourn times, as we explained in some detail in Sec.~\ref{Sojourn time operators}, are average quantities, defined as sums over probabilities of presence which, being pure probabilistic statements, remain consistent also when a classical notion of trajectory is not available.

\subsection{Non-spatiality}\label{Non-spatiality}

An interesting aspect is that one can exploit the very notion of sojourn time to understand in what sense the classical notion of trajectory wouldn't be anymore available in quantum physics. For this, consider a free particle, described by the state $|\varphi_t\rangle$, and a time-interval $[t_1,t_2]$, with $-\infty\leq t_1 < t_2\leq\infty$. Then, 
\begin{align}
\label{free sojourn-time finite time}
T^0_\varphi(B_r;[t_1,t_2]) &= \int_{t_1}^{t_2}dt\, {\cal P}_{\varphi_t}(B_r) =\langle \varphi |T^0(B_r;[t_1,t_2])|\varphi\rangle\nonumber\\
&\leq t_2 - t_1
\end{align} 
is the average time spent by the particle inside $B_r$, during the time-interval $[t_1,t_2]$, with 
\begin{equation}
\label{sojourn-time operator finite time}
T^0(B_r;[t_1,t_2]) = \int_{t_1}^{t_2}dt\, e^{\frac{i}{\hbar}H_0t}P_r e^{-\frac{i}{\hbar}H_0t}
\end{equation}
the free sojourn time operator restricted to the time-interval $[t_1,t_2]$, which, as we know, is a bona fide self-adjoint operator~\cite{Jaw}. 

Thus, according to the spectral theorem, we also know that there exist a projection-valued measure $F^0(B_r;[t_1,t_2]; \cdot)$, such that (\ref{sojourn-time operator finite time}) can be written in the diagonal form 
\begin{equation}
\label{sojourn-time operator finite time, spectral resolution}
T^0(B_r;[t_1,t_2]) = \int_{\mathbb{R}} F^0(B_r;[t_1,t_2]; dt)\, t.
\end{equation}

Now, although $T^0(B_r;[t_1,t_2])$ is  self-adjoint, and is therefore a well-defined observable, it is certainly not an observable in the conventional sense, as it doesn't correspond to an instantaneous measurement, but, rather, to a continuous measurement in the limit of zero-disturbance, as we have seen in Sec.~\ref{Physical clocks}, where we have shown how to implement the sojourn time measurement by means of physical clocks, in the zero-field limit. 

However, there are no a priori reasons not to extend the usual Born rule also to $T^0(B_r;[t_1,t_2])$, taking seriously its interpretation as a time of sojourn (or of permanence) observable, and therefore interpret the associated projection-valued measure in the usual probabilistic sense. 

More precisely, given a (Borel) subset $\Delta \subseteq \mathbb{R}$, $F^0(B_r,[t_1,t_2];\Delta)$ is to be interpreted as the projection operator into the set of states that, in the course of their free evolution, spend inside $B_r$, during the time interval $[t_1,t_2]$, amounts of time whose values are in $\Delta$. 

In other terms,
\begin{equation}
\label{probability projection valued measure}
{\cal P}_{\varphi}(B_r;[t_1,t_2];\Delta) =\langle \varphi |F^0(B_r;[t_1,t_2];\Delta)|\varphi\rangle
\end{equation}
is the probability that the quantum free evolving particle, described by the initial state $|\varphi\rangle$, sojourns in $B_r$, during the time-interval $[t_1,t_2]$, an amount of time whose value is in the set $\Delta$. 

Therefore, setting $\Delta =\{0\}$, ${\cal P}_{\varphi}(B_r;[t_1,t_2];\{0\})$ has to be understood as the probability for the free particle, during the time-interval $[t_1,t_2]$, of spending a \textit{zero} amount of time in $B_r$. Said it differently, it corresponds to the probability for the particle to not enter, for any measurable amount of time, the spatial region $B_r$. 

The puzzling result that was proved by Jaworski~\cite{Jaw}, is that for any choice of $|\varphi\rangle$ and time-interval $[t_1,t_2]$, such a probability is always equal to zero. In other terms, there are no eigenstates of the sojourn time operator (\ref{sojourn-time operator finite time}) corresponding to the zero eigenvalue. This means that the particle will always spend (with probability $1$) some time in $B_r$, during whatever time-interval $[t_1,t_2]$, and this independently of the choice of its initial condition.

So, if we take seriously the interpretation of (\ref{free sojourn-time finite time}) as a measure of the time spent by the particle inside $B_r$, and if we assume that the particle is a \textit{spatial} entity, that is, an entity existing and evolving inside our three-dimensional Euclidean space, we are faced with an apparent paradox. Indeed, if the particle is a \textit{local} corpuscle, then, by taking a ball $B_r$ of arbitrary small radius $r$, a time-interval $[t_1,t_2]$ with $t_2$ arbitrary close to $t_1$, and an initial state at time $t_1$ localized at an arbitrary astronomical distance far away from the origin, we clearly expect that, however strange, erratic and speedy would be the free particle displacements in space, under these conditions the time it spends in $B_r$, during the infinitesimal time-interval $[t_1,t_2]$, should be equal to zero. But, as we said, this expectation is false, and therefore the hypothesis that the quantum particle is a \textit{local} entity is not tenable. 

Then, let us assume that, on the contrary, it is a \textit{non-local} entity, i.e., an entity that, somehow, is spread all over space. In this case it becomes relatively easy to understand why zero cannot be an eigenvalue of the sojourn time operator, as the quantum particle would be able to be  present, in every moment, in every region of space. 

However, setting $t_1=-\infty$ and $t_2=\infty$, we would expect in this case the sojourn time (\ref{free sojourn-time finite time}) to be always infinite. But again, we know this is not the case, as it is a bounded operator.

Considering the above, we must conclude that the crucial point is not about the locality or non-locality of the quantum particle, but about its presumed \textit{spatiality}. The only possible conclusion is that if a microscopic particle can manifest as a non-local entity, it is because it is first of all a \textit{non-spatial} entity, i.e., an entity that sojourns most of its time in a space that is not our ordinary three-dimensional Euclidean space~\cite{Aer, Aer2, Sas3}. 

For this reason, a microscopic quantum entity shouldn't be called ``particle,'' as to be such it should possess at least the attribute of spatiality. In fact, a quantum particle doesn't possess many other fundamental attributes usually associated to a particle, like for instance the one of individuality~\cite{Sas3}. 

If a quantum entity doesn't possess, in general, a position in space, as ``having a spatial position'' is just a property (most of the time ephemeral) that is created during a measurement process, it is clearly improper to refer to (\ref{free sojourn-time finite time}) as a sojourn (or permanence, residence, transit) time, as the term ``sojourn'' refers to the property of remaining (or sojourning) in the \textit{spatial} region $B_r$, whereas the quantum entity is a \textit{non-spatial} entity, that is, an entity that doesn't sojourn in physical space! 

But then, if we nevertheless consider that the self-adjoint observable (\ref{sojourn-time operator finite time}) is telling us something about the reality of the quantum world, what is it exactly? In other terms, how should we interpret the sum (\ref{free sojourn-time finite time})?

\subsection{Total availability}\label{Total availability}

Following Aerts' terminology~\footnote{The conceptual language developed by Aerts in his ``creation-discovery view'' is much richer and subtler than what can be appreciated by our brief mention here, and we refer the interested reader to~\cite{Aer, Aer2} for a complete exposition of it.}, we can say that our reality consists of all those entities that are \textit{available} to us, in the sense of being available to our \textit{experiments} (and more generally to our experiences), which are essentially \textit{creation-discovery} processes. 

Typically, what we call classical observations, are experiences of pure discovery (i.e., of discovery of what is already manifest), whereas quantum observations are experiences of pure creation (i.e., of creation of what isn't manifest prior to the observation). And in between these two limit cases, we have all kind of possible ``quantum-like'' intermediary observational processes, where both aspects of creation and discovery can be simultaneously present. 

In the case of our concern, the entity in question is a quantum entity (a microscopic pseudoparticle) and the creative aspect of the experience is the one of manifesting a spatial localization, by interacting with a local macroscopic measuring apparatus.  

The important point  to be emphasized here is that, contrary to the case of a classical object, the spatial localization of the quantum entity doesn't exist prior to the observational process (or it exists, but only in a potential sense). Consequently, when measuring the spatial localization of the non-spatial quantum entity, we may or may not succeed in manifesting its presence, i.e., its temporary spatial existence. And the \textit{relative frequency of success} with which we can do this is obviously a measure of the (degree of) \textit{availability} of the quantum entity in participating in such a spatial experience and produce a successful result. 

More precisely, the probability ${\cal P}_{\psi_t}(B_r)$ has to be understood as a measure of the (degree of) \textit{availability}, at time $t$, of the non-spatial quantum entity described by the state $|\psi_t\rangle$, in lending itself to an interaction  with a measuring apparatus in order to manifest (i.e., to create) a temporary spatial localization inside the region $B_r$. 

Therefore, the proper interpretation of the sum 
\begin{equation}
\label{total spatial availability}
T_\varphi(B_r) = \int_{-\infty}^\infty dt\, {\cal P}_{\psi_t}(B_r)
\end{equation}
is the following: it is not the time spent, on average, by the scattering particle inside $B_r$, but its \textit{total availability} in $B_r$, that is, its \textit{total availability in lending itself to the creation of a spatial localization inside $B_r$, by means of an interaction with a measuring apparatus}~\cite{Sass-avail}.

This means that the classical concept of time of sojourn, or time of permanence, has to be replaced by the more general quantum concept of \textit{total availability}, that is, the total availability of a quantum entity in being part of an experience the outcome of which is the creation of a temporary localization in a given region of space. 

This also means that the classical concept of time-delay, which is the difference of the total availability in $B_r$ between an interacting and free entity, in the limit $r\to\infty$, has to be interpreted as the \textit{total (spatial) availability shift} experienced by the quantum entity, as a consequence of the interaction~\cite{Sass-avail}. 

In conclusion, if it is true that our conception of time is dependent upon our classical observation of macroscopic entities moving along trajectories in the three-dimensional physical space, and if it is also true, as hypothesized by Aerts, that~\cite{Aer}: ``[$\cdots$] quantum entities are not permanently present in space, and that, when a quantum entity is detected in such a non-spatial state, it is `dragged' or `sucked up' into space by the detection system,'' then we are forced to recognize that time-concepts like ``time of sojourn,'' ``time of permanence,'' ``duration,'' and so on, are classical notions that need to be upgraded in order to remain fully consistent also in relation to non-spatial quantum entities.

\begin{acknowledgements}
I dedicate this article to Philippe A. Martin, from whom I had the pleasure to learn (when in the nineties I was his student at the EPFL, in Lausanne) the theory of time-delay, one of the traditional subjects of research of the Swiss school of mathematical physics; a subject that Philippe, in turn, was able to learn directly from one of its founders: Josef M. Jauch.
\end{acknowledgements}


\begin{thebibliography}{58}

\bibitem{Wig} E. P. Wigner, ``Lower Limit for the Energy Derivative of the Scattering Phase Shift,'' Phys. Rev. \textbf{98}, No. 1, 145--147 (1955).

\bibitem{Eis} L. Eisenbud, ``PhD. thesis, Princeton University,'' (unpublished) (1948).

\bibitem{Smi} F. T. Smith, ``Lifetime Matrix in Collision Theory,'' Phys. Rev. \textbf{118}, 349--356 (1960). 

\bibitem{Gold} M. Goldberger and K. M. Watson, ``Collision Theory,'' John Wiley Sons, New York (1964).

\bibitem{Jau-Mar} J. M. Jauch and J.-P. Marchand, ``The Delay Time Operator for Simple Scattering Systems,'' Helv. Phys. Acta \textbf{40}, 217--229 (1967).

\bibitem{Jau-Sin-Mis} J. M. Jauch, K. B. Sinha, B. N. Misra, ``Time-Delay in Scattering Processes,'' Helv. Phys. Acta \textbf{45}, 398--426 (1972).

\bibitem{Bol-Osb} D. Boll\'e and T. A. Osborn, ``Equivalence between time-dependent and time-independent formulations of time delay,'' Phys. Rev. D \textbf{11}, No. 12, 3417--3423 (1975).

\bibitem{Mar2} Ph. A. Martin, ``Time delay of quantum scattering processes,'' Acta Phys. Austriaca, Suppl. XXIII, 157--208 (1981).

\bibitem{AC} W. O. Amrein and M. B. Cibils , ``Global and Eisenbud-Wigner time delay in scattering theory,'' Helv. Phys. Acta \textbf{60}, 481--500 (1987).

\bibitem{ACS} W. O. Amrein, M. B. Cibils and K. B. Sinha,  ``Configuration space properties of the S-matrix and time delay in potential scattering,'' Ann. Inst. Henri Poincaré \textbf{47}, No. 4, 367--382 (1987).

\bibitem{Sas1} M. Sassoli de Bianchi and Ph. A. Martin, ``On the definition of time delay in scattering theory,'' Helv. Phys. Acta \textbf{65}, 1119--1126 (1992). 

\bibitem{Ger} C. G\'erard and R. Tiedra de Aldecoa, ``Generalized definition of time delay in scattering theory,'' J. Math. Phys. \textbf{48}, 122101 (2007).

\bibitem{Rich} S. Richard and R. Tiedra de Aldecoa, ``Time delay is a common feature of quantum scattering theory,'' eprint arXiv:1008.3433 (2010)

\bibitem{Sale} H. Salecker and E. P. Wigner, ``Quantum Limitations of the Measurement of Space-Time Distances,'' Phys. Rev. \textbf{109}, No. 2, 217--228 (1958).

\bibitem{Baz} A. I. Baz', ``Lifetime of intermediate states,'' Sov. J. Nucl. Phys. \textbf{4}, 182--188 (1967).

\bibitem{Ryba} V. F. Rybachenko, ``Time of penetration of a particle through a potential barrier,'' Sov. J. Nucl. Phys. \textbf{5}, 635--639 (1967).

\bibitem{Har} T. E. Hartman, ``Tunneling of a wave packet,'' J. Appl. Phys. \textbf{33}, No. 12, 3427--3433 (1962).

\bibitem{Hau} E. H. Hauge and J. A. Stovneng, ``Tunneling times: a critical review,'' Rev. Mod. Phys. \textbf{61}, No. 4, 917--936 (1989).

\bibitem{Olk} V. S. Olkhovsky and E. Recami, ``Recent developments in the time analysis of tunneling processes,'' Phys. Rep. \textbf{214}, No. 6, 339--356 (1992).

\bibitem{Olk2} V. S. Olkhovsky and E. Recami, ``Time as a Quantum Observable, Canonically Conjugated to Energy, and Foundations of Self-Consistent Time Analysis of Quantum Processes,'' Advances in Mathematical Physics, \textbf{2009}, 83 pages (2009). Doi:10.1155/2009/859710.

\bibitem{Olk3} V. S. Olkhovsky, E. Recami and J. Jakielc, ``Unified time analysis of photon and particle tunnelling,'' Phys. Rep. \textbf{398}, No. 3, 133--178 (2004).

\bibitem{Rec} E Recami, V. S. Olkhovsky and S. P. Maydanyuk, ``On non-self-adjoint operators for observables in quantum mechanics and quantum field theory,'' International Journal of Modern Physics A, \textbf{25}, No. 9, 1785--1818 (2010).

\bibitem{Rec2} V. S. Olkhovsky ``On time as a quantum observable canonically conjugate to energy,'' Phys. Usp. \textbf{54} (8) (2011).

\bibitem{Lan} R. Landauer and Th. Martin, ``Barrier interaction time in tunneling,'' Rev. Mod. Phys. \textbf{66}, No. 1, 217--228 (1992).

\bibitem{Muga} J. Muga \textit{et al.} (Eds.), ``Time in Quantum Mechanics - Vol. 1,'' Lect. Notes Phys. \textbf{734}, Springer, Berlin Heidelberg, Second Edition (2008).

\bibitem{Muga-vol2} J. Muga \textit{et al.} (Eds.), ``Time in Quantum Mechanics - Vol. 2,'' Lect. Notes Phys. \textbf{789}, Springer, Berlin Heidelberg, (2009).

\bibitem{JawWard} W. Jaworski and D. Wardlaw, ``Time delay in tunneling: transmission and reflection time delays,'' Phys. Rev. A \textbf{37}, 2843--2854 (1988).

\bibitem{JawWard2} W. Jaworski and D. Wardlaw, ``Time delay in tunneling: Sojourn-time approach versus mean-position approach,'' Phys. Rev. A \textbf{38}, No. 10, 5404--5407 (1988).

\bibitem{JawWard3} W. Jaworski and D. Wardlaw, ``Sojourn time, sojourn time operators, and perturbation theory for one-dimensional scattering by a potential barrier,'' Phys. Rev. A \textbf{40}, No. 11, 6210--6218 (1989).

\bibitem{JawWard4} W. Jaworski and D. Wardlaw, ``Connection between complex interaction times and the sojourn-time operator,'' Phys. Rev. A \textbf{43}, No. 9, 5137--5140 (1991).

\bibitem{Sas2} M. Sassoli de Bianchi, ``Conditional time delay in scattering theory,'' Helv. Phys. Acta \textbf{66}, 361--377 (1993). 

\bibitem{Mar3} Ph. A. Martin and M. Sassoli de Bianchi, ``On the theory of Larmor clock and time delay,'' J. Phys. A: Math. Gen. \textbf{25}, 3627--3647 (1992).

\bibitem{Narnhofer} H. Narnhofer, ``Another definition for time delay,'' Phys. Rev. D \textbf{22}, 2387--2390 (1980).

\bibitem{Temple} This rule is however not without difficulties, see for instance: G. Temple, Nature, \textbf{135}, p. 957 (1935). 

\bibitem{Hilgevoord} J. Hilgevoord, ``Time in quantum mechanics,'' Am. J. Phys. \textbf{70}, 301--306 (2002).

\bibitem{Allc} G. R. Allcock, ``The time of Arrival in Quantum Mechanics,'' Ann. of Phys. \textbf{53}, 253--285 (1969).

\bibitem{Wern} R. Werner, ``Screen observables in relativistic and nonrelativistic quantum mechanics,'' J. Math. Phys. \textbf{27}, No. 3, 793--803 (1986).

\bibitem{Kong} K. Kong Wan, R. H. Fountain, Z. Y. Tao, ``Observables, maximal symmetric operators and POV measures in quantum mechanics,'' J. Phys. A  \textbf{28}, 2379--2393 (1995).

\bibitem{Muga-Lea} J.G. Muga, C.R. Leavens ``Arrival time in quantum mechanics, '' Physics Reports \textbf{338}, 353--438 (2000).

\bibitem{AJS} W. O. Amrein, J. M. Jauch and K. B. Sinha,  \textsl{Scattering Theory in Quantum Mechanics}, Benjamin, Reading (1977)

\bibitem{Je} A. Jensen, ``Time-delay in potential scattering theory,'' Comm. Math. Phys. \textbf{82}, (1981), 435--456. 

\bibitem{Martin-spin} Ph. A. Martin and M. Sassoli de Bianchi, ``Spin Precession Revisited,'' Found. of Phys. \textbf{24}, N. 10, 1371--1378 (1994).

\bibitem{JawWard5} W. Jaworski and D. Wardlaw, ``Sojourn-time operator approach to interaction time in quantum scattering: General formulation for arbitrary scattering systems,'' Phys. Rev. A \textbf{45}, 292--306 (1992).

\bibitem{Gol} R. Golub et al., ``A modest proposal concerning tunneling times,'' Phys. Lett. A \textbf{148}, 27--30 (1990).

\bibitem{Mar1} Ph. A. Martin, ``Scattering theory with dissipative interaction and time delay,'' Il Nuovo Cimento \textbf{30B}, N. 2, 217--238 (1975).

\bibitem{Jaw} W. Jaworski, ``The concept of a time-of-sojourn operator and spreading of wave packets,'' J. Math. Phys. \textbf{30}, 1505--1514 (1989).

\bibitem{Lavine} R. Lavine, ``Absolute continuity of positive spectrum for Schr\"odinger operators with long-range potentials,'' J. Funct. Anal. \textbf{12}, 30--45 (1973).

\bibitem{Dambo} J. A. Damborenea, I. L. Egusquiza, J. G. Muga, ``Asymptotic behaviour of the probability density in one dimension,'' Am. J. Phys. \textbf{70}, 738--740 (2002).

\bibitem{Farina} J. E. G. Farina, ``An elementary approach to quantum probability,'' Am. J. Phys. \textbf{61}, 466--468 (1992).

\bibitem{Mar4} Ph. A. Martin, ``On the time-delay of simple scattering systems,'' Comm. Math. Phys. \textbf{47}, No. 3, 221--227 (1976)

\bibitem{Bol-Osb2} D. Boll\'e and J. D'Hondt, ``On the Hilbert-space approach to classical time delay,'' J. Phys. A: Math. Gen. \textbf{14}, 1663--1674 (1981).

\bibitem{Narn-Thir} H. Narnhofer and W. Thirring, ``Canonical scattering transformations in classical mechanics,'' Phys. Rev. A \textbf{23}, No. 4, 1688--1697 (1981).

\bibitem{OlkA} V. S. Olkhovsky, Sov. J. Part. Nucl. \textbf{15}, 130 (1984); Nukleonika \textbf{35}, 99 (1990); Mysteries, Puzzles and Paradoxes in Quantum Mechanics, ed. R. Bonifaccio, pp. 272–-276 (AIP, 1998).

\bibitem{OlkB} V.S.Olkhovsky, E.Recami and A. I. Gerasimchuk,``Time Operator in  Quantum Mechanics. I: Nonrelativistic case,'' Il Nuovo Cimento \textbf{22} A, No. 2, 263--278 (1974).

\bibitem{Ver} B. J. Verhaar et al., ``On the lifetime of the intermediate system in quantum mechanical collisions,'' Physica \textbf{91A}, 119--132 (1978)

\bibitem{Fal} J. P. Falck and E. H. Hauge, ``Larmor clock re-examined,'' Phys. Rev. B \textbf{38}, 3287--3297 (1988)

\bibitem{Sas} D. S. Saraga and M. Sassoli de Bianchi, ``On the One-Dimensional Scattering by Time-Periodic Potentials: General Theory and Application to Two Specific Models,'' Helv. Phys. Acta \textbf{70}, 751--779 (1997).

\bibitem{Yaf} D. R. Yafaev, ``On the quasi-stationary approach to scattering for perturbation periodic in time,'', in \textit{Recent Developments in Quantum Mechanics}, edited by A. Boutet de Monvel et al., Kluwer Academic Publishers, Netherlands, 367--380 (1991).

\bibitem{Mar5} Ph. A. Martin and M. Sassoli de Bianchi, ``On the low- and high-frequency limit of quantum scattering by time-dependent potentials,'' J. Phys. A: Math. Gen. \textbf{28}, 2403--2427 (1995).

\bibitem{Aer} D. Aerts, ``The Stuff the World is Made of: Physics and Reality,'' p. 129, in ``The White Book of `Einstein Meets Magritte','' Edited by Diederik Aerts, Jan Broekaert and Ernest Mathijs, Kluwer Academic Publishers, Dordrecht, 274 pp. (1999). 

\bibitem{Aer2} D. Aerts, ``The entity and modern physics: the creation-discovery view of reality,'' p. 223, in ``Interpreting Bodies, Classical and Quantum Objects in Modern Physics,'' Edited by Elena Castellani, Princeton University Press, Princeton, 332 pp. (1998). 

\bibitem{Sas3} M. Sassoli de Bianchi, ``Ephemeral Properties and the Illusion of Microscopic Particles,'' Foundations of Science, 16, No. 4 pp. 393--409 (2011); doi: 10.1007/s10699-011-9227-x. An Italian translation of the article is also available: `` Propriet\'a effimere e l'illusione delle particelle microscopiche,'' AutoRicerca, Volume 2, pp. 39--76 (2011).

\bibitem{But} M B$\ddot{\text{u}}$ttiker, ``Larmor precession and the traversal time for tunneling,'' Phys. Rev. B \textbf{27}, N. 10, 6178--6188 (1983).

\bibitem{Boll2} D. Boll\'e and T. A. Osborn, ``Time delay in $N$-body scattering,'' J. Math. Phys. \textbf{20}, N. 6, 1121--1134 (1979).

\bibitem{Amr and Jac} W. O. Amrein and P. Jacquet, ``Time delay for one-dimensional quantum systems with steplike potentials,'' Phys. Rev. A \textbf{75}, N. 2, 022106 (2007)

\bibitem{Ald} R. Tiedra de Aldecoa, ``Time delay and short-range scattering in quantum waveguides,'' Ann. Henri Poincar\'e \textbf{7}, N. 1, 105--124 (2006). (Note that when dealing with scattering in waveguides, $P_r$ is not anymore the projection onto the set of states localized in the ball $B_r$, of radius $r$, but in a cylinder $\Omega_r = \Sigma \times [-r,r]$, where $\Sigma$ is a given bounded open connected set.)

\bibitem{Sassoli} M. Sassoli de Bianchi, ``Levinson's theorem, zero-energy resonances, and time-delay in one-dimensional scattering systems,'' J. Math. Phys. \textbf{35}, N. 6, 2719--2733 (1994).

\bibitem{Sass-avail} M. Sassoli de Bianchi, ``From permanence to total availability: a quantum conceptual upgrade,'' to appear in Foundations of Science, doi: 10.1007/s10699-011-9233-z.

\end{thebibliography}
\end{document}